\documentclass[physics , phd , 12pt , final]{uleththesis}

\usepackage{amsfonts}
\usepackage{amsmath}
\usepackage{amssymb}
\usepackage{amstext}
\usepackage{amsthm}
\usepackage{array}
\usepackage{bm}
\usepackage{booktabs}
\usepackage{braket}
\usepackage[english]{babel}
\usepackage{cancel}
\usepackage[tableposition=top]{caption}
\usepackage[usenames,dvipsnames]{color}
\usepackage{colortbl}
\usepackage{dcolumn}
\usepackage{enumerate}
\usepackage{float}    \floatstyle{plaintop} \restylefloat{table} 
\usepackage{graphicx}
\usepackage{hyperref}
\usepackage{latexsym}
\usepackage[mathlines]{lineno}
\usepackage{listings}
\usepackage{marvosym}
\usepackage{multirow}
\usepackage{slashed}
\usepackage{soul}
\usepackage{subfigure}
\usepackage{tabularx}
\usepackage{tikz} 
\usepackage{tikz-feynman}
\tikzfeynmanset{compat=1.1.0}
\usepackage[normalem]{ulem}
\usepackage{url}
\usepackage{xspace}

\renewcommand{\bar}[1]{\overline{#1}}    

\definecolor{Gray}{gray}{0.9}

\title{Relativistic Generalized Uncertainty Principle and its Implications}
\author{Vasil Nikolaev Todorinov}

\degreeyear{2020}

\prevdegrees{Bachelor of Science, Sofia University ``St. Kliment Ohridsky", 2015\\
             Master of Science, Sofia University ``St. Kliment Ohridsky", 2017}

\defensedate{9th December 2020}

\addsignatureline{Dr. Saurya Das}{Ph.D.}{Professor}{Thesis Supervisor}
\addsignatureline{Dr. Mark Walton}{Ph.D.}{Professor}{Thesis Examination Committee Member}
\addsignatureline{Dr. Kent Peacock}{Ph.D.}{Professor}{Thesis Examination Committee Member}
\addsignatureline{Dr. Pasquale Bosso}{Ph.D.}{Assistant Professor}{Internal External Examiner \newline Department of Physics \& Astronomy \newline Faculty of Arts and Science}
\addsignatureline{Dr. J. Gegenberg}{Ph.D}{Professor}{ External Examiner \newline Department of Mathematics and Statistics\newline University of New Brunswick }

\begin{document}

\frontmatter
 
\ulethtitle
\ulethapproval


\begin{abstract}
\indent\indent The fundamental physical description of the Universe is based on two theories: Quantum Mechanics and General Relativity. Unified theory of Quantum Gravity (QG) is an open problem. Quantum Gravity Phenomenology (QGP) studies QG effects in low-energy systems. The basis of one such phenomenological model is the Generalized Uncertainty Principle (GUP), which is a modified Heisenberg uncertainty relation and predicts a deformed position-momentum commutator. 

Relativistic Generalized Uncertainty Principle (RGUP) is proposed in this thesis, which gives a Loerentz invariant minimum length and resolves the composition law problem. RGUP modified Klein--Gordon, Schr\"odinger and Dirac equations with QG corrections to several systems are presented. The Lagrangians of Quantum Electrodynamics for the gauge, scalar, and spinor fields are obtained. The RGUP corrections to scattering amplitudes are then calculated. The results are applied to high energy scattering experiments providing much needed window for testing minimum length and QG theories in the laboratory.  
\end{abstract}

\acknowledgments{
\indent\indent First and foremost, I would like to express about my deep gratitude for my supervisor Dr.
Saurya Das whose expertise and attention to both detail and the big picture, provided
invaluable guidance for my research, all the while being so friendly and straightforward
during our discussions. He remains my principal mentor, from whom I learned
about various useful techniques and some of the important research problems in quantum
gravity. The material presented here could not have been developed without his assistance
for which I am extremely grateful.

 I would like to thank Dr. Mark Walton for being so supportive and for his useful
comments during the PhD program. I am greatly indebted for his useful suggestions in my
meetings with the advisory committee.

 I should be thanking Dr. Kent Peacock for his kind advice and useful comments during
the PhD program. I really appreciate that he kindly drew my attention to some of the more fundamental
questions about and kept me mindful of the bigger picture. 

 I would also like to express my gratitude to Dr. Pasquale Bosso, which have been an essential part of my research and results, as well as a good friend and guide in the world of academia. 
 
 My gratitude also goes to the University of Lethbridge and in particular the members of the Physics \& Astronomy department of the faculty of Arts and Science for providing a stimulating environment and useful feedback and ideas.
} 

\tableofcontents
\listoftables
\listoffigures

\mainmatter

\chapter{Introduction}\label{Ch1:Introduction}
\begin{quote}
	``If I have seen further it is by standing on the shoulders of Giants."
	
	\begin{flushright}
		 ``Letter from Sir Isaac Newton to Robert Hooke" --  Isaac Newton 1675
	\end{flushright}
\end{quote}

\section{Need for Quantum Gravity}\label{Sec:NeedForQG}
 Almost every physical phenomenon in the Universe can be described by two fundamental theoretical frameworks: Quantum Mechanics and General Relativity. These theoretical frameworks contain the theories and models that describe matter and its fundamental interactions, and predict 
many experimentally observed results. In order to put the results presented in this thesis in the proper context, this section will provide a basic introduction to the nature of the basic principles of those frameworks.

\subsection{Quantum Mechanics}
  Quantum Mechanics (QM) describes the interactions between atoms, molecules, elementary particles such as electrons, muons, photons, quarks and their anti-particles at very small scales, of the order of $10^{-10} - 10^{-19}$ m.
  
 The birth of QM is credited to Max Planck in 1901 with his hypothesis of quanta in the spectrum of black body  radiation \cite{Plank1901}. 
Similarly to Rayleigh in 1900 \cite{rayleigh1900vi},  Planck assumed that the black body is composed of many oscillating atoms and postulated that energy is absorbed and emitted through electromagnetic radiation. Planck postulated that the emission and absorption of energy happens in fundamental portions of energy $E$, called quanta
\begin{equation}
    E=h\nu\,,
\end{equation}
where $\nu$ is the frequency of the photon associated with each quantum, and 
$h=6.626\times10^{-19}\text{J $\cdot$ s}$ is the Planck constant. Additionally, he postulated that the energy of the oscillators comprising the black body can only take discrete values $E_n$
\begin{equation}
    E_n=nhf\,,
\end{equation}
where $n$ is an integer quantum number, $f$ is the frequency of the oscillator, and $h$ is Planck's constant.

The spectral energy distribution derived by Planck fit the experimental results exceptionally better than its predecessors. While previous models fit well to a particular frequency range of the spectrum and did not match anywhere else, Planck's law fit the experimental data throughout the entire observed frequency range. Further confirmation of Planck hypothesis was provided by A. Einstein in 1905 with the explanation of the photoelectric effect \cite{EinsteinPhoton1905}. 
The following years provided further strides in the development of Quantum Mechanics: by N.Bohr in 1913 with the presentation of Bohr's model of the atom \cite{Bohr1913a,Bohr1913b,Bohr1913c}; by L. de Broglie in 1925 with the proposition of the wave-matter duality \cite{Broglie1925}; by Dirac in 1928-1930, with the discovery of antimatter particles \cite{Dirac1928,Dirac1930}. Although the set of influential papers presented above lays the foundations of QM, there are many other works with comparable impact on the QM description of the world.

For example, the main tool of QM was provided by  E. Schr\"odinger in 1926 by an equation that bears his name  
\begin{equation}
        i\hbar \frac{\partial \psi(x,t)}{\partial t}=\left[-\frac{\hbar^2}{2m}\nabla^2+V(x)\right]\psi(x,t)\,,
\end{equation}
where $\hbar=\frac{h}{2\pi}$ is the reduced Planck constant, $V(x)$ is the potential function and  $ \psi(x,t)$ is an object called a wave-function, which is related to the probability of finding the particle at a particular location at a given time \cite{Schrodinger1926}.

   One of the main differences between quantum and classical mechanics is the fact that observables in QM, such as position $\hat{x}$, and momentum $\hat{p}$, are described by operators acting on the Hilbert space of states. A direct consequence of this fact is the Heisenberg Uncertainty Principle (HUP). 
The Uncertainty Principle was proposed by
W. Heisenberg in 1927 \cite{Heisenberg1927} as a physical argument based microscope thought experiment. Later  Schrödinger was the first, showed that the Heisenberg relations can be derived quite generally from non-commutativity, and can be expressed as
an inequality between the uncertainties of measurements of the position and momentum operators
\begin{equation}\label{Eq:HeisenbergUncertaintyRelation}
\Delta x \Delta p \geq \frac{1}{2}|\langle[ x,p]\rangle|~~
~\text{or}~~
\Delta x\,\Delta p\geq \frac{\hbar}{2}\,.
\end{equation}
One can easily see from the above that a physical experiment that simultaneously measures the eigenvalues of position and momentum operators with infinite precision cannot exist. However, on further examination, one can see that both uncertainties can be made arbitrary small by making the other arbitrary large.
Therefore, if one assumes that the uncertainty in position and momentum has its origin in their non-commutativity as operators, then one can conclude from Eqs.\eqref{Eq:HeisenbergUncertaintyRelation} above that $[x,p]=i\hbar$. 
Similar conclusions can be drawn for any two self-adjoined operators $A$ and $B$ and their commutator 
$[A,B]$. The discovery of QM and HUP changed the way physicists view the world, mainly by showing that physical systems are not deterministic but probabilistic. 

 The principles of  QM can be found among the foundations of Solid State Physics \cite{OReilly2002}, Thermodynamics \cite{OConnel2005}, Optics \cite{Garrison2008}, Quantum Chemistry \cite{Lewars2011} and many other fields of study. Perhaps the most important application of QM is when it is applied to Classical Field Theory to formulate the theoretical framework of Quantum Field Theory (QFT). The most prominent example of a model formulated with the use of QFT is the Standard Model (SM) of particle physics. SM successfully describes the behaviour of elementary particles under three of the four fundamental interactions of nature, namely electromagnetism, weak and strong nuclear forces. Additionally, SM successfully predicts the way leptons and weak bosons acquire mass. SM is responsible for the prediction and discovery the existence of a plethora of particles, the most recent of which is the Higgs boson, theorized in 1964
simultaneously by three independent groups: R. Brout and F. Englert; P. Higgs; G. Guralnik, C. R. Hagen, and T. Kibble \cite{Brout1964,Higgs1964,Kibble1964} and observed by the CMS experiment in LHC \cite{CMSLHSHiggs2012}.

 Despite its many successes, QM and its extensions such as the SM, 
 cannot be considered as the ultimate theory that describes the Universe. This is due to a few shortcomings, namely the fact that they fail to describe the behaviour of elementary particles under the influence of the last of the four fundamental interactions: gravitational interactions. This is a result of the treatment of spacetime as a fixed flat background, which 
 General Relativity teaches us is not the case,
 as shown in the following section. 
\subsection{General Relativity}
  The behaviour of matter under the influence of the fourth fundamental force of nature, namely the gravitational force, is described by the theory of General Relativity (GR), proposed by A. Einstein 
 in 1915
 \cite{Einstein1911GR,Einstein1913GR}. GR was build on two postulates:
\begin{itemize}
    \item {\it Principle of Relativity}: All frames of reference (including those that are accelerating) are equivalent with respect to the fundamental laws of nature. This results in the speed of light $c$ in a vacuum being the same for all observers, regardless of the motion of the light source or observer.
    \item {\it Equivalence principle}: The weak principle states that the local effects of motion
in a curved space (gravitation) are indistinguishable from those of an accelerated observer in flat space, without exception. Or in other words that the gravitational and inertial masses are equal $M_{\text{inertial}}=M_{\text{gravitational}}$. \\The strong principle states that the outcome of any local experiment 
(gravitational or otherwise) in a freely falling laboratory
is independent of the velocity of the laboratory and its location in spacetime.
\end{itemize}
The these two postulates form the base on which based on which in 1905 Einstein formulated the precursor to GR, the theory of Special Relativity (SR) \cite{Einstein1905SR}. In SR, Minkowski showed that space and time can be treated on an equal footing as coordinates of a four dimensional spacetime. Additionally there exists a frame independent maximum speed, namely the speed of light $c$. Taking this into consideration Einstein formulated the transformation laws for vectors in spacetime known as already known at the time as Lorentz–FitzGerald deformation hypothesis.  

In contrast with QM, GR treats spacetime as a dynamical entity, interacting with matter and energy.
According to the Einstein equations, matter or energy curves spacetime
\begin{equation}\label{Eq:EinsteinEquations}
     R_{\mu\nu}-\frac{1}{2}R\,g_{\mu\nu}+\Lambda g_{\mu\nu}=\frac{8\pi G}{c^4}T_{\mu\nu}\,,
\end{equation}
where the Greek letter indices are the spacetime indices that run from zero to three $\nu,\mu\in \{0,1,2,3\}$, $R_{\mu\nu}$ is the curvature (Ricci) tensor, $g_{\mu\nu}$ is the spacetime metric, $T_{\mu\nu}$ is the energy-momentum tensor, $\Lambda$ is a constant known as the cosmological constant, and $R=g_{\mu\nu}R^{\mu\nu}$ is the curvature (Ricci) scalar. In turn, the curvature of spacetime determines the dynamics of point-particles through the geodesic equation
\begin{equation}
    \frac{d^2x^{\mu}}{ds^2}+\Gamma^{\mu}_{\alpha\beta}\frac{dx^{\alpha}}{ds}\frac{dx^{\beta}}{ds}=0\,,
\end{equation}
where $\Gamma^{\mu}_{\alpha\beta}$ are the Christoffel symbols
(first derivatives of the metric). 

GR has been tested to a high degree of accuracy in astronomical and cosmological observations, e.g. 
in the perihelion precession of the orbit of Mercury, gravitational lensing,  gravitational red shift of light and the accelerated expansion of our Universe. In fact, gravity 
is the only relevant interaction at large distances, and with the addition of dark matter and dark energy it provides a satisfactory 
description of the contents and evolution of the Universe 
from the beginning to
its final fate.

\section{Quantum Gravity} \label{Sec:QuantumGravty}
  As  discussed in previous sections, QM considers spacetime as a flat stationary background, while GR considers it to be a dynamical object that interacts with matter and energy. On the other hand GR fails to take into account the probabilistic nature of observables described by QM. 
 Therefore, one needs to come up with a new theory that will give rise to QM and GR in their respective domains. Based on prior experience, one hopes that a physical theory that encompasses both QM and GR, and describes all fundamental interactions of matter will be a QFT. 
 
 GR describes the gravitational interaction as a classical field theory with the following action 
\begin{equation}\label{Eq:EinsteinHilbertAction}
    \mathcal{S}=\int \left[\frac{c^4}{16\pi G}\left(R-2\Lambda\right)+\mathcal{L}_{\text{matter}}\right]{\sqrt{-g}}\,\mathrm{d}^{4}x\,,
\end{equation}
where $g=\det (g_{\mu\nu})$, $R$ is the Ricci scalar and $\Lambda$ is the cosmological constant.   The equations of motion derived
from this action are the Einstein equations Eq.\eqref{Eq:EinsteinEquations}. One can notice, however, that unlike other QFTs the coupling constant of GR is  dimensionful, {\it i.e.} Newton's constants $G$. The dimensionful coupling constant gives rise of a length scale, and means that when one tries to quantize GR using the standard methods prescribed in the framework of QFT, one runs into difficulties. In particular, experimentally measurable quantities such as scattering cross-sections of processes involving gravity, not unexpectedly turn out to be infinite. The infinite nature of such results is common and is usually tamed by renormalization. The application of renormalization techniques to perturbatively quantized gravity, on the other hand, fail to obtain a finite measurable answer. Therefore, one can conclude that a new different approach to QG is needed.

Since its conception various the approaches and theories tacking the problem of quantizing gravity have come to be known as theories of Quantum Gravity (QG). Some examples of such theories include String Theory (ST),
Loop Quantum Gravity (LQG), and Doubly Special Relativity (DSR). These are just few of
attempts to solve this problem, all of which have met with various degrees of success.
In the following section the basic principles behind the best candidates for theories of QG are reviewed .
\subsection{String Theory}
   ST was originally developed in late 1960s as an attempt to describe the behaviour of strongly interacting particles: hadrons such as protons and neutrons. In 1961, G. Chew and S. Frautschi \cite{chew1961principle} discovered that the mesons make families called Regge trajectories \cite{regge1959introduction}. They noticed that there was a relationship between the hadrons masses and spins.  This relationship was later understood by Y. Nambu, H. B. Nielsen and L. Susskind to be the relationship expected from rotating strings \cite{Susskind1969,Susskind1970,Nambu:1997wf} that there exists a correlation between the  masses and spins of mesons, which is identical to the relationship expected from rotating strings.
   
   While working with experimental data, R. Dolen, D. Horn and C. Schmid developed some sum rules for hadron exchange. They found  that in the data, there were two peaks ``stealing" from the background. The authors interpreted this as saying that the t-channel contribution was dual to the s-channel one, meaning both described the whole amplitude and included the other \cite{dolen1968finite}.
   This result prompted G. Veneziano to construct a scattering amplitude that had the property of Dolen–Horn–Schmid duality, later renamed world-sheet duality \cite{veneziano1968construction}. Later C. Lovelace calculated a loop amplitude, and noted that there is an inconsistency unless the dimension of the theory is 26 \cite{lovelace1971pomeron}.
   
  In 1969–70 Y. Nambu \cite{nambu1995quark}, H. B. Nielsen \cite{nielsen1969almost} and L. Susskind \cite{susskind1970structure}
   they recognized that the theory could be given a description in space and time in terms of strings. Up to this point the theory contained only bosonic states. Then in 1971, P. Ramond added fermions to the model, which led him to formulate a two-dimensional supersymmetry \cite{ramond1971dual}. Inspired by Ramonds work J. Schwarz and A. Neveu added another sector to the fermi theory a short time later reducing the critical dimension to 10 \cite{neveu1971tachyon}. 
   
   Since then ST theory has seen a couple of revolutions, introducing concepts like the AdS/CFT  correspondence \cite{Banks:1996vh} and M-theory. These developments have elevated ST from a model of strong interacting particles to one of the strongest candidates for not only full theory of QG, but a Grand Unified Theory as well. 
   
ST proposes that instead of point particles the world is comprised of one-dimensional objects of finite length, the strings. These strings propagate in a $D$-dimensional pseudo-Riemannian spacetime background called target space and described by the $X^\mu$ coordinates.
A point particle traveling trough time and space has a trajectory or a world line. Similarly strings  traveling through the target space ``sweep" a two dimensional surface called a world sheet. This world sheet geometry is described by a world sheet metric $h_{ab}$, which is treated as a dynamical object in the theory, and acts as a ground state around which all modes of vibration of the string oscillate. These excitations of modes of oscillation of the string are the elementary particles.

The main purpose of ST is to consistently describe all interaction of matter with in one unified theory. This goal cannot be achieved without the description of gravity in a quantum framework.

To demonstrate that ST is a viable theory of quantum gravity one can consider  a calculation made by  T. Yoneya, and  J. Schwarz and J. Scherk \cite{SCHWARZ1982,SCHERK1974}, and reviewed in \cite{Zwiebach2004}.  Considering the action for a closed string 
\begin{equation}
    \mathcal{S}=\frac{1}{2\alpha^\prime}\int d^2\sigma \sum_{i\,j\,\mu} \eta_{ij} \frac{dX^i}{d\sigma_\mu} \frac{dX^j}{d\sigma_\mu}\,,
\end{equation}
where $\alpha^\prime$ is the Regge slope, or inverse of the string tension, $X^i$ are the coordinates of the worldsheet in the background spacetime, and $\sigma_\mu$ are the coordinates on the worldsheet. This expression can be generalized from a flat background metric to an arbitrary one. This generalization is done by substituting the flat spacetime metric $\eta_{ij}$ with an arbitrary one defined by the coordinates of the background spacetime 
\begin{equation}
\eta_{ij}\rightarrow g_{ij}(X)\,,
\end{equation}
where the indices $i,j \in \{0,1,2,\ldots,D-1$\}, and $D$ is the dimensions of the background spacetime.
Then one gets the following expression for the action
\begin{equation}
    \mathcal{S}=\frac{1}{2\alpha^\prime}\int d^2\sigma \sqrt{-h}h_{ab} g_{ij}(X) \frac{dX^i}{d\sigma_a} \frac{dX^j}{d\sigma_b}\,.\label{Eq:Polaykov}
\end{equation}
This is called Polyakov action and it was proposed by A. Polyakov in 1981 \cite{Polyakov1981}. This action describes a two dimensional field theory as the two dimensional path of the string through spacetime (worldsheet). By introducing a slowly varying field background dilaton filed $\Phi(X)$, and imposing conformal (scaling) symmetry on the worldsheet , one arrives at the following
\begin{equation}
    \mathcal{S}=\frac{1}{2\alpha^\prime}\int d^2\sigma \sqrt{-h}\left(h_{ab} g_{ij}(X) \frac{dX^i}{d\sigma_a} \frac{dX^j}{d\sigma_b}+\alpha^\prime R \Phi(X)\right)\,,
\end{equation}
where $R$ is the Ricci scalar given by the metric $g_{ij}(X)$, one can calculate the local functionals for the coupling functions $g_{\mu\nu}$ and $\Phi$. These functionals $\beta_{\mu\nu}^{g}$ take the following form 
\begin{equation}
\beta_{\mu\nu}^g +8\pi^2g_{\mu\nu}\frac{\beta^{\Phi}}{\alpha^\prime}=- T_{ij}^{\text{matter}}+\alpha^\prime \left(R_{ij}-\frac{1}{2}g_{ij}R\right) + \mathcal{O}(\alpha^{\prime\,2})\,.
\end{equation}
By imposing conformal symmetry on the worldsheet one can reduce the local functionals to zero $\beta_{\mu\nu}^{g} \rightarrow 0$, which effectively gives Einstein equations in 26 dimensions. This theory can then be compactified to the observed four dimensions. 

  So far it has been proven that the theory of gravity arises naturally in string theory. To prove that it is a quantum theory let, one consider Eq.\eqref{Eq:Polaykov}. With the coordinate change $u=\tau+\sigma$ and $v=\tau-\sigma$, where $\tau=\sigma_0$ and $\sigma=\sigma_1$, the equations of motion for the closed string are of the form 
 \begin{align}\label{Eq:StringEoM}
    X^i_L=\frac{1}{2}x_{0L}^i +\sqrt{\frac{\alpha^\prime}{2}}
\bar\alpha_0^i u+ i\sqrt{\frac{\alpha^\prime}{2}}\sum_{n\neq 0}\frac{\bar\alpha_n^i}{n}e^{-inu}\,,\\
 X^i_R=\frac{1}{2}x_{0R}^i +\sqrt{\frac{\alpha^\prime}{2}}
\alpha_0^i v+ i\sqrt{\frac{\alpha^\prime}{2}}\sum_{n\neq 0}\frac{\alpha_n^i}{n}e^{-inv}\,,
\end{align}
 where $ X^i_L$ and $ X^i_R$ denote the left and right moving strings, and $\bar\alpha_n^i$ and $\alpha_n^i$ are the transverse oscillation modes of the string. To quantize these equations of motion one substitutes $\bar\alpha_n^i$ and $\alpha_n^i$  with operators, which obey the following commutation relations
 \begin{equation}
     [\alpha_n^i,\alpha_m^j]=m\delta_{m+n,0}\eta^{ij}\,,\,\,\,\,
          [\bar\alpha_n^i,\bar\alpha_m^j]=m\delta_{m+n,0}\eta^{ij}\,,\,\,\,\,\,
               [\bar\alpha_n^i,\alpha_m^j]=0\,.
 \end{equation}
Through these  operators a new set of operators can be defined 
\begin{equation}
  \bar L_n^{\perp}=\frac{1}{2}\sum_{p\in\mathbb{Z}}\bar\alpha_n^i\bar\alpha_{n-p}^i\,,\quad L_n^{\perp}=\frac{1}{2}\sum_{p\in\mathbb{Z}}\alpha_n^i\alpha_{n-p}^i\,.
\end{equation}
These operators are known as Virasoro operators and they obey the following algebra
\begin{equation}
     [L_n^{\perp},L_m^{\perp}]=(n-m)L_{m+n}^{\perp} +\frac{\mathcal{C}}{12}(n^3-n)\delta_{n+m,0}\,,
\end{equation}
%
%
  where $\mathcal{C}$ is the central charge of the algebra, 
  which commutes with every other operator in the algebra. Virasoro operators $\bar L_n^{\perp}$  obey the same algebra and commute with the $L_n^{\perp}$  operators. For the closed string Virasoro operators acting on the physical states $\ket{phys}$ must obey the following condition $L_0^{\perp}\ket{phys}=\bar L_0^{\perp}\ket{phys}=0$. This condition is known as the level-matching condition.
  Using the above operators and the level-matching condition  one can express the mass for the excitation states of the closed strings
  \begin{equation}
      M^2=\frac{2}{\alpha^\prime}\left(N^{\perp}+\bar N^{\perp} -2\right)\,,\label{Eq:Spectrum}
  \end{equation}
  where $N^{\perp}$ and $\bar N^{\perp}$ are number operators which can be expressed in terms of the Virasoro operators as follows 
  \begin{equation}
      N^{\perp}=L_0^{\perp}-\frac{\alpha^\prime}{4}p^ip^i,\quad \bar N^{\perp}=\bar L_0^{\perp}-\frac{\alpha^\prime}{4}p^ip^i\,.
  \end{equation}
  By applying this to  
  Eq.\eqref{Eq:Spectrum} one sees that the first excited state will give massless  excited states. One can show that they further split into two massless vector fields which correspond to electrodynamics, and a massless traceless tensor field, which is the graviton, mediating gravitational interactions. 
  
 The most significant results of ST are also its weakest points. For example ST is the only theory that successfully predicts the number of dimensions of the Universe however the number of those dimensions is $D=26$ for a bosonic string,
 $D=10$ for supersymmetric ST, and $D=11$ for M-theory after taking into account the T-duality and S-duality. However, there are only three spatial and one temporal dimension experimentally observed so far. This leads to a discrepancy between the observations and predictions. One of the solutions to this problem was proposed by using compactification to make the extra dimensions compact and with a Calabi–Yau topology. The Calabi–Yau manyfolds support massless, chiral spinors, so that realistic masses and chiral fermions can be described in 4 dimensions. However, they’re not the only possibility. A problem with this solution is the fact that there are more than $10^{100}$ different six dimensional Calabi–Yau manifolds, introducing an ambiguity known as the vacuum selection problem.
 
 This leads to the next point. String theory relies heavily on Supersymmetry \cite{Martin1998} for the inclusion of fermions in the standard theory. Supersymmetry is the symmetry connecting bosons and fermions, realized by introducing supersymmetric partners for all the particles in the theory. This effectively makes ST very rich in terms of particles that it predicts. However, all the experiments searching for these super-partners have turned out empty handed. Thus ST remains a tentative theory, without any experimental evidence to support it so far.
  \subsection{Loop Quantum Gravity}
     All theories describing the gravitational interaction agree with GR and its interpretation that spacetime geometry and the force of gravity are closely related. Further expanding on that line of reasoning, the quantization of gravity and the quantization of the metric tensor are one and the same. In that case, as canonical quantum gravity proposes, the spacetime metric should arise as an expectation value of a wave functional from the Hilbert space of a  background independent non-perturbative quantum theory. The dynamics of the theory in this approach are governed by a Hamiltonian operator, which in turn is given by the Wheeler–DeWitt equation \cite{DeWitt:1967yk}. LQG arises as a way to address some of the probelms canonical quantum gravity has, such as the fact that the Wheeler–DeWitt equation is ill defined in the general case.
     
  The Hamiltonian formulation of GR used by LQG was developed in 1959 by R. Arnowitt, S. Deser and W. Misner \cite{Arnowitt1959}. This formulation was obtained through studying the evolution
 of canonical variables defined classically through a splitting of spacetime on hypersurfaces $\Sigma$ at a given point in time $t\in \mathbb{R}$. The resulting manifold is constructed as a direct product of the time and hypersurfaces $\mathcal{M}=\Sigma\times\mathbb{R}$. The Hamiltonian that one gets from  Einstein-Hilbert action Eq.\eqref{Eq:EinsteinHilbertAction} in the this construction, also known as ADM Hamiltonian, has the form
 \begin{equation}\label{Eq:ADMHamiltonian}
    H=-\frac{c^{4}}{16\pi G }\sqrt{g}\left[^{(3)}R+g^{-1}\left(\frac{1}{2}\pi^{2}-\pi^{ij}\pi_{ij}\right)\right]\,,
 \end{equation}
 where $^{(3)}R$ is the Ricci curvature of the 3D hypersurfaces $\Sigma$, and $\pi_{ij}$ is defined as follows
 \begin{equation}
     \pi^{ij}=\sqrt{^{(4)}g}\left({^{(4)}}\Gamma _{pq}^{0}-g_{pq}{^{(4)}}\Gamma _{rs}^{0}g^{rs}\right)g^{ip}g^{jq}\,,
 \end{equation}
 where $g=\det (g_{\mu\nu})$. Note that all Latin indices signify spatial coordinates which run through $i,j,p,q,k,l\in \{1,2,3\}$. In 1982, A. Sen successfully wrote the Hamiltonian Eq.\eqref{Eq:ADMHamiltonian} in terms spinor fields \cite{SEN1982}. Ten years prior in 1971 R. Penrose explored the rise of space from a quantum combinatorial structure. This investigations resulted in the development of spin networks \cite{Penrose1971}. 
 
  The fundamental canonically conjugate variables of spinor gravity were identified by A. Ashtekar in 1986 \cite{Ashtekar1986new,Ashtekar1987new}. This allowed  for an unusual way of rewriting the metric on three-dimensional spatial slices in terms of an $SU(2)$ gauge field and its complementary variable. In these variables the Einstein-Hilbert action written in the form 
      \begin{equation}
	\mathcal{S} = \int \mathrm{d}^4 x ~ E_{\mu I} E_{\nu J} ~ {}^4 F^{IJ}_{\tau \sigma} ~ \varepsilon^{\mu\nu\tau\sigma}~, \label{Eq:LoopAction}
\end{equation}
  where ${}^4 F^{IJ}_{\tau \sigma}$ is the field strength tensor. This formulation introduces a set of four orthogonal  vector fields $E_I^\mu$ called a tetrad. These fields have the following properties
  \begin{equation}
      \delta_{IJ}=g_{\mu\nu}E_I^\mu E_J^\mu\,,\quad\text{and}\quad  E_I^\mu E^I_\nu=\delta^\mu_\nu\,,
  \end{equation}
  where $\delta_{IJ}$ and $\delta^\mu_\nu$  are the Kronecker delta function, and  the indices $\mu,\nu$ are the spacetime indices  and behave like indices in regular curved spacetime, and the capital letter indices $I,J$ are internal indices which behave like a flat space indices. Then one can write the metric in terms of those  Ashtekar variables
  \begin{equation}
      (\det g)g^{\mu\nu}=\sum_{I=1}^3 \tilde{E}_I^\mu \tilde{E}_I^\nu\,,
  \end{equation}
  where
 \begin{equation}
     \tilde{E}_I^\mu=\sqrt{\det g}E_I^\mu\,.
 \end{equation}
  The choice of tetrads is not unique, and in fact one can perform a local rotation in space with respect to the internal indices $I,J$ without changing the metric. This means that ADM Hamiltonian expressed in the Ashtekar variables  contains a $SU(2)$ gauge symmetry/invariance. These tetrads transform under the $SO(1,3)$ group, {\it i.e.} a local Lorentz transformation as
 \begin{equation}
     (E_I^\mu)^\prime=\Lambda_I^J(x)E_J^\mu\,.
 \end{equation}
  The gauge connection associated with the local Lorentz symmetry is defined as follows 
\begin{equation}\label{Eq:LorentzConnection}
    W^{IJ}_\mu=E^I_\nu\partial_\mu E^{\nu J}+E^I_\nu E^{\sigma J} \Gamma^{\nu}_{\sigma\mu}\,,
\end{equation}
where $\Gamma^{\nu}_{\sigma\mu}$ is the Levi-Civita connection. One can easily check that the Ashtekar tetrads are invariant under the $SU(2)$ group. The complex connection can be expressed in terms of the local Lorentz connection  Eq.\eqref{Eq:LorentzConnection} as follows
\begin{equation}\label{Eq:SU(2)connection}
{}^4A^{IJ}_\mu[W]=W^{IJ}_\mu-\frac{1}{2}i\varepsilon^{IJ}_{KL}W^{KL}_\mu\,,
\end{equation}
with $\epsilon^{IJ}_{KL}$ the completely antisymmetric tensor. As evident from the expression above, the $SU(2)$ connection is self-dual. The covariant derivative is 
\begin{equation}
    D_\rho E_{\mu I} = \partial_\rho E_{\mu I} -E_{\mu J} ~ {}^4 A^J_{\rho I}\,.
\end{equation}
The Yang--Mills field strength for the connection presented in Eq.\eqref{Eq:SU(2)connection} is  
\begin{equation}
    {}^4 F^{IJ}_{\mu\nu} = \partial_\mu ~ {}^4A^{IJ}_\nu - \partial_\nu ~ {}^4 A^{IJ}_\mu + {}^4 A^{IM}_\mu ~ {}^4 A^J_{\nu M} - {}^4 A^{IM}_\nu ~ {}^4 A^J_{\mu M}.
\end{equation}
The ADM Hamiltonian  written in these Ashtekar variables makes possible the background independent quantization of GR. In particular a quantization was based on procedures developed for the study of Quantum Chromodynamics (QCD) in terms of Wilson loops, developed by K. G. Wilson in 1974 \cite{Wilson1974}. These Wilson loops formed a basis for the non-perturbative quantization of gravity. The fundamental element of this basis is the Wilson loop of the connection ${}^4A^{IJ}_a$, defined as the trace of the holonomy of the same connection on a closed curve $\alpha$
    \begin{equation}
	\mathcal{T}[\alpha] = - \mathrm{Tr} \left[U_\alpha\right] = \mathrm{Tr} \left[\mathcal{P} \exp \left(\oint_\alpha A\right)\right]\,.
    \end{equation}
Between 1988–1990, C. Rovelli and L. Smolin  obtained an explicit basis of states of quantum geometry \cite{Rovelli1988,Rovelli1990}. This basis of states turned out to be spin networks developed by R. Penrose in 1971 \cite{Penrose1971}, by exploring the rise of space from a quantum combinatorial structure. Following this, in 1994, they showed that there are quantum operators in the theory related to area, volume, and length. Furthermore, as discussed in further sections, LQG shows that these operators have a discrete spectrum \cite{Rovelli1994}. 

Further, during 2007-2008, J. Engle, R. Pereira and C. Rovelli found a covariant formulation of the theory. This formulation is called spin foam theory. In this framework one can obtain an expression for the Schwarzschild black hole entropy. 

One can conclude that LQG is a robust candidate for theory of QG. However, just as ST, LQG is a tentative theory without experimental data to support it. Furthermore, the question on how the theory couples to the Standard Model, and if there are some constraints on the matter content of the theory, still remains unanswered.

As mentioned before, a unified theory of QG 
must reproduce the existing theories, namely the Standard Model and GR. This remains one of the open problems of LQG.
 \subsection{Doubly Special Relativity}
   DSR is the most recent of the theories of QG reviewed here. A clarification worth mentioning is that DSR is cannot be considered as a full fledged theory of QG as ST and LQG. DSR is a model focusing on the kinematic aspects of the existence of an additional Lorentz invariant scale. DSR considers effects on transformation laws between observers and symmetries of spacetime. Unlike ST and LQG it does not attempt to formulate a full theory of QG. Rather DSR tries to address some of the problems which arise in the process of quantization. The reason DSR is discussed here is to demonstrate the diversity of ideas existing in the QG family.
   
 The main premises of the theory is was proposed by G. Amelino-Camelia, J. Magueijo, and L. Smolin in 2000-2002. They attempted to modify special relativity by introducing an observer-independent length \cite{AmelinoCamelia:2000mn,AMELINOCAMELIA2001255,Magueijo:2001cr,Magueijo:2002am}.  
 This was achieved through the inclusion of an additional postulate in SR. Namely
 \begin{itemize}
     \item There is an observer independent scale which has the of dimension of mass $\kappa$
(or length $\lambda=\kappa^{-1}$), identified with the Planck mass (or Planck length).
 \end{itemize}
 This postulate is the reason for the term ``Doubly'', due to the fact there are two observer independent scales, the speed of light $c$ and  Planck mass or Planck length, which are defined as follows 
\begin{subequations}
\begin{align}
    M_{\text{Pl}}\equiv\sqrt{\frac{\hbar c}{G}}\approx 2.2\times 10^{-8}\text{Kg}\\
    l_{\text{Pl}}\equiv\sqrt{\frac{\hbar G}{c^3}}\approx 1.6\times 10^{-35}\text{m}\, .
\end{align}
\end{subequations}
These scales are generally considered to be the scales at which the QG effects become relevant. 
It is assumed that the standard Special Relativity is recovered from DSR, by taking the limit $\kappa \rightarrow\infty$.

Symmetries of spacetime in SR form a ten
dimensional group, the Poincar\'e group, with generators corresponding to rotations $J_I=(\varepsilon_{IJK} M^{JK})/2$, boosts $K_I=M_{0I}$, and
translations $P_{\mu}$, which have the following algebra
\begin{subequations}
\begin{align}
\label{Eq:PoincarePP} \left[P^\mu, P^{\nu}\right]&=0\,,\\
   \label{Eq:PoincarePM}\left[P^\mu, M^{\nu\rho}\right]& =  i\hbar\left(P^{\nu}\delta^{\mu\rho}-P^{\rho}\delta^{\mu\nu}\right)\,,\\
    \label{Eq:PoincareMM}\left[M^{\mu\nu},M^{\rho\sigma}\right] &= i\hbar\left(\eta^{\mu\rho}M^{\nu\sigma}
    -\eta^{\mu\sigma} M^{\nu\rho}-\eta^{\nu\rho}M^{\mu\sigma}+\eta^{\nu\sigma}M^{\mu\rho}\right)\,.
\end{align}
\end{subequations}
This algebra has two operators which commute with every other operator in the algebra. These operators are called Casimir invariants. They are $
P_\mu P^\mu$ and $W^\mu W_\mu$,
where $P_\mu$ is the operator of translations, and 
\begin{equation}
W_\mu\equiv \frac{1}{2}\varepsilon_{\mu \nu \rho \sigma }M^{{\nu \rho }}P^{\sigma}\,,
\end{equation}
is the Pauli–Lubanski pseudovector.
The first Casimir, {i.e.} the square of the four-momentum, is a manifestation of the Einstein's dispersion relation 
\begin{equation}
    E^{2}-\vec{P}^{2}c^{2}=m_{0}^{2}c^{4}\,.
\end{equation}
As a consequence of the introduction of a frame independent mass/length scale, one can expect that there exists a general deformed algebra, which gives the  Poincar\'e algebra in the appropriate limits. Algebras which possess these qualities are generally referred to as $\kappa$-Poincar\'e algebra. $\kappa$-Poincar\'e algebra is an Poincar\'e algebra deformed in a Hopf algebra \cite{Majid:1994cy}. Hopf algebras arise in algebraic topology, where they originated and are related to the H-space concept in group scheme theory \cite{bergman1985everybody}. One explicit expression of $SO_q(3,1)$ algebra is given below
\begin{subequations}
\begin{align}\label{Eq:k-Poincare}
[J_i, J_j] &= i\, \varepsilon_{ijk} J_k, \quad [J_i, K_j] = i\, \varepsilon_{ijk} K_k\,,\\
 [K_i, K_j] &= -i\, \varepsilon_{ijk} J_k,\\
  [J_i, P_j] &= i\, \varepsilon_{ijk} P_k, \quad [J_i, P_0] =0,\\
   \left[K_{i}, P_{j}\right]& = i\,  \delta_{ij} \left( \frac{\kappa}{2} \left(1 -e^{-2\frac{P_{0}}{\kappa}}
\right) + \frac{1}{2\kappa} \vec{P}\,{}^{ 2}\, \right) - i\, \frac{1}{\kappa}
P_{i}P_{j}\,,\\
\left[K_{i}, P_{0}\right]& = i\, P_i\,.
\end{align}  
\end{subequations}
Since it turns out that the action of symmetry generators must be deformed in DSR, one may refer to the theory as ``Deformed Special Relativity''. The modified dispersion relations derived from the algebra presented in Eq.\eqref{Eq:k-Poincare}  has the following form
\begin{equation}\label{Eq:DSRCasimir}
\kappa^2 \cosh \frac{P_0}{\kappa} - \frac{\vec{P}{}^2}2\, e^{P_0/\kappa} = m_{0}^{2}c^{4}.
\end{equation}
One can easily see that this dispersion relation is no longer linear in terms of the square of the momentum. This leads to the fact that the action of the $\kappa$-Poincar\'e group is not linear as well. When one considers two particles $A$ and $B$
\begin{equation}
    \Lambda(P_A+P_B)\neq \Lambda(p_a)+\Lambda(p_b)\,,
\end{equation}
where $\Lambda$ are the deformed generators of the Lorentz transformations.
The implications of the above are known as the Composition law problem\footnote{Also known informally as  The Soccer-ball problem.}.  The Composition law problem arises when one considers many body system all parts of which are moving at the same velocity. In that case the corrections to the total energy and momentum of the system depends on the number of small bodies it comprised by. Therefore, the magnitude of the QG correction rises are discussed and shown in the Appendix \ref{app:CLP}. 
Additionally it can be shown that the $\kappa$-Poincar\'e structure is a non-associative algebra, which leads to the following 
\begin{equation}
    (P_A+P_B)+P_C\neq P_A+(P_B+P_C)\,,
\end{equation}
which contradicts the experiments. 

DSR and its Lorentz invariant mass/length scale, predict a dependence of the speed of light on the energy it carries. This model is known as Rainbow Gravity and provides an avenue for experimental tests of QG. Up to the accuracy of the measurements, the experimental results do not agree with the model. However, with the advancement of experiments  and accumulation of data such effects are likely to be found.    

\section{Minimum length and Quantum Gravity}\label{Sec:MinimumLengthAndQG}
  One common aspect of all the theories mentioned above and many other theories of QG, as explained in \cite{Garay:1994en}, is that they all predict a minimum measurable length. In ST this length is the string length. In LQG the expectation value of the length operator. And postulated Lorentz invariant fundamental scale in DSR. In order to show the diversity of arguments supporting the existence of minimum length, this section is a brief overview of how a minimum length arises in several different cases in the context of QG.
  
A really simple case in which the concept of minimal uncertainty in position  arises when considering an ultra-relativistic particle, $pc \gg mc^2$, for which one can ignore the mass. Therefore  $E/c=p$, which in turn leads to  uncertainty in momentum $\Delta p=\Delta E/c$. Then the uncertainty relation for an ultra-relativistic particle can be written in terms of its position and energy, as
$\Delta x\Delta E\geq c\hbar/2$. 
In this relation, one sees that the uncertainty in position can be indefinitely small. 
In that case the uncertainty in position can  be made smaller than its Compton wavelength $c\hbar/E$,
i.e. $\Delta x\leq c\hbar E^{-1}$.
However this implies that its uncertainty in  energy will exceed the rest energy of the particle, contradicting the definition for a  particle itself, as a localized concentration of energy. This is because, as one knows, the definition of a particle in QFT is an excited state of a field in particular region of space. Therefore, in the attempt to infinitely localize a state in a region of space smaller than its Compton length, the uncertainty in energy will exceed its rest mass, and one can no longer be certain that there is indeed a particle in that region of space. From this argument one can conclude that an experiment cannot measure the position of a particle with accuracy more than its Compton wavelength. One sees that a minimum length scale arises even when considering extremely simplified lines of reasoning. 

\subsection{Microscope thought experiments}

 Another argument presented by Hossenfelder in \cite{Hossenfelder:2012jw} that can further clarify and support the existence of
a minimum length, is constructed when one considers the following thought experiment. In order to measure the position of a target one bombards it with photons and detects the scattered photons, an experimental construction also known as Heisenberg microscope. There is an intrinsic uncertainty due to the random nature of the direction of the scattering, given by $\Delta x \geq \hbar/2\Delta p$.
However, there is another factor that adds to the uncertainty: 
assume that the particle is attracted to the photon by a Newtonian gravitational force (per unit mass of the target) equal to $l_{Pl}^2 c \varepsilon/ \hbar r^2$, where $\varepsilon$ is the energy of the photon, and $r$ its distance from the particle. The argument here is made under the assumption that the energy and matter density is not high enough to cause large curvatures.  This force will cause the particle to accelerate in the direction of the scattered photon, as result of which it travels a distance $l_{Pl}^2\varepsilon/c$. 
Since the direction of photon scattering is unknown, this effectively introduces a second uncertainty $\Delta x \geq l_{Pl}^2\Delta p/\hbar$. Combining the two uncertainties, one gets
\begin{equation}
    \Delta x\geq \text{max}\left(\frac{\hbar}{\Delta p},\frac{ l_{Pl}^2}{\hbar}\Delta p\right)\geq l_{Pl}\,,
    \label{Eq:MinimumUncertaintyInPosition}
\end{equation}
where $l_{Pl}=\sqrt{\hbar G/c^3}\approx 10^{-35}\,m$ is the Planck length. In other words, while one needs photons with high energy to get a high spatial resolution, this will result in a significantly greater gravitational interaction with the target, therefore producing more uncertainty. 

Therefore, a new uncertainty relation, which accommodates the existence of minimum length will give rise to the so-called Generalized Uncertainty Principle (GUP) and a modification of the standard Heisenberg algebra. Since the goal is to eventually construct a covariant version of GUP, an extension of Eq.\eqref{Eq:MinimumUncertaintyInPosition} to a generalized time-uncertainty is made. 

The properties of this can be explained by considering a similar thought experiment with two clocks, which one would like to synchronize using photons. The accuracy of the synchronization will be determined by two factors: first, the uncertainty in the emission of photons, according to the energy-time uncertainty relation, is $\hbar/\Delta\varepsilon$; second, the gravitational interaction between the photon and the clock. According to the Einstein equations, the photon will generate its own gravitational field. Assuming that the clock is stationary before the interaction, and the radius of the interaction is $r$, the duration of the interaction between the clock and the photon will be $\sqrt{g_{00}}\,r/c$, where $g_{00}$ is the time-time component of the metric generated by the photon. It can be shown that it will take the form of 
\begin{equation}
    g_{00}= 1 -\frac{4l_{Pl}^2\varepsilon}{\hbar c r}\,.
\end{equation}
This results in the following uncertainty in time measurement due to the gravitational interaction
\begin{equation}
  \Delta t \geq   \frac{2l_{Pl}^2\Delta \varepsilon}{\hbar c^2\sqrt{1-4l_{Pl}^2\varepsilon/\hbar c r}}\geq\frac{2l_{Pl}^2}{\hbar c^2}\Delta \varepsilon\,.
\end{equation}
Therefore, the final uncertainty in the two synchronizing clocks will have the form
\begin{equation}
    \Delta t\geq \text{max}\left(\frac{\hbar}{\Delta \varepsilon},\frac{l_{Pl}^2\Delta \varepsilon}{\hbar c^2}\right)\geq \frac{l_{Pl}}{c}\,.
\end{equation}
It is worth clarifying that these are low energy approximations and idealizations  of an exact corrections that a full theory of QG is expected to give.

\subsection{Minimum length and field theory }

  As mentioned before, attempts to quantize GR using the methodology of QFT produce non-renormalizable divergences, which has lead one to believe that the usual quantization prescriptions cannot be applied to GR. However, these methods contain a valuable insight into the quantum nature of gravity. Mainly the rise of a minimum uncertainty in position or minimum length. 
  
  It is reasonable for one to believe that classical GR should arise as a low energy limit of a larger theory of QG. If that assumption is true, one can conclude that a theory of QG's vacuum state should correspond to a classical solution of the Einstein equations. For simplicity in this argument one chooses the vacuum state to be the Minkowski spacetime. Additionally, one of the simplest quantum
fluctuations of the metric  is considered  {\it i.e.} fluctuations in  its conformal factor. The quantization prescription  used can be found in \cite{narlikar1983quantum}. 
The conformal fluctuations over a
flat metric $\eta_{\mu\nu}$ will have the following form
\begin{equation}
g_{\mu\nu}(x)=[1+\phi(x)^2]\eta_{\mu\nu}\,,\label{Eq:Metric}
\end{equation}
where $\phi(x)$ are conformally invariant metric fluctuations. The corresponding generating functional for the path integral for these
fluctuations can be written as
\begin{equation}
  \mathcal{Z}= \int \mathcal{D}\phi\,\,\text{exp}\left(\frac{-i}{l_{Pl}^2}\int d^4x~\eta_{\mu\nu}\nabla^{\mu}\phi\nabla^{\nu}\phi\right)\,.
\end{equation}
 Because of the quadratic nature of path integral, the above be evaluated in a closed form, under the assumption that the conformal fluctuations behave like a scalar field, giving rise to the corresponding probability amplitude \cite{srednicki2007quantum}
\begin{equation}
    \mathcal{A}(\phi)=\left(\frac{1}{2\pi\, D^2}\right)^{1/4}\text{exp}\left(-\frac{1}{2D^2}\phi^2\right)\,,\label{Eq:Fluctuations}
\end{equation}
where 
\begin{equation}
    D^2=\frac{l_{Pl}^2}{4\pi^2 l^2}\,,
\end{equation}
and $l$ is the interval over which our measuring apparatus averages the conformal fluctuations. 
More detailed version of the calculation can be found in \cite{PADMANABHAN198538,Padmanabhan:1985jq,Padmanabhan_1986,Padmanabhan_1987}.
One can see from the above that for large averaging interval,
i.e. $l\gg l_{Pl}$, the amplitude of the conformal fluctuations $\mathcal{A}(\phi)$ vanishes, effectively giving us a flat metric and classical gravity. In other words, the conformal fluctuations of the metric occur only at length scales close to the Planck length $l_{Pl}$.

  However, to put a proper lower limit on the distance between two events, let us consider two such events separated by a proper distance $l(x,y)$. Using 
Eq.\eqref{Eq:Metric} the  probability of two events to be separated by that distance is given by
\begin{equation}
\braket{ ds^2}=\bra{0}[1+\phi(x)^2]\eta_{\mu\nu}\ket{0}=\lim_{x\rightarrow y}\langle l^2(x,y)\rangle=\lim_{x\rightarrow y}[1+\langle\phi(x)\phi(y)\rangle]l_0^2(x,y)\,,
\end{equation}
where $l_0^2(x,y)$ is the interval given by the flat metric $\eta_{\mu\nu}$. Using the usual rule of path integral, this leads to 
\begin{equation}
    \langle\phi(x)\phi(y)\rangle=\frac{l^2_{Pl}}{l_0^2(x,y)}\,.
\end{equation}
It is evident from the above that if the proper interval between two events is smaller than the Planck length, the
fluctuations are no longer perturbative and 
the classical limit cannot be reached. This effectively shows that Planck length $l_{Pl}$ is the minimum measurable length.

\subsection{Minimum length from String Theory}\label{Sec:Stringtheory}

  In order to understand how a minimal length arises from ST, one must first consider one of its symmetries, namely the momentum-winding symmetry, also known as T-duality. To understand T-duality one must consider a simple analogy.  
  
Two string theories, one of which describes strings propagating in a spacetime shaped like a circle of radius $R$.  And another theory which describes strings propagating on a spacetime shaped like a circle of radius proportional to $\alpha^\prime/R$.
There exists a correspondence between the two theories. They are equivalent in the sense that all observable quantities in one description are identified with quantities in the dual description. For example, the momentum of a string on spacetime with radius $R$ is quantized as 
\begin{equation}
p=n\hbar/R\,,
\end{equation}
where $n$ is an integer. Respectively the energy of the string (the Kaluza-Klein modes) will be 
\begin{equation}
    E_n=\frac{c\hbar |n|}{R}\,.
\end{equation}
The same string can be wrapped around a circle $m$ times and it will have energy (the winding modes)
\begin{equation}\label{Eq:WindinModes}
    \tilde E_n =\frac{\hbar c |m| R}{\alpha^\prime}\,.
\end{equation}
This duality between momentum or Kaluza-Klein modes and winding modes
lead to interesting implications.
The Kaluza-Klein modes obey the Heisenberg uncertainty relation   $\Delta x\sim \hbar / \Delta p$, when the winding modes have a more interesting commutation relation. One can qualitatively read the uncertainty relation for the winding modes from Eq.\eqref{Eq:WindinModes}
\begin{equation}
\Delta p \approx \frac{\Delta E}{c} \approx \frac{\hbar}{\alpha^\prime} \Delta x\,.
\end{equation}
This leads to an interesting result. In the low energy regime the strings obey the Heisenberg uncertainty relation. However, due to T-duality at high energies the strings obey a different uncertainty relation. This led to the proposition of a Generalized Uncertainty relation that works in both regimes \cite{AMATI198941,Chang:2011jj,KONISHI1990276}, {\it i.e.}
\begin{equation}\label{Eq:STGUP}
    \Delta x\geq \frac{\hbar}{2\Delta p}+\alpha^\prime\frac{\Delta p}{\hbar}\,.
\end{equation}
The first term is the well known Heisenberg term and dominates when the momentum is small, while the second one corresponds to the fuzziness introduced by ST and dominates when one tries to localize the string. This form of the uncertainty principle implies a global minimum in the position uncertainty $l_s\sim \sqrt{\alpha^\prime}$. One can see that this global minimum is the string length.

\subsection{Minimum length from Loop Quantum Gravity}
To understand how minimum length arises from LQG one must first study the area and volume operators, studied in \cite{ROVELLI1995593,Ashtekar_1997}.
  In order to obtain the spectrum of the area operator  one needs to start by formulating the area operator in terms of  Ashtekar variables. 
The regularised classical area operator is defined as 
\begin{equation}\label{Eq:AreaOperator}
[A_S]^\epsilon:=k\gamma\int_S d^2u \left|[P^\perp_a]^\epsilon (u)[P^\perp_b]^\epsilon (u)\delta^{ab}\right|^{\frac{1}{2}}\,,
\end{equation}
 where the lower case indices denote spacetime $a,b,c\in \{0,1,2,3\}$ and the uppercase indices $I,J,K\in\{1,2,3\}$ denote the indices in the $SU(2)$ manifold. The 
$P^\perp_a$ is the projection of $P_a^J$ onto the normal of $\Sigma$, where  $P_a^J$ is defined as follows
\begin{equation}
P_a^J=\frac{E_a^J}{k\gamma}\,.
\end{equation}
The classical area operator shown in Eq.\eqref{Eq:AreaOperator} is then promoted to quantum operator acting on an eigenstate of the LQG Hamiltonian. One can show that the final form of the quantum area operator is  
\begin{equation}
\hat{A}_S\ket{s^{\vec{j}}_{\alpha,\vec{m}\vec{n}}}=4\pi \gamma l_p^2 \sum_{\substack{v\in V(\alpha)\\
                             v\in I(S)}}\left|\left(\sum_{\substack{e\in E(\alpha)
                  e\cap v \neq \emptyset}}
                  \kappa(e,S)\hat{Y}^{(v,e)}_j \right)^2\right|^\frac{1}{2}\ket{s^{\vec{j}}_{\alpha,\vec{m}\vec{n}}}\,,
\end{equation}
where $\hat{Y}^{(v,e)}$ are functions defining the edges of the studied area.
Studying this explicit form of the quantum area operator, one finds that the smallest area eigenvalue, also known as area gap, is given by 
\begin{equation}
 \alpha_0=2\pi \gamma l_p^2\sqrt{3}\,.
\end{equation}
This result has interesting implications. For example, it is easily noticeable that the smallest eigenvalue of the area operator is proportional to the square of the Planck length. This hints the existence of a minimum length scale in the Universe.

Attempts have been made to repeat the above line of reasoning in constructing the classical expression for the length of a curve, in order to obtain the spectrum of the length operator. The length of a curve in LQG is expressed as follows
\begin{equation}
l(c)=\int_0^1\sqrt{g_{ab}(c(t))\dot{c}^a(t)\dot{c}^b(t)} dt=\int_0^1\sqrt{e^I_a(c(t))e^J_b(c(t))\dot{c}^a(t)\dot{c}^b(t)\delta_{I}} dt,
\end{equation}
where one expresses the metric in terms of Ashtekar variables 
\begin{equation}
g_{ab}=\frac{k}{4}\varepsilon_{aCD}\varepsilon_{b}^{ef}\varepsilon^{ijk}\varepsilon_{iMN}\frac{P^C_j P^D_k P^M_e P^N_f}{\det P}\,.
\end{equation}
The smallest eigenvalue of the length operator proposed above cannot be obtained through the methodology used for the area operator.  Instead, a several of different alternatives have been proposed. For example a different form of the length operator. Alternative forms of the length operator can be found with the use of the Thiemann identity \cite{thiemann1998length}. Tikhonov regularization for the inverse $\hat{V}_{v,RS}$ operator \cite{Bianchi2009length}, in terms of other geometrical operators and the volume operator \cite{ma2010new}. The eigenvalue of the length of a curve operator was found to be 
\begin{equation}
    l_{\text{min}}=\left(8\pi\gamma l_{\text{Pl}}^2\right)^{1/2}\,,
\end{equation}
where $\gamma$ is the Immirzi parameter. This implies the existence of minimum measurable length. Furthermore, one can see that similarly to string theory the minimum length found in LQG is proportional to the Planck length. 

\section{Summary}
   The physical theories of QM and GR were reviewed in section \ref{Sec:NeedForQG} and they were shown to be wanting in regard to the question of quantization of gravity. This apparent lack of consistency between QM and GR serves as evidence that a new and more fundamental theory of nature is required. The lack of such theory at the moment has left a vacuum in our understanding of the Universe. Numerous attempts at filling this gap in understanding has lead to the development of the various theories of QG. 

  In an attempt to show the diversity of ideas, methods, assumptions and approaches contained in the field of QG, section \ref{Sec:QuantumGravty} reviews the most popular examples of QG theories and their basic principles, interesting results and weaknesses. 
It is apparent that there exists a commonality amongst all the theories of QG.  Regardless of the approach and initial assumptions considered by current theories, all of them have something in common. They all predict some form of length or energy scale at which our models of the universe break down and QG effects become relevant. A few of the examples of minimum length arising from theories of QG are discussed in section \ref{Sec:MinimumLengthAndQG}. The diversity of ideas and methods which give the same result provides a robust evidence for the existence of minimum length, and pose the question of how one goes about experimentally proving its existence. The answer to this question is contained in the next chapter, where a brief introduction to the field of QG Phenomenology is presented.

\chapter{Quantum Gravity Phenomenology}\label{Ch2:QuantmGravityPhenomenology}
\begin{quote}
	What we observe is not nature in itself but nature exposed to our method of questioning.
	
	\begin{flushright}
		Physics and Philosophy -- Werner Heisenberg
	\end{flushright}
\end{quote}

  QG has been a subject of study for theoretical physicists for almost 70 years. So far there has not been single experiment which supports or refutes any theory of QG, although it is well known that in regions of extremely high curvature, classical GR is bound to fail. 
Therefore, there must exist a (quantum) theory whose low energy limit is GR. On the other hand, there is a possibility that such a theory doe not exist and classical GR is an effective field. However, the search for such a theory can be done in two ways: first, by considering the deep fundamental and philosophical questions, for example renormalization, loss of unitarity, and the nature of spacetime itself; and the second, by searching already existing experimental data or proposing new experiments in order to find unobserved low-energy effects predicted by various QG theories. 

The second approach is known as  Quantum Gravity Phenomenology (QGP).
In order to understand what the field of QGP is, one needs to first understand its name. The first two words of its name, as presented in Chapter \ref{Ch1:Introduction} stand for the different theories and approaches taking on the task of quantizing gravity. The third part of the name is phenomenology and
according to the Oxford English dictionary is “the division of any science which is concerned with the description and classification of its phenomena, rather than causal or theoretical explanation” \cite{dictionary1989oxford}. Applying phenomenology to physics, one makes models out of already existing theories and then applies these models to current or future experiments, in search of new and interesting phenomena. In other words, phenomenology bridges a gap between theory and experiment.

Guided by extensive reviews \cite{AmelinoCamelia:2008qg,Bosso:2017hoq} the author of this thesis introduces the field of QGP. 
The particular field of QGP is concerned with modeling effects of the quantization of gravity and spacetime,
and the phenomena that these quantizations imply. QGP is a relatively new field of study, which has already produced interesting results and constraints on some parameters and scales in which QG effects are relevant.
\section{COW experiment} \label{sec:COW}
  A precursor to the field of QGP was a work by Colella, Overhauser, and Werner published in 1975 \cite{PhysRevLett.33.1237,Colella:1975dq}. This work examines the validity of the Schr\"odinger equation
\begin{equation}
        i\hbar \frac{\partial \psi(x,t)}{\partial t}=\left[-\frac{\hbar^2}{2m}\nabla^2+M_G \,\phi(\vec{r})\right]\psi(x,t)\,,
\end{equation}
when describing the dynamics of matter with a wave function $ \psi(x,t)$  in the presence of the Earth's gravitational potential $\phi(\vec{r})$. Such models are in short called Schr\"odinger–Newton models.  The COW experiment uses neutrons due to the fact that they have no charge and thus the effects of the gravitational potential are not drowned by the electromagnetic field of the Earth. The experiment is designed to see how their interference pattern is influenced by the presence of the Earths gravitational field.  
The weakness of the effects of the gravitational interaction a single particle experiences is mitigated by the large number of particles that form the interference pattern and the overall effects of the gravitational fields is enough
to induce observable effects. The COW experiment and Newton-Schrodinger equation \cite{PhysRev.187.1767} prove that the Earth's gravitational field coherently changes the phase of the neutron wave function. 

The COW experiment is the first experiment explicitly proving that gravity has observable effects on quantum systems. Further, the COW experiments show that the effects of the gravitational force can be amplified  using statistical systems to produce observable results. However, the models use a Newtonian description of gravity, which leaves much room for improvement. 

By showing that gravitational effects are observable in quantum experiments, Colella, Overhauser, and Werner's results inspired further attempts to improve the models and experiments. Their results laid the foundation for the development of the field of QGP. 

The COW experiment gives rise to another interesting question about the scale on which QG effects will be relevant. In chapter \ref{Ch1:Introduction}, the most promising candidates for a theory of QG were presented. In section \ref{Sec:MinimumLengthAndQG} it was shown that they share a parameter, namely the minimum measurable length scale. The minimum length scale is assumed to be of the order of Planck length
\begin{equation}
    l_{\text{Pl}}\equiv\sqrt{\frac{\hbar G}{c^3}}\approx 1.6\times 10^{-35}m\,.
\end{equation}
The energy domain in which effects described by QG theories will be dominant is energies of the order of the Planck energy
\begin{equation}
E_{\text{Pl}}\sim 10^{19}GeV\,.
\end{equation}
Although current experiments are still orders of magnitude away from probing such high energies, analogous to the COW experiments it is reasonable to expect that there will be some low energy artefacts of the QG effects which show up in current experiments. In the following chapter a few of the systems that show the most promise in providing experimental evidence of QG effects are discussed.

\section{The analogy with Brownian motion}\label{sec:BrownianMotion}

  One of the hypotheses about quantum spacetime is that space and/or time is discrete at the Planck scale. In order to directly confirm such hypothesis one needs an experiment that can probe spacetime at extremely high energy. However, it may be recalled that the atomic hypothesis was confirmed  by the observations of effects from the random motion of atoms. This was the great work by Einstein on Brownian motion in 1905 \cite{Einstein1905}. This motion typically consists of random fluctuations in a particle's position inside a fluid. The motion is a result of the collisions between the  molecules of the fluid and the Brownian particle.  Importantly, the statistical nature of the motion of the atoms  and the large number of collisions per second $10^{14}$ acting as an amplifier allowed the resulting Brownian motion to be observed at scale 10 orders of magnitude larger than the atomic length scale. Therefore, something similar can happen for QG: QG effects may be manifest at 10 or 15 orders of magnitude higher than the putative QG scale (the Planck scale). This is what makes QGP an attractive option for exploring and detecting QG effects at lower (accessible) energy scales and thereby strengthen or refute a theory of QG. 

Analogously to the quantum hypothesis and Brownian motion, the effects of discreteness of space and/or time on a fundamental level are expected to show up in random processes.  For example, fluctuations on the propagation of light or kinematics of elementary particles. 

Discreteness of spacetime, however, is not the only effect that can influence stochastic processes and thus have an influence in well studied classical systems.
In fact, there exists a conjecture that 
gravitons in a squeezed vacuum state will introduce metric fluctuations. To illustrate the argument one must consider a metrics of the following form
\begin{equation}
    ds^2=dt^2-dx^2+h_{\mu\nu}dx^\mu dx^\nu\,,
\end{equation}
where $h_{\mu\nu}$ is a tensor representing the metric fluctuation. The interval between two events in that case can be written as 
\begin{equation}
    \sigma(x-x^\prime)=\sigma_0 +\sigma_1(h_{\mu\nu}) +\mathcal{O}(h_{\mu\nu}^2)\,,
\end{equation}
where $\sigma_0=1/2(x-x^\prime)^2$, and $\sigma_1$ is the first order shift in the distance. 
The metric-averaged retarded Green’s function for a massless scalar field becomes \cite{Ford1995}
\begin{equation}\label{Eq:LightConeFluctuations}
    \braket{G_{\text{ret}}(x,x^\prime)}=\frac{\theta(x-x^\prime)}{8\pi}\sqrt{\frac{\pi}{2\braket{\sigma_{1}^2}}}\text{exp}\left[-\frac{\sigma_0^2}{2\braket{\sigma_{1}^2}}\right]\,,
\end{equation}
where  $\theta(x-x^\prime)$ is the generalized theta function. One can see that the averaged retarded propagator for the scalar field is 
a Gaussian function which is nonzero both inside and outside of the classical light cone. As one knows the speed of light is the slope of the light cone. According to Eq.\eqref{Eq:LightConeFluctuations} the light cone seems to fluctuate around the classical one. Therefore the speed of photon propagation fluctuates around the classical speed of light \cite{Ford1995}. Since by its very nature noise is a 
random process present in all physical experiments, it is reasonable to expect the random effects of quantum gravity to contribute in the  noise of the incredibly sensitive interferomters used to measure gravitational radiation \cite{AmelinoCamelia:1999gg,Hogan:2008zw}. 

An important study of the noise of such detectors was presented by P. Bosso in 2018 \cite{BOSSO2018498}. In his work P. Bosso explores effects due to the existence of minimum length on the shot noise present in LIGO experiment. He concludes that the corrections to the noise are  $10^9$ times smaller than the shot noise itself. That result effectively puts an upper bound on the scale on which QG effects are expected to show up.

\section{Symmetries of spacetime and large extra dimensions}\label{sec:Symmetries}

As discussed in a previous chapter a common feature that some QG theories share is the existence of minimum measurable length proportional to the Planck length. Therefore, QG effects are relevant on very high energy scales. It is important to note that length itself is not a Lorentz invariant entity. Furthermore, the classical spacetime symmetries do not account for the existence of such length. Thus, most theories of QG propose that at Planck scales the fundamental symmetries of spacetime introduced by SR, such as homogeneity\footnote{The laws of physics are the same at every point in spacetime.}, isotropy\footnote{There is no preferred direction in spacetime.}, and the fact that spacetime is continuous and smooth manifold, are either violated, modified or generalized. 
   
According to SR there is one frame independent scale that governs the transformation laws between observers, {\it i.e.} $c$ the speed of light. These transformations form the Poincar\'e group and its subgroup, the Lorentz group. Therefore, one can say that the symmetries of a flat spacetime are governed by the speed of light. 

If one introduces additional scale or structure in a flat spacetime, its
symmetries will be accordingly affected. This can be seen by studying  Planck-scale discretization of spacetime proposed by 't Hooft in 1996 \cite{tHooft:1996ziz}. Continuous symmetry transformations such as the Lorentz transformations are clearly incompatible with a discrete network
of points. Lorentz symmetry is also often at odds with spacetime non-commutativity or fuzziness. Spacetime non-commutativity introduces a minimal measurable length similarly to the HUP. In particular, the
non-commutativity  and minimum length scale \cite{AmelinoCamelia:2000ge} affect the laws of transformation between inertial observers. The infinitesimal symmetry transformations are actually described in terms of the
language of Hopf algebras \cite{Majid:1994cy,Lukierski:1993wx}, rather than by the Poincar\'e algebra. 
In a large number of recent studies of non-commutative spacetimes it has
indeed been found that the Lie-algebra formed by the Poincar\'e symmetries is either broken
to a smaller Lie algebras or deformed into Hopf-algebra \cite{mattingly2005modern,Liberati:2013xla}. This allows for the rotations operators to remain unchanged as well as momenta operators to remain commutative.

The nature of spacetime and Lorentz symmetry, and in particular its violation will impact QM and its extensions too. According to QFT, Lorentz symmetry is closely related with CPT symmetry, which stands for:  charge conjugation (C); parity transformation (P); and time reversal (T). So a QG theory that violates  Lorentz symmetry is expected to break CPT symmetry. Theories utilizing generalizations of the Heisenberg uncertainty relation or ones modifying the Einstein dispersion relation usually modify the Lorentz generators as well. Due to this  that  will violate CPT symmetry and in turn Lorentz covariance.
In modeling of systems with modified symmetries of spacetime, one can use effective field theories with added Lorentz violating terms. For example, one can consider the following modified dispersion relation
\begin{equation}\label{Eq:DSRDispersion}
 \frac{E^2}{c^2}-p^2=m^2c^2 +\sum_{n=1}^N \eta^{(n)}\frac{p^n}{(M_{Pl}c)^{n-2}}\,,
\end{equation}
where $\eta^{(n)}$ is dimensionless parameter which is used for fixing the theory. With the terms within the summation in the RHS are additional ones to the standard relativistic dispersion relation. 
This modification is similar to that proposed in DSR theories.
Using the Lorentz violating modified dispersion relation Eq.\eqref{Eq:DSRDispersion}, one can derive modifications of the reaction thresholds for particles in Ultra-High-Energy Cosmic Rays. The reason one uses reaction thresholds is that they rely not on the corrections for the total energy of the particle but on its mass. Therefore the terms violating the Lorentz symmetry become relevant when they are of the order of magnitude of the mass of the heaviest particle. 

This effectively sets a threshold for the critical energy 
\begin{equation}
   E_{cr}\sim \left(m^2 M_{Pl}^{n-2}\right)^{1/n} \,,
\end{equation}
where $m$ is the mass of the heaviest particle in the decay. Above the critical energy $E_{\text{cr}}$ Lorentz violationg effects are expected to show up.

Considering an effective field theory at energies close to $E_{cr}$, one can calculate cross-sections for the interactions involving Ultra-High-Energy Cosmic Rays. Comparing it to the unmodified case one can obtain the following bounds on $\eta$,
\begin{equation}
-10^{-3}\leq\eta\leq 10^{-6}\,
\end{equation}
where $\eta$ is considered to be the same for all terms in Eq.\eqref{Eq:DSRDispersion}. 

 Another hypothesis is that the principle of relativity is not obeyed at the Planck scale $E_{Pl}$, and therefore there is a preferred state of motion and rest.
 If SR is violated the speed of light would be no longer an invariant.  One can then look for a variation in $c$ proportional to $E/E_{Pl}$, where $E$ is the
energy of a photon.  
One looks for an energy dependent speed of light of the form $v=c(1\pm \kappa E/E_{Pl} )$, where $a$ is a dimensionless parameter to be determined.  This effect can potentially found in  light coming from astrophysical sources such as gamma ray bursts and supernovae. The reason for this is the fact that even if QG effects are of the order of the Planck length  $\sim 10^{-19}GeV$, the distance traveled by the photons acts as an amplifier. The stacking of these small effects will result in a arrival time of a photon being offset by a few seconds.  
Thus far, such an effect has not been observed, and so one must conclude that the parameter $\kappa\ll 1$ \cite{Ackermann:2009aa}. 

The Fermi space telescope has detected a number of very highly energetic gamma ray bursts bigger than what was  previously expected. Furthermore, several of the documented  bursts show  the higher energy photons ($>$GeV) arrive with a delay of more than 10 seconds after the onset of the burst has been observed in the low energy range (keV-MeV). It is still unclear what is the cause of this delay of the high energetic photons. More statistics and further analysis will eventually allow the dependence of speed of light on the energy of the photons to be put to the test.  \cite{AmelinoCamelia:2009pg,Maccione:2007yc}.

The principle of relativity can be tested by using the prediction that very high energy cosmic ray protons interact with the cosmic microwave background (CMB).  Such interactions were predicted to take place at an energy of above $10^{19}$~eV, and cause the protons to lose energy. This leads to the so called Greisen-Zatsepin-Kuzmin (GZK) cutoff on the ultra high energy cosmic rays spectrum \cite{PhysRevLett.16.748}.  The above states that cosmic ray protons coming from further away than $75$ megaparsecs should not be observed. The constraint on that distance is the mean free path for this interaction. This prediction was confirmed recently by observations at the AUGER cosmic ray detector \cite{Abraham:2008ru}. 

The results presented here and other analyses as well \cite{Stecker:2011ps} make it seem unlikely the principle of relativity breaks down at Planck scales, at least at order $E/E_{Pl}$.  The results to date do however supports a more subtle hypothesis such as DSR \cite{AmelinoCamelia:2000ge,Magueijo:2001cr}.  Additionally, there is a possibility that the principles of SR break down in a way that can only be observed by experiments sensitive to effects of the order of $(E/E_{Pl})^2$.

The hypothesis that QG effects break P-symmetry has implications for observations of the CMB.  It leads to predictions for a signal in the CMB spectrum which would show up in correlations between the temperature fluctuations and certain polarization modes, called $B$-modes \cite{Lue:1998mq}. The Planck satellite is expected to probe this effect. 

 A completely different category of models studies the possibility that quantum gravitational effects could be much stronger than usually thought due to a modification of the gravitational interaction at shortest distances. Such a modification occurs in scenarios with large additional spatial dimensions, 
the existence of which is predicted by string theory, 
and has the consequence that quantum gravity could become observable in Earth-based collider experiments, such as the Large Hadron Collider (LHC).

If this should turn out to be a correct description of Nature, one would see the production of gravitons and black holes at the LHC \cite{Landsberg:2008ax}. The gravitons themselves would not be captured in the detector and would lead to a missing energy signal, the missing particles having spin 2. 
Ideally, the distribution of decay products would allow for the determination of the parameters of the model; the number and size of the extra dimensions. Black hole production and decay would be a striking signature, and allow us to examine the fate of black hole information during the evaporation process in the laboratory.

\section{Strong gravitational fields}

  When talking about strong gravitational field one can not help but think of compact objects such as black holes and neutron stars. It turns out that objects such as these play a major role in the field of QGP. In particular thought experiments in which the radius of the apparent horizon of a microscopic black hole is measured show that the position-momentum uncertainty relation must be modified to allow for the existence of minimum measurable length \cite{MAGGIORE199365,PhysRevD.49.5182}. 
  
Another avenue for the observation of QG effects is the detection of Gravitational waves coming from merging black holes detected by the LIGO experiments. Due to the methodology used, the observation of gravitational waves relies greatly on the underlying theoretical model. In particular, GR shows that gravitational waves will obey 
the usual Einstein dispersion relation. However, as mentioned above, many theories of QG 
suggest that this dispersion relation is modified. Thus, if one models Gravitational wave detection with this modified dispersion relations and then compares the signal to noise ratio of the cases with and without QG correction, one find the scale of such corrections.

The Cosmic Microwave Background (CMB) is another physical phenomenon which can be used to test QG effects in strong gravitational fields. The energy density in the early Universe suggests that the gravitational fields at this early stage of the evolution of the Universe were very strong. Therefore, it is reasonable to believe that QG effects amplified by the strong gravitational fields in the early Universe have been encoded in the CMB. 
In light of some of the results presented in this work, the role of CMB in the field of QGP will be mentioned again when future research and implications of the results are discussed in chapter \ref{Ch5:Conclusion}.

\section{Generalized Uncertainty Principle }\label{sec:GeneralizedUncertaintyPrincile}
   In Chapter \ref{Ch1:Introduction}, it was discussed how minimal length and a minimum uncertainty in position follow from different approaches to quantizing the force of gravity. Arguments presented were based mainly on the assumption that at high energies used to probe the small length scales would actually disturb the structure of spacetime  with their gravitational effects.
   
As mentioned before, a fuzzy spacetime gives rise to a minimum measurable length similarly to the GUP. The modification of the HUP  to accommodate for the existence of minimum measurable length, instead of considering non-commutative spacetime, is not a significant leap.  In the 1995 paper by 
Kempf, Mangano, and Mann \cite{Kempf1995}, the authors apply the previously mentioned principles to QM. The particular form of the generalized uncertainty relation has been proposed independently by ST \cite{AMATI198941,Chang:2011jj,KONISHI1990276} and black hole thought experiments \cite{MAGGIORE199365,PhysRevD.49.5182}. 
The corresponding position-momentum commutator is to be given by
\begin{equation}\label{Eq:KMMGUP}
    [x,p]=i\hbar(1+\beta p^2)\,.
\end{equation}
 This particular commutator gives rise to an uncertainty relation of the type 
\begin{equation}
    \Delta x \Delta p \geq \frac{\hbar}{2}(1+\beta(\Delta p)^2)\,, \label{Eq:MinimumLength}
\end{equation}
more commonly known as Generalized Uncertainty Principle (GUP), where $\beta$ is positive and independent of $\Delta x$ and $\Delta p$, and $\beta$ will have dimensions of inverse momentum squared \cite{Kempf1995}. There are models utilizing negative $\beta$, however, in the following work the author has c. It is easy to see that this uncertainty relation will give a minimal uncertainty in position $\Delta x_0 =\hbar\sqrt{\beta}$. This minimum uncertainty in position is proportional to the
Planck length, which as seen previously,
can be expressed as
a combination of the fundamental constants in nature, namely $G,c$ and $\hbar$. In fact it is
evident from Eq.\eqref{Eq:MinimumLength} that there is a threshold below which, even if  the uncertainty in momentum is increased, the uncertainty in position will increase instead of approaching infinity as HUP suggests.

 One can naturally ask as to what happens to the Hilbert space representation of QM in this case. In order of ensuring that the a minimum uncertainty in position  exists. One must generally require physical states to be normalizable,
to have well defined expectation values of position and momentum, and also
well defined uncertainties in these quantities. Furthermore the physical states in that case will be constrained to the common domain of the symmetric operators $\bf{x},\,\bf{p},\,\bf{x^2},$ and $\bf{p^2}$.

By introducing GUP one introduces a minimal uncertainty in position $\Delta x\geq \Delta x_0$, where  $\Delta x_0$ is a non zero positive number. Therefore one can write the standard deviation of the position operator
\begin{equation}
   (\Delta x)^2=\bra{\psi}\left({\bf x}-\bra{\psi}{\bf x}\ket{\psi}\right)^2\ket{\psi}\geq \Delta x_0~~\forall \psi\,.
\end{equation}
Position eigenstates have zero uncertainty $\Delta x=0$, and therefore are infinitely localizable. Similarly to QM, it naturally follows that the physical states are not eigenstates of the position operator. Since the physical states all belong to the Hilbert space and the eigentates of all self-adjoined oparators form an orthotormal basis in the Hilbert space, it is easy to conclude that position operator is no longer self-adjoint, only symmetric. The symmetry of the position operator ensures that all position expectation values are real. 

Additionally, it is no longer possible to approximate eigenvectors by series of physical states with $\Delta x $ decreasing to zero. Therefore, according to all the reasons mentioned above, uncertainty relation of the type presented in  Eq.\eqref{Eq:KMMGUP} implies that the Heisenberg algebra will no longer find a representation in the Hilbert space of position wave functions. 

  However, one can still define a quasi-position representation in which one defines maximal localization states $\ket{\psi^{\text{ml}}_\xi}$, for which 
  \begin{equation}
      \bra{\psi^{ml}_{\xi}}{\bf x}\ket{\psi^{ml}_{\xi}}=\xi\,.
  \end{equation}
  The standard deviation of the position operator over such states is $\Delta x_0$.
The momentum ${\bf p}$ and position ${\bf x}$ operators acting on the quasi-position wave function gives
\begin{align}
{\bf p}\cdot\psi({\xi}) &= \frac{\tan(-i\hbar\sqrt{\beta}\partial_{{\xi}})}{\sqrt{\beta}} 
\psi({\xi})\\\label{Eq:Quasy-position}
    {\bf x}\cdot \psi(\xi)&=\left(\xi +i\hbar\beta\frac{\tan (-i\hbar\sqrt{\beta}\partial_{\xi})}{\sqrt{\beta}}\right)\psi(\xi)\,.
\end{align}
It follows from the above that the position and momentum operators ${\bf x}$ and  ${\bf p}$ can be expressed
in terms of the multiplication and differentiation operators $\xi$ and $-i\hbar\partial_{\xi}$ which obey the
canonical commutation relation. Additionally, in the limit $\beta \rightarrow 0$ from the quasi-position representation one recovers the usual position representation. However, the position operator defined in Eq.\eqref{Eq:Quasy-position} is not diagonalizable nor symmetric in any part of the domain of $\bf{x^2}$. 

Note that the position spectrum in this case is not discrete. Space is a continuous flat background, but with a measurable minimum. On the other hand, momentum is also a continuum but does not have a measurable minimum or maximum. 
Equivalently, the momentum eigenfunctions are Dirac delta functions in momentum space. However, in \cite{Bosso:2020aqm} P. Bosso showed that a proper position representation is not possible in theories with minimum length, due to the absence of position eigenfunctions. 

Applying the methodology proposed \cite{Kempf1995} in  to the standard case of linear harmonic oscillator, one can get a corrected energy spectrum
as follows
\begin{equation}
 E_n=\hbar\omega \left(n+\frac{1}{2}\right)\left(\frac{1}{4\sqrt{r}}+\sqrt{1+\frac{1}{16r}}\right)   +\hbar\omega\frac{1}{4\sqrt{r}}n^2\,,
\end{equation}
where $E_n$ is the energy of the $n$-th level, and $r=(2\beta\hbar m\omega)^{-2}$.

 Eq.\eqref{Eq:KMMGUP} can be generalized to include $N$ space dimensions
\begin{equation}\label{Eq:KMMGUP3D}
 [x_i,p_j]=i\hbar\delta_{ij}(1+\beta \vec p^2)\,,
\end{equation}
where the indices $i,j,k\in \{1,2,3,\ldots,N\}$ \cite{Kempf1995}.
Because one would like to have the Hilbert space representation for the momentum operators to remain the same, i.e. to allow an infinite localization of momentum.  The momentum operators must be commutative. The position operators on the other hand, do not commute, and one has the following commutation relation:
\begin{equation}
    [x_i,x_j]=2i\hbar\beta(p_ix_j-p_jx_i)\,.
\end{equation}
The authors of \cite{Kempf1995} have shown that the above  non-commutativity of space is essential to have a modification of the position and momentum commutator. Using the above commutator one can derive the modified rotation generators
\begin{equation}
L_{ij}=\frac{1}{1+\beta \vec p^2}(p_ix_j-p_jx_i)\,.
\end{equation}
According to \cite{bosso2017generalized} it can be proven that for the above, the algebra of the rotation group will be modified leading to interesting effects.

In conclusion, the existence of minimal length have implications for even the simplest non-trivial case. Numerous studies have been conducted in this paradigm and reinforced the above. 

\subsection{Universality of Quantum Gravity effects}

  Gravity is a fundamental force
and is universal; this means that it affects and is affected by all systems which possess mass or energy. This is despite  the extreme smallness of the gravitational (or Newton's) constant $G$ which makes quantum gravity effects normally too weak to be measurable.
In their paper \cite{Das2008}, Saurya Das and Elias Vagenas argued 
that quantum gravity corrections can be estimated for any system with a well defined Hamiltonian. Effects which result from quantum gravity corrections of these systems are proportional to the Planck length squared $l_{Pl}^2=G\hbar/c^3$. 
Thus far, it was discussed how a minimal measurable length and the modified uncertainty relation arise in various theories and some of its implications. 
Here, some of their potential experimental signatures in some well known quantum systems will be reviewed.

The parameter $\beta$ introduced in the previous sections can be written 
as a product of two factors, an arbitrary dimensionless $\beta_0$ used to fix the model and dimensional $1/(M_{Pl}c)^2=l_{Pl}^2/\hbar^2$, where $M_{Pl}$ is the Planck mass. The dimensionless parameter is often assumed to be of the order of unity, which means the minimum measurable uncertainty in position is of the order of Planck length. In other words, the GUP parameter can be written as 
\begin{equation}
    \beta=\frac{\beta_0}{(M_{Pl}c)^2}\,.
\end{equation}
However, instead of 
imposing such a limitation, one can obtain upper bounds on $\beta_0$ from  experiments, as shown in 
\cite{Das2008}.  
A given value of $\beta_0$
corresponds to the existence of minimum length scale  $l_{min}=\sqrt{\beta_0}l_{Pl}$. This length scale cannot exceed that corresponding to electroweak interaction
($\approx 10^{-19}$ m), as otherwise it would already have been observed, say in the Large Hadron Collider (LHC).
This gives a bound that $\beta_0\leq10^{34}$.

As done in \cite{Ali:2010yn}, one can consider the most general form of the position-momentum commutator
\begin{equation}\label{Eq:GeneralGUP}
    [x_i,p_j]=i\hbar F(\vec{p})\,,
\end{equation}
where $F(p)$ is an arbitrary function of momentum. After an expansion of that function in Taylor series one gets 
 \begin{equation}\label{Eq:ADVcommutator}
     [x_i, p_j] = i \hbar \left(  \delta_{ij}-\beta\left( p \delta_{ij}+\frac{p_i p_j}{p} \right)+\beta^2 \left( p^2 \delta_{ij}  + 3 p_{i} p_{j} \right)  \right)+\mathcal{O}(\beta^3,p^2\vec{p})\,,
 \end{equation}
 where
$\beta = {\beta_0}/{M_{Pl}c} = {\beta_0 \ell_{Pl}}/{\hbar},$
$M_{Pl}=$ Planck mass, $\ell_{Pl}\approx 10^{-35}~m=$ Planck length,
and $M_{Pl} c^2=$ Planck energy $\approx 10^{19}~GeV$. It is evident from Eq.\eqref{Eq:ADVcommutator} that the position and momentum operators are not canonically conjugate. Therefore one introduces a pair of auxiliary canonically conjugate variables $x_0$ and $p_0$ for which 
\begin{equation}
    [x_{0 i},p_{0 j}]=i\hbar \delta_{ij}.
\end{equation}
The physical position and momentum operators then can be expressed as functions of those auxiliary variables. An example of an explicit expression of the physical in terms of the auxiliary variables is as follows 
\begin{equation}
    x_i = x_{0 i}\quad,\quad p_i = p_{0i} \left( 1 - \beta p_0 + 2\beta^2 p_0^2 \right)\,,
\end{equation}
where the $x,p,x_0$ and $p_0$ without an index or vector sign signify the expectation value of the respective operator. 

  One can see from Eq.\eqref{Eq:ADVcommutator} that every correction to the commutator is of higher order in momentum  is multiplied by the same order of the coefficient $\beta$, which is very small. Therefore in terms of these
  newly defined physical momentum operators, one can write any general Hamiltonian. One can than expand the general Hamiltonian in Taylor series with respect of $\beta$. The  result is a leading order Hamiltonian proportional to $\beta^0$,  plus perturbation terms proportional to $\beta^n$ where $n\in\{1,2,3,\ldots\}$. In order of illustrating this one can consider the correction terms up to  $n=2$ 
 \begin{equation}
     H=H_0+H_1+H_2\mathcal{O}(\beta^3)\,,
 \end{equation}
 where 
 \begin{equation}\label{Eq:UniversalCorrections}
     H_0=\frac{p_0^2}{2m}+V(t)\qquad \text{and} \qquad H_1+H_2=-\frac{\beta}{m} p_0^3+ \frac{5\beta^2}{2m}~p_0^4\,.
 \end{equation}
The corrections to the Hamiltonian are corrections to the kinetic part of the Hamiltonian and they do not depend on the potential. Therefore, system with non-zero kinetic energy will experience QG effects. Furthermore, one can use this methodology to obtain QG corrections to any system with a well defined Hamiltonian.

One can apply this line of thinking to a quantum system such as the hydrogen atom, and one finds the correction due to GUP effects for Lamb shift of the energy levels \cite{Brau:1999uv}.

 The perturbation of the Hamiltonian for the hydrogen atom with GUP has the following form
 \begin{equation}
     H_1+H_2=(4\beta m)\left[H_0^2+k\left(\frac{1}{r}H_0+H_0\frac{1}{r}\right)+\left(\frac{k}{r}\right)^2\right]\,,
     \end{equation}
 where $k$ is the 
 fine-structure constant. The Lamb shift for the $n$th energy level, which one can calculate from the leading order of the Hamiltonian is
 \begin{equation}
    \Delta E_n=\frac{4\beta^2}{3m^2}\left(\ln\frac{1}{\beta}\right)|\psi_{nlm}(0)|^2\,.
 \end{equation}
The GUP corrections for the Lamb shift can be obtained by taking the expectation value of the perturbed Hamiltonian \cite{Ali2011}. Comparing the leading term and the GUP corrections one gets
 \begin{equation}\label{Eq:LambShiftRelativeCorrection}
     \frac{\Delta E_{0(GUP)}}{\Delta E_{0}}\approx10\beta_0\frac{m E_0}{M_{Pl}^3}\,.
 \end{equation}
 The most recent experiments have not yet observed effects from the perturbation of the Hamiltonian, therefore the GUP effects must be smaller than their uncertainty. Using the accuracy of the most recent experiments, one gets the following upper bound on the dimensionless parameter:  $\beta_0\leq10^{36}$. 
 Similarly from the Landau levels of a charged particle in magnetic field, one gets  $\beta_0\leq10^{50}$, while the potential barrier in a scanning tunnel microscope (STM) yields the bound $\beta_0\leq10^{21}$.
More work needs to be done in this direction,
because QG theories tell that $\beta_0$ should be of the order of 1. Many other studies yielding corrections to the energy levels of various systems have been conducted \cite{Bosso2018,Ali2010,Das2011,Das2008,Das2014,bushev2019testing}.

Noticeably, the bounds on the $\beta_0$ parameters are extremely large. However one can see that the relative size of the GUP corrections as presented in Eq.\eqref{Eq:LambShiftRelativeCorrection} are linearly proportional to the energy $E_0$ and mass $m$ of the studied system. This result suggests that there are two ways one can improve on these bounds, based on the parameters of the studied system. The first is to consider highly energetic systems, such as electromagnetic scattering of accelerated particles, and the second arises from studying  more massive systems. 

Additionally, it is expected that future experiments will improve on these bounds by increasing the accuracy. Furthermore, in the present work by considering a frame independent minimum length and relativistic generalized uncertainty principle, the author was able to improve the bound on the $\beta_0$ parameter by almost 10 orders of magnitude. This persuades one that further study and development of the idea might lead to promising new results.

\section{Summary}
  Phenomenology is the branch of physics responsible for  making models out of already existing theories and then applying these models to current or future experiments, in search of new and interesting phenomena. In chapter \ref{Ch2:QuantmGravityPhenomenology} above the field of QGP is introduced as a relatively new field which tries to close the gap between current experiments and the family of theories attempting to quantize the force of gravity.
  
In section \ref{sec:COW} the COW models and experiments are introduced as a precursor to the field of QGP. The experiment shows that the phase of a neutron wave function is affected by the Earth's gravitational field. Many consider the 1975 experiment to be a precursor to the field of QGP \cite{AmelinoCamelia:2008qg}. 

In section \ref{sec:BrownianMotion} the author reviewed the similarly of QG corrections to the way Brownian motion was detected at a length scale 10 orders of magnitude bigger than the atomic radius. According to this analogy, low energy effects originating from QG effects at Planck scale should be present in the noise of current high precision interferometers.

Potential tests of the fundamental properties of spacetime  close to Planck energies is discussed in section \ref{sec:Symmetries}.

Since this work is a phenomenological study of existence  minimum measurable length through the generalization of the Heisenberg uncertainty relation, much attention is paid to the GUP as a way of introducing minimum uncertainty in position to quantum mechanical systems. In section \ref{sec:GeneralizedUncertaintyPrincile} a brief discussion of some of the more interesting consequences of the  GUP effects is presented. Furthermore the author presented evidence that the GUP approach to QGP is very general and can be applied to a wide array of quantum systems and effects. Lastly, experimental bounds on the scale of the GUP parameters are put into place showing plenty of room for improvement and prompting further study. 

\chapter{Relativistic Generalized Uncertainty Principle}
\label{Ch3:RGUP}
\begin{quote}
	“Not only is the Universe stranger than we think, it is stranger than we can think.”
	
	\begin{flushright}
		 ― Werner Heisenberg,  Physics and Philosophy: The Revolution in Modern Science 1958
	\end{flushright}
	\end{quote}

As seen from chapter \ref{Ch1:Introduction}, theories of QG, such as ST, LQG, as well as DSR, have one thing in common: 
they all predict a minimum measurable length or a scale in spacetime.
This is in direct contradiction with the Heisenberg Uncertainty Principle \cite{Heisenberg1927}, since the latter allows for infinitely small uncertainties in position.

Chapter \ref{Ch2:QuantmGravityPhenomenology} discussed phenomenological approaches to QG. The majority of the phenomenological studies dealing with the existence of minimum length and its effects on experiments use an extension of the Heisenberg uncertainty principle to a new so-called Generalized Uncertainty Principle (GUP). 
The minimum length, normally considered to be of the order of 
Planck length, $\ell_{Pl}=10^{-35}$ m, signifies the scale at which QG effects would manifest. While this is a valid approach if one models low energy systems described by QM, when considering applications to higher energy scales one is faced with the problem that theories incorporating a minimum length break Lorentz covariance, simply because a measured length is not a Lorentz invariant quantity. The above mentioned Lorentz symmetry breaking implicitly implies a special 
choice of a frame of reference, effectively 
bringing back the notion of an aether, to circumvent the results which, Einstein obtained in the Theory of Special Relativity!  
Due to this difficulty, GUP models so far have mainly been considered in non-relativistic context. 

There has been a few attempts applying GUP to relativistic theories and quantum field theory \cite{Szabo:2006wx,Chaichian:2004za,snyder1947quantized,Quesne:2006is,Faizal:2017map,Faizal:2014dua,Pramanik:2014zfa,Deriglazov:2014yta,Pramanik:2013zy,Husain:2013zda,Kober:2010sj,Hossenfelder:2006cw,Capozziello:1999wx,zakrzewski1994quantum,Kober:2011dn}. 
However, the minimum measurable length proposed by them is still not a Lorentz invariant quantity. Furthermore, 
other difficulties such as the Composition Law problem are met. The Composition law problem \cite{Hossenfelder:2014ifa,amelino2011relative,hossenfelder2013comment,amelino2013reply}, outlined in Appendix \ref{app:CLP}, arises when one considers a dispersion relation containing terms of higher than second order in momentum. If one considered a system of a large number of non-interacting particles, naturally one expects that the energy of the whole system is equal to the sum of all the particles in it. However, when one considers a  modified dispersion relation of the form 
\begin{equation}\label{Eq:modifiedDispersionRelation}
    D^{2}(p) = E^2-  \vec{p}^2 -  \eta \frac{E}{M_{\text{Pl}}} \vec{p}^2+ \cdots = m^2\,,
\end{equation}
where $\eta$ is a dimensionless numerical coefficient, and the speed of light is considered to be one $c=\hbar=11$. The non-linear corrections to the Einstein's dispersion relation between energy and momentum arise from a
 choice of a non-trivial connection on momentum space, $\cal P$.  
In the case of plastic collision $A + B \rightarrow C$,  the composition of momenta takes the form
\begin{equation}
    p^{(C)}_\mu = \left(p^{(A)} \oplus p^{(B)}\right)_\mu\,.
\end{equation}
The above relies on the assumption that there exist a nice set of coordinates on the momentum space $\cal P$, which allows for the expansion of the composition of two momenta,
\begin{equation}\label{Eq:MomentumComposition}
\left(p^{(A)} \oplus p^{(B)}\right)_\mu =p^{(A)}_\mu + p^{(B)}_\mu -
\frac{1}{M_{\text{Pl}}}  \tilde{\Gamma}_\mu{}^{\alpha\beta}\, p^{(A)}_\alpha\,
p^{(B)}_\beta+ \cdots\,.
\end{equation}
${\tilde \Gamma}$ denotes the
 connection coefficients on momentum space evaluated
at the origin $p_{\mu}=0$. The
 connection coefficients
 on momentum space $\tilde{\Gamma}_\mu{}^{\alpha\beta}(0)$ is dimensionless.
One can write the dispersion relation as the usual quadratic term plus leading order corrections. This is justified in the
case of elementary particles, because even for most energetic cosmic
rays the ratio of their energies to $M_{\text{pl}}$ is of order of $10^{-8}-10^{-9}$,
and the higher order terms can be safely neglected.

When one considers a composite system of a huge number $N$ of elementary non-interacting particles with identical masses $m$ and
momenta $p_\mu$ and obeying the same dispersion relation shown in Eq.\eqref{Eq:modifiedDispersionRelation}. Then, the total mass of the system is
$M_{\text{system}}= N\, m$ and its total momentum is $\mathbf{P}_{\text{system}\,\,\mu} =
N\, p_\mu$. Substituting this to Eq.\eqref{Eq:modifiedDispersionRelation} it can be easily shown that
\begin{equation}\label{Eq:TotalDispersionRelation}
    E_{\text{system}}^2=  \vec{\mathbf{P}}_{\text{system}}^2 + M_{\text{system}}^2 + N\eta\, \frac{E_{\text{system}}}{ M_{\text{Pl}}} \vec{\mathbf{P}}_{\text{system}}^2+ \ldots
\end{equation}
After comparing Eq.\eqref{Eq:modifiedDispersionRelation} with Eq.\eqref{Eq:TotalDispersionRelation} one notices that considering the composite system makes the QG corrections grow linearly with the number of parts $N$, which implies that QG should have been detected in statistical systems. 

Presented in this chapter is a study proposing a Lorentz 
covariant or relativistic GUP, which provides a frame independent minimum length and reduces to its familiar non-relativistic versions at low energies. Furthermore, the symmetries of spacetime expressed through the Poincar\'e group are modified in a way which preserves the quadratic nature of Einstein dispersion relation. In this way the composition law problem is avoided. The result of the study are summarized and published in \cite{Todorinov:2018arx}.

\section{Lorentz invariant minimum length}

  Evidently, from Eq.\eqref{Eq:KMMGUP3D}, while $x^{i}$ and $p^{i}$ are the physical position and momentum, they are not canonically conjugate. Therefore, the expression of the connection between the two operators becomes less straightforward.  
To remedy this in the following calculations, 
two new $4$-vectors, 
$x_0^\mu$ and $p_0^\mu$ are introduced, such that
these auxiliary variables are 
canonically conjugate. That is,
\begin{align}
\label{Eq:MathVar}p_0^{\mu} =  -i\hbar\frac{\partial}{\partial x_{0\,\mu}}, \quad
[x_0^{\mu},p_0^{\nu}] =  i\hbar\eta^{\mu\nu}.
\end{align}
One can then make the assumption that the physical positions $x^\mu$ and momentum $p^\mu$ are functions of the auxiliary ones $x_0^\mu$ and $p_0^\mu$,
\begin{equation}\label{Eq:PositionMomentumFunctions}
 x^\mu=x^\mu(x_0,p_0),\quad
 p^\mu=p^\mu(x_0,p_0)\,.
 \end{equation}
Operating under the assumption made in Eq.\eqref{Eq:PositionMomentumFunctions}, one makes the Taylor expansion of the two physical operators in terms of the auxiliary ones, using the following formula 
\begin{multline}\label{Eq:Taylor}
T(x_{0}^{0},\ldots ,x_{0}^{d},p_{0}^{0},\ldots ,p_{0}^{d})=\\\sum _{n_{0}=0}^{\infty }\cdots \sum _{n_{d}=0}^{\infty }\sum _{m_{0}=0}^{\infty }\cdots \sum _{m_{d}=0}^{\infty }
{\frac {(x_{0}^{0}-a_{0})^{n_{0}}\cdots (x_{0}^{d}-a_{0})^{n_{d}}(p_{0}^{0}-b_{0})^{n_{0}}\cdots (p_{0}^{d}-b_{d})^{n_{d}}}{n_{0}!\cdots n_{d}!\,m_{0}!\cdots m_{d}!}}\\\left({\frac {\partial ^{n_{0}+\cdots +n_{d}\,m_{0}+\cdots +m_{d}}f(x_{0}^{0},\ldots ,x_{0}^{d},p_{0}^{0},\ldots ,p_{0}^{d})}{\partial x_{0}^{0\,n_{0}}\cdots \partial x_{0}^{d\,n_{d}}\,\partial p_{0}^{0\,n_{1}}\cdots \partial p_{0}^{d\,n_{d}}}}\right)(a_{0},\ldots ,a_{d},b_{0},\ldots ,b_{d})\,,
\end{multline}
where $f(x_{0}^{0},\ldots ,x_{0}^{d},p_{0}^{0},\ldots ,p_{0}^{d})$ is a general function of the auxiliary momentum and position. The auxiliary position  $x_0^\mu$  and momentum  $p_0^\mu$  are non-commutative operators, which means that when taking the Taylor expansion one needs to be mindful of their order. 
The formula for the Taylor expansion presented above in Eq.\eqref{Eq:Taylor}, is then applied to the general form of the position and momentum operators given in Eq.\eqref{Eq:PositionMomentumFunctions}. Certain conditions applied to the position and momentum operators are discussed in the following paragraphs.

Commutator between two momentum operators different from zero, introduces ``fuzzyness'' in momentum space or minimum momentum uncertainty. Analogously the case discussed in chapter \ref{Ch2:QuantmGravityPhenomenology} section \ref{sec:GeneralizedUncertaintyPrincile} minimum uncertainty in momentum implies that no physical states are eigenstates of the momentum operator. Then the same is true for the Hamiltonian or energy operator, as it is a function of the momentum operator.  If the Schr\"odinger equation is to be true, any physical state must be an eigenstate of the Hamiltonian operator with the systems energy as an eigenvalue. Therefore, if one wants to have well defined eigenstates and eigenvalues of the energy operator one must make the following restriction on the momentum commutator
\begin{equation}
    [p^\mu,p^\nu]= 0\,,
\end{equation}
{\it i.e.} the physical momentum operators commutate with each other. In other words, momentum eigenvalues exist and can be measured with arbitrary accuracy. From this condition, a conclusion can be drawn that the physical momentum operator cannot be a function of the auxiliary position, since 
\begin{equation}
 [f(x_0,p_0),g(x_0,p_0)]\approx \frac{i\hbar^3}{(l_{\text{Pl}})^2}[p_0,p_0]+\frac{i\hbar^3}{(M_{\text{Pl}}c)^2}[x_0,x_0]+[x_0,p_0]+\mathcal{O}(x_0^2,p_0^2)\,.
\end{equation}
The auxiliary position-momentum commutator is never equal to zero.  eTherefore one can conclude that in order of the physical momentum operators to be commutative they cannot be functions of the auxiliary position only of the auxiliary momentum.

The goal is to construct a Relativistic Generalized Uncertainty Principle (RGUP) which gives Lorentz invariant or frame independent minimum uncertainty in position. Therefore, our position operator must obey the Lorentz/Poincar\'e symmetries, or equivalently the physical position and momentum operators must be four vectors transforming under the Poincar\'e algebra. Therefore, terms in the Taylor expansion containing any more or less than one free index must be zero. 
When all of the conditions described above are met, the physical position and momentum operators reduce to infinite polynomials of the auxiliary variables
\begin{subequations}
\begin{align}
 \nonumber  x^\mu(x_0,p_0)&=C_1\,x_{0}^{\mu} + C_2 p_{0}^{\mu}+C_3\,x_{0}^{\mu}x_{0}^{\nu}x_{0\nu}+C_4\,x_{0}^{\nu}x_{0\nu}p_{0}^{\mu}\\\label{Eq:GeneralPosition} &+C_5\,x_{0}^{\mu}x_{0\nu}p_{0}^{\nu}+C_6\,x_{0}^{\mu}p_{0}^{\nu}p_{0\nu}+C_7\,x_{0}^{\nu}p_{0}^{\mu}p_{0\nu}+C_{8}\,p_{0}^{\mu}p_{0}^{\nu}p_{0\nu}\ldots\\
    p^\mu(p_0)&=D_1\,p_{0}^{\mu} +D_2\,p_{0}^{\mu}p_{0}^{\nu}p_{0\nu}+\ldots\,\,.
\end{align}
\end{subequations}
These infinite polynomials are truncated such that the coefficients multiplying every term to have dimensions of up to inverse momentum squared.  
This condition leads to the following 
\begin{equation}
    C_{8}=0\,.
\end{equation}
Which is in place due to the fact that every term in Eq.\eqref{Eq:GeneralPosition} must have the dimensions of length, which means that $C_{8}$  is proportional to $\hbar/(c M_{\text{Pl}})^4$. Since the calculations presented below are perturbative and are meant to serve as a phenomenological study, one can safely assume for all intents and purposes a term like this is undetectable using current and near future experiments. Furthermore, due to the commutativity of the auxiliary momentum operators the term proportional to $C_{8}$ in Eq.\eqref{Eq:GeneralPosition} does have no input into the final expression of the position-momentum commutator.

The spatial part of the commutator between the physical position $x^\mu$ and physical momentum $p^\nu$ is computed and compared to the quadratic part of the general GUP commutator Eq.\eqref{Eq:ADVcommutator}. This allows one to draw conclusions for the rest of the coefficients:
\begin{subequations}\label{Eq:XPRepresentation}
\begin{align}
\label{Eq:GUPx}x^{\mu}&=x_0^{\mu}-\kappa\gamma p_0^{\rho}p_{0\rho}x_0^{\mu}+\beta\gamma p_0^{\mu}p_0^{\rho}x_{0\rho}+\xi\hbar\gamma p_0^{\mu}, \\
\label{Eq:GUPp}p^{\mu}&=p_0^\mu\,(1+\varepsilon\gamma p_0^{\rho}p_{0\rho})\,.
\end{align} 
\end{subequations}
The dimensionality of the corrections in the Eq.\eqref{Eq:GUPx} above, was separated into the dimensional parameter $\gamma$, having the dimension of inverse momentum squared, and the dimensionless parameters $\kappa$, $\beta$, $\epsilon$m and $\xi$ in Eq.\eqref{Eq:XPRepresentation} above. The quantity $\gamma$ is defined as $\gamma=\frac{1}{(M_{Pl}\,c )^2}$,
which helps in clarifying the subsequent analysis. 
The final form of the most general quadratic Relativistic Generalized Uncertainty Principle (RGUP), allowing the existence of Lorentz invariant minimum measurable length, is:
\begin{equation}
\label{Eq:GUPxp}[x^{\mu},p^{\nu}]=i\hbar\,\left(1+(\varepsilon-\kappa)\gamma p^{\rho}p_{\rho}\right)\eta^{\mu\nu}+i\,\hbar(\beta+2\varepsilon)\gamma p^{\mu}p^{\nu}\,.
\end{equation}
 Assuming that the proposed
 modifications arise at the Planck scale, the dimensionless parameters are normally considered of order unity.
The minimal length which follows from the algebra Eq.\eqref{Eq:GUPxp} 
is of the order of the Planck length
\footnote{The parameters used in \cite{Kempf1995} and \cite{Quesne2006} are related to the ones used here as follows: 
$\beta_1=(\kappa-\varepsilon)\gamma$ and $\beta_2=(\beta+2\varepsilon)\gamma$.}. 
Note that Eq.\eqref{Eq:GUPxp} reduces in the non-relativistic ($c\rightarrow \infty$) limit to 
Eq.\eqref{Eq:KMMGUP3D}. Furthermore when one takes the non-GUP limit
($\gamma \rightarrow 0$) on top of the non-relativistic one obtains the standard Heisenberg  algebra. 

One can recall that the classical position and momentum in SR are four-vectors in spacetime and Lorentz transformations act on them as follows
\begin{align}
    X^{\prime\mu}&=\Lambda^\mu_\nu X^\nu\,,\\
     P^{\prime\mu}&=\Lambda^\mu_\nu P^\nu\,,
\end{align}
where $\Lambda^\mu_\nu$ is an orthogonal matrix for which $\Lambda^T=\Lambda^{-1}$. The Minkowski metric $\eta^{\mu\nu}$ and the scalar product of any two four-vectors are invariant under the Lorentz transformations.

In relativistic QM, position and momentum are promoted to operators and the Lorentz transformations $\Lambda$ are represented by a unitary operator $U(p^\nu,M^{\rho\sigma})$, {\it i.e.} $U^*=U^{-1}$. Where $p^\nu$ is the generator of translation and $M^{\rho\sigma}$
is the generator of rotations of the Poincar\'e group. The position and momentum in this case transform as follows 
\begin{align}
    x^{\prime\mu}&=U(p^\nu,M^{\rho\sigma})x^\mu U^{-1}(p^\nu,M^{\rho\sigma})\,,\\
     p^{\prime\mu}&=U(p^\nu,M^{\rho\sigma})p^\mu U^{-1}(p^\nu,M^{\rho\sigma})\,.
\end{align}
The operator $U(p^\nu,M^{\rho\sigma})$ can be expressed from the generators of the Poincar\'e algebra as
\begin{equation}
    U(p^\nu,M^{\rho\sigma})=\exp\left[i a_\nu p^\nu\right]\exp\left[i \frac{\omega_{\rho\sigma}M^{\rho\sigma}}{2}\right]\,,
\end{equation}
where the coordinate system is translated by a vector $a_\mu$, and is rotated by $\omega_{\rho\sigma}$. 

The position-momentum commutator transforms as follows
\begin{equation}\label{Eq:TransformedComutator}
    [x^{\prime\mu},p^{\prime\nu}]=U[x^{\mu},p^{\nu}] U^{-1}\,.
\end{equation}
Substituting $[x^{\mu},p^{\nu}]$ for its expression as given in Eq.\eqref{Eq:GUPxp}, one gets 
\begin{subequations}
\begin{align}
   [x^{\prime\mu},p^{\prime\nu}]&=U \left\{i\hbar\,\left(1+(\varepsilon-\kappa)\gamma p^{\rho}p_{\rho}\right)\eta^{\mu\nu}+i\,\hbar(\beta+2\varepsilon)\gamma p^{\mu}p^{\nu}\right\} U^{-1}\\
    &=i\hbar\,\left(1+(\varepsilon-\kappa)\gamma p^{\prime\rho}p^{\prime}_{\rho}\right)\eta^{\mu\nu}+i\,\hbar(\beta+2\varepsilon)\gamma U p^{\mu} U^{-1}Up^{\nu} U^{-1}\\
      &=i\hbar\,\left(1+(\varepsilon-\kappa)\gamma p^{\prime\rho}p^{\prime}_{\rho}\right)\eta^{\mu\nu}+i\,\hbar(\beta+2\varepsilon)\gamma p^{\prime\mu} p^{\prime\nu} \,.
\end{align}
\end{subequations}
Therefore, the commutator between position and momentum has the same form in every frame. One can then safely conclude that every frame will observe the same minimum measurable length.

According to Eqs.\eqref{Eq:GUPx} and \eqref{Eq:GUPp}, 
one gets for the commutator between two position operators  
\begin{equation}
    \label{Eq:GUPxx}[x^{\mu},x^{\nu}]=
i\hbar\gamma\frac{-2\kappa+\beta}{1+(\varepsilon-\kappa)\gamma p^{\rho}p_{\rho}}\left(x^{\mu}p^{\nu}-x^{\nu}p^{\mu}\right).
\end{equation}
Eq.\eqref{Eq:GUPxx} is a direct relativistic expansion of a similar commutator between two position operators  in the quasi-position representation presented \cite{Kempf1995}. It is worth noticing that
in both cases one arrives at a non-commutative ``fuzzy" spacetime.
However, the bigger issue is the fact that the above algebra does not close leading one to question if spacetime can be described by a single algebra.

Utilizing the position and momentum expressions presented in Eqs.\eqref{Eq:GUPx} and \eqref{Eq:GUPp} together with Eqs.\eqref{Eq:GUPxp} and \eqref{Eq:GUPxx}, one can find the following relations between the parameters used in this work and the parameters used in \cite{Quesne2006}.
Another fact worth mentioning is that the last two terms in Eq.\eqref{Eq:GUPx} break isotropy of spacetime
by introducing the preferred direction of $p_0^\mu$.
Since this violates the principles of relativity, 
in further calculations it will  be assumed that $\beta=\xi=0$.

\section{\label{sec:LorentzPoicare}Lorentz and Poincar\'e algebra}
  So far in this chapter relativistic expression of GUP with the corresponding Lorentz invariant minimum length is formulated.  However, a study of the effects of RGUP on the symmetries of spacetime is required, especially due to the ``fuzzy'' non-commutative nature of spacetime. In the process, it will be proved that the RGUP presented in this work resolves the composition law problem. 
  
In SR, the symmetries of spacetime are represented by the Lorentz algebra, reflecting the rotational symmetry or isotropy and the boost symmetry. 
The Poincar\'e algebra combines isotropy and homogeneity. Since Lorentz algebra is a sub-algebra of the Poincar\'e algebra, the first step is to formulate the Lorentz operators. The generators of the Lorentz group are constructed using Eqs.\eqref{Eq:GUPx} and \eqref{Eq:GUPp}, from the following expression
\begin{equation}
    M^{\mu\nu} = p^{\mu}x^{\nu}-p^{\nu}x^{\mu}
    = \left[1+(\varepsilon-\kappa)\gamma p_0^{\rho}p_{0\,\rho}\right]\tilde{M}^{\mu\nu}\,,
\end{equation}
where $\tilde{M}^{\mu\nu}=p_0^{\mu}x_0^{\nu}-p_0^{\nu}x_0^{\mu}$ are the Lorentz generators constructed out of the canonical variables $x_0$ and $p_0$. The auxiliary  position $x_0$, momentum  $p_0$, and the unmodified Lorentz operator constructed out of them, $\tilde{M}^{\mu\nu}$, are the generators of the standard Poincar\'e  algebra, presented below
\begin{subequations}
\begin{align}
[x_0^\mu,\tilde{M}^{\nu\rho}]&=i\hbar\left(x_0^{\nu}\eta^{\mu\rho}-x_0^{\rho}\eta^{\mu\nu}\right)\,,\\
[p_0^\mu,\tilde{M}^{\nu\rho}]&=i\hbar\left(p_0^{\nu}\eta^{\mu\rho}-p_0^{\rho}\eta^{\mu\nu}\right)\,,\\
 [\tilde{M}^{\mu\nu},\tilde{M}^{\rho\sigma}]&=i\hbar\left(\eta^{\mu\rho}\tilde{M}^{\nu\sigma}-\eta^{\mu\sigma}\tilde{M}^{\nu\rho}-\eta^{\nu\rho}\tilde{M}^{\mu\sigma}+\eta^{\nu\sigma}\tilde{M}^{\mu\rho}\right)\,.
\end{align}
\end{subequations}
Which is the usual form of the Poincar\'e algebra  for canonically conjugate position and momentum. 

One is now in a position to compute the Poincar\'e algebra for the physical operators.
The expression of the modified Poincar\'e algebra for the physical position and momentum, and by extension the physical Lorentz generators, is given by the following commutators
\begin{subequations}
\begin{align}
    [x^{\mu},M^{\nu\rho}]& =  [1 + (\varepsilon - \kappa) \gamma p_0^{\rho} p_{0\,\rho}] [x_0^{\mu},\tilde{M}^{\nu\rho}] 
    + i\hbar 2 (\varepsilon - \kappa) \gamma p_0^{\mu} \tilde{M}^{\nu\rho}\,,\\
    [p^{\mu},M^{\nu\rho}]& =  [1 + (2 \varepsilon - \kappa) \gamma p_0^{\rho} p_{0\,\rho}] [p_0^{\mu},\tilde{M}^{\nu\rho}]\,,\\
    [M^{\mu\nu},M^{\rho\sigma}]& =  [1 + (\varepsilon - \kappa) \gamma p_0^{\rho} p_{0\,\rho}]^2 [\tilde{M}^{\mu\nu},\tilde{M}^{\rho\sigma}]\,.
\end{align}
\end{subequations}
One can then express the operators $x_0^{\mu}$, $p_0^{\mu}$, and $\tilde{M}^{\mu\nu}$ in terms of the physical ones. After series expansion and truncation to first order in $\gamma$, one arrives at the following commutators, expressing the physical Poincar\'e algebra
\begin{subequations}\label{Eq:PoincareAlgebra}
\begin{align}
\label{Eq:xM}  [x^\mu,M^{\nu\rho}] &=  i\hbar[1 + (\varepsilon - \kappa) \gamma p^{\rho} p_{\,\rho}]\left(x^{\nu}\eta^{\mu\rho}-x^{\rho}\eta^{\mu\nu}\right) + i\hbar 2 (\varepsilon - \kappa) \gamma p^{\mu} M^{\nu\rho}\,,\\
   \label{Eq:pM}[p^\mu, M^{\nu\rho}]& =  i\hbar[1 + (\varepsilon - \kappa) \gamma p^{\rho} p_{\,\rho}]\left(p^{\nu}\eta^{\mu\rho}-p^{\rho}\eta^{\mu\nu}\right),\\
\label{Eq:MM}  [M^{\mu\nu},M^{\rho\sigma}]& = i\hbar\left(1 + (\varepsilon - \kappa) \gamma p^{\rho} p_{\,\rho}\right)\left(\eta^{\mu\rho}M^{\nu\sigma}-\eta^{\mu\sigma} M^{\nu\rho} 
  -\eta^{\nu\rho}M^{\mu\sigma}+\eta^{\nu\sigma}M^{\mu\rho}\right)\,.
\end{align}
\end{subequations}
It is worth noting that the Lorentz algebra does not close on its own, due to the addition of the RGUP correction.
However, it is possible to close it by including momenta, {\it i.e.}, by considering the Poincar\'e algebra.
Additionally, as assumed above the physical momentum $p^{\mu}$ still forms an abelian subgroup.
From the algebra presented in Eqs.\eqref{Eq:PoincareAlgebra}, one can see that the Poincar\'e algebra has four quadratic operators which commute with every other operator in the algebra, also known as Casimir invariants. The four quadratic Casimirs are:  
    \begin{align}
  \label{Eq:PhysicalCasimirs}  p^\mu p_\mu~&\text{and}~ W^\mu W_\mu\\
       p_0^\mu p_{0\,\mu}~&\text{and}~ W_0^\mu W_{0\,\mu}\,,
\end{align}
where $W_\mu$ and $W_{0\,\mu}$ are the physical and auxiliary Pauli-Lubanski pseudovectors defined as follows
\begin{align}
    W_\mu&\equiv \frac{1}{2}\varepsilon_{\mu \nu \rho \sigma }M^{{\nu \rho }}p^{\sigma}\\
     W_{0\,\mu}&\equiv \frac{1}{2}\varepsilon_{\mu \nu \rho \sigma }\tilde{M}^{{\nu \rho }}p_0^{\sigma}\,.
\end{align}
Exploring spacetime symmetries as described by SR one can notice the Casimir operators of the Poincar\'e group are related to Einstein's dispersion relation \cite{ohlsson2011relativistic}. Since the  RGUP modified Poincar\'e group has both the auxiliary momentum squared $p_0^\mu p_{0\,\mu}$ and the physical momentum squared as its Casimir invariants, the dispersion relation for both the physical and auxiliary momentum are quadratic. Therefore, the formulation of RGUP presented in this work completely avoids the Composition law problem, and both the auxiliary and physical energies and momenta are summed up linearly as they should. 

However, only one pair of Casimirs, namely the ones presented in Eq.\eqref{Eq:PhysicalCasimirs}, correspond to actual physical quantities like $(mc)^2$. Since $x^\mu$, $p^\mu$, and $M^{\mu\nu}$ are the physical quantities one measures in an experiment, the physical momentum squared $p^{\rho}p_{\rho}$ must give the physical quantity $(mc^2)^2$, so the Casimir operators that actually carry information are $p_{\rho}p^{\rho}$ and $ W^\mu W_\mu$. 
 The remaining ones are are a convenient mathematical construction used for calculations. Calculations testing the associativity of spacetime, momentum spacetime, and the Lorentz algebra in addition to explicit calculations for the representations of the Poincar\'e algebra can be found in Appendix \ref{app:Poincare}. 
 
Proceeding further and inspecting 
Eqs.\eqref{Eq:PoincareAlgebra},
one notices that  there exists a line in parameter space, namely
$\varepsilon=\kappa$ for the following reason. 
The RGUP theory lying on that line is non-trivial, in the sense it is a genuine GUP effect (distinct from HUP) and gives rise to a Lorentz invariant minimum measurable length, yet it leaves the Poincar\'e algebra
{\it unmodified} . 
For these reasons, further calculations in this chapter will be restricted to this line and the corresponding 
choice of parameters. 
Then the RGUP algebra
and the non-commutativity of spacetime following from 
Eqs.\eqref{Eq:GUPxp} and \eqref{Eq:GUPxx} respectively, assume the following form:
\begin{align}
[x^{\mu},p^{\nu}]&=i\hbar\,\left(\eta^{\mu\nu}+2\kappa\gamma p^{\mu}p^{\nu}\right)\,, \label{Eq:RGUPk=e} \\
[x^{\mu},x^{\nu}]&=-
2i\hbar\kappa\gamma\left(x^{\mu}p^{\nu}-x^{\nu}p^{\mu}\right)\,, \label{Eq:STNCk=e}
\end{align}
where in order to have a minimum uncertainty in position $\kappa>0$ is required. The commutation relations given by Eqs. \eqref{Eq:RGUPk=e} and \eqref{Eq:STNCk=e} are distinct from the simplest case of non-commutative topology, i.e. the ``fuzzy" sphere. The way they differ is that the coordinates do not form a closed algebra, which makes defining a topology difficult. 

Notice that for any one-dimensional space, Eq.\eqref{Eq:RGUPk=e} above reduces to Eq.\eqref{Eq:KMMGUP} \cite{Kempf1995} and the quadratic part of Eq.(1) of \cite{Ali2011} (up to an unimportant numerical factor). 
It is interesting to note that algebras 
Eqs.\eqref{Eq:RGUPk=e} and \eqref{Eq:STNCk=e}
have similarities to the ones proposed 
in \cite{Snyder:1946qz}.
In the following sections, a study of the applications and phenomenological applications of the results above are presented.
\section{\label{sec:Applications}Applications}
  One can easily check that for the case 
$\varepsilon=\kappa$, Eqs.\eqref{Eq:GUPx} and \eqref{Eq:GUPp} take the following form,
\begin{subequations}
\begin{align}
\label{Eq:X}    x^{\mu}&=x_0^{\mu}(1-\kappa\gamma p_0^{\rho}p_{0\rho})\,,\\
    \label{Eq:P} p^{\mu}&=p_0^\mu\,(1+\kappa\gamma p_0^{\rho}p_{0\rho})\,.
\end{align}
\end{subequations}
As shown earlier, since now the definition of $M^{\mu\nu}$ in terms of the physical position and momentum, $x^{\mu}$ and $p^{\mu}$, as well as the Poincar\'e algebra, remains unchanged, the squared physical momentum $p^{\rho}p_{\rho}$ is a Casimir invariant, commuting with every other operator in the group.
The Klein-Gordon (KG) equation can be derived from Einsteins dispersion relation. Additionally, according to SR, the dispersion relation is related to the Casimir invariant of the Poincar\'e group.
The KG equation can be written in the following form
\begin{equation}
    p^{\rho}p_{\rho}=-(mc)^2,
\end{equation}
or, in terms of the variables $p_0^{\mu}$,
\begin{equation}
    \label{Eq:ModKG}
    p_0^{\rho}p_{0\rho}(1+2\kappa\gamma p_0^{\sigma}p_{0\sigma})=-(mc)^2 \,,
\end{equation}
where $m$ is the mass of the particle.
Observe that using Eq.\eqref{Eq:MathVar},
the KG equation now is a fourth order differential equation (as opposed to second order in case of the standard KG equation) with four 
linearly independent solutions.
However, in spherical coordinates the presence of mixed-derivative terms (time-space derivatives and space-derivatives on different coordinates) rules out analytical solutions \cite{hebey2008introduction}.
Therefore, to reduce the order of the equation one can solve Eq.\eqref{Eq:ModKG} as a quadratic equation for 
$p_0^{\rho}p_{0\rho}$. 
Eq.\eqref{Eq:ModKG} 
can be reduced to a KG-like second order equation for $p_{0\rho}p_0^{\rho}$  with a modified left hand side (LHS).
This is done by writing Eq.\eqref{Eq:ModKG} in the following form
\begin{equation}
    \frac{p_0^{\rho}p_{0\rho}}{2\kappa\gamma}-(p_0^{\rho}p_{0\rho})^2-\frac{(mc)^2}{2\kappa\gamma}=0\,.
\end{equation}
A term $\frac{1}{(4\kappa\gamma)^2}$ is added and subtracted, in order to compleate the square. One then gets
\begin{equation}\label{Eq:KGReductionStep}
    \left(p_0^{\rho}p_{0\rho}-\frac{1}{4\kappa\gamma}\right)^2=\frac{1}{(4\kappa\gamma)^2}-\frac{(mc)^2}{2\kappa\gamma}\,.
\end{equation}
The next step is to take the square root of Eq.\eqref{Eq:KGReductionStep}. The result is the following equation
\begin{equation}\label{Eq:ModPar0}
    p_0^{\rho}p_{0\rho}=-\frac{1}{4\kappa\gamma}\pm\sqrt{\frac{1}{\left(4\kappa\gamma\right)^2}-\frac{(mc)^2}{2\kappa\gamma}}\,.
\end{equation}
Epand the LHS of Eq.\eqref{Eq:ModPar0} in Taylor series for the small parameter $\gamma$. The leading order of such expansion must be $(mc)^2$, which comes from an unmodified dispersion relation. Expanding the right hand side (RHS) of Eq.\eqref{Eq:ModPar0} in Taylor series one gets
\begin{align}
\label{Eq:minus}\mu_-=&-\frac{1}{2 \kappa  \gamma}+ (m\,c)^2+2\kappa\gamma(m\,c)^4+O\left(\gamma^2\right),\\
\label{Eq:plus}\mu_+=&-( m\,c)^2- 2\kappa\gamma(m\,c)^4+O\left(\gamma ^2\right)\,.
\end{align}
It is easy to notice that the lowest order in the Taylor series expansion for the plus sign solution \eqref{Eq:plus} gives us the unmodified rest energy  $(mc)^2$ as the leading term. On the other hand, it is clear that the leading term in Eq.\eqref{Eq:minus} is of the order $\gamma^{-2}$. Therefore, in the $\gamma \rightarrow 0$ limit it  gives us a solution with infinite mass and energy. Additionally, one can see that Eq.\eqref{Eq:minus} has the wrong sign in front of the mass. Therefore, the solutions of Eq.\eqref{Eq:KGReductionStep} presented in Eq.\eqref{Eq:minus}  will be dismissed as nonphysical since they do not replicate the solution of the unmodified KG equation in the low energy limit.
A GUP-corrected effective
second order KG equation is then obtained from  \eqref{Eq:plus}:
\begin{eqnarray}
\label{Eq:ModPar}
p_0^{\rho}p_{0\rho}&&=-\frac{1}{4\kappa\gamma}+\sqrt{\frac{1}{\left(4\kappa\gamma\right)^2}-\frac{(mc)^2}{2\kappa\gamma}}\ \\
&& 
\simeq - ( m\,c)^2-2\kappa\gamma(m\,c)^4-{\cal O}\left(\gamma ^2\right)\,,\label{Eq:KGMassApproximation}
\end{eqnarray}
where the other solution has been discarded since it does not reduce to $(mc)^2$ in the $\gamma\rightarrow 0$ limit.
In this process, two of the remaining solutions of the fourth order equation Eq.\eqref{Eq:ModKG} are lost.
As was shown in \cite{Ali2011},
including those solutions introduced very small corrections, and therefore they can be ignored when conducting phenomenological study.
However, such an exclusion of solution is not considered in the following chapters. Thus in chapter \ref{Ch4:QFTWML}, the full form of the KG Eq.\eqref{Eq:ModKG} equation is considered when calculating the RGUP modified field Lagrangian.

The following three sections contain a study of the applications of the effective RGUP modified KG  Eq.\eqref{Eq:ModPar} above, as well as the effective GUP-modified Dirac equation for a number of quantum systems, which the author believes to be the best candidates for manifestion low energy artefacts of QG effects.  

\subsection{Klein-Gordon equation}

  Writing Eq.\eqref{Eq:KGMassApproximation} as an operator equation acting on the wavefunction $\Psi$ and using Eq.\eqref{Eq:MathVar}, one  obtains the following GUP-modified differential form of the Klein--Gordon equation up to ${\cal O}\left(\gamma ^2\right)$ 
\begin{equation}\label{Eq:ModKGDiff}
    \frac{1}{c^2}\frac{\partial^2}{\partial t_0^2}\Psi-\nabla_0^2\Psi+\frac{1}{\hbar^2}\left[  (m\,c)^2+2 \kappa  c^4 \gamma ^2 m^4 \right]\Psi=0\,.
\end{equation}
It is worth noticing that in the limit $\gamma\rightarrow 0$, the above equation reduces to the standard KG equation.
Moreover, the solutions of the modified equation have the same form as the standard one but with modified parameters \cite{gordon1926comptoneffekt,klein1986quantentheorie}.

\subsection{Energy spectrum for relativistic Hydrogen atom}
  The simplest model of the hydrogen atom has the
  energy spectrum given by considering the proton and the electron as two elementary particles interacting electromagnetically. In order to covariantly introduce the electromagnetic interaction in systems described by Eq.\eqref{Eq:ModKGDiff}, one must minimally couple the derivatives to the four-potential $A^{\rho}$. The procedure used and the discussions of its validity and merits can be found in  
\cite{Bosso:2018uus,Das:2009hs}, where the four-potential $A^{\rho}$ is coupled to the auxiliary momentum $p_{0\,\mu}$, as follows
\begin{equation}\label{Eq:MinimalCoupling}
    \frac{\partial}{\partial x_{0\rho}}\rightarrow \frac{\partial}{\partial x_{0\rho}}+ \frac{ie}{\hbar}\,A^{\rho}\,.
    \end{equation}
This choice of minimal coupling given a theory which is locally gauge invariant allows freedom of gauge choice. Therefore, for this particular application, the Hydrogen atom nuclear potential in the Coulomb gauge was chosen
\begin{equation}\label{Eq:Coulomb}
    A_{\rho} = \left\{\frac{e}{4\pi\varepsilon_0\,r },0,0,0\right\}\,.
\end{equation}
Minimally coupling Eq.\eqref{Eq:ModKGDiff} according to the procedure described above,
one obtains the GUP modified Klein-Gordon equation for the Hydrogen atom, namely
\begin{equation}\label{Eq:HA}
 -\left( i\frac{\partial }{\partial t_0}+ c\frac{\alpha}{r}\right)^2\Psi-c^2\nabla_0^2\Psi 
    +\frac{c^2}{\hbar^2}\left[ (m\,c)^2+2\kappa\gamma(m\,c)^4 \right]\Psi=0,
\end{equation}
In the above, $\alpha={e^2/4\pi\epsilon_0 \hbar c}$ is the fine structure constant.
The solution of Eq.\eqref{Eq:HA} in spherical coordinates is of the form 
\cite{Griffiths2018}
\begin{equation}
    \Psi_{nlm}=R_{nl}(r)\,Y_{lm}(\theta,\phi)\,e^{-E_{0\,nl}t/\hbar}\,,
\end{equation}
where the $R_{nl}(r)$ are  spherical Bessel functions and $Y_{lm}(\theta,\phi)$ the spherical harmonics. The energy
levels for these solutions are given as
\begin{equation}\label{Eq:HAenergy}
    E_{0\,nl}=\left( m\,c^2+\kappa\gamma m^3c^4 \right)\left[1+\frac{\alpha^2}{(n+1-\eta)^2}\right]^{-1/2}\,,
\end{equation}
where
\begin{equation}
    \eta=\frac{1}{2}\pm\sqrt{\left(l+\frac{1}{2}\right)^2-\alpha^2},
\end{equation}
where positive sign predicts a much higher than observed binding energy. And the minus sign recovers the standard 
Hydrogen atom spectrum 
from Eq.\eqref{Eq:HAenergy} in the non-relativistic limit, with the identification
$N\equiv n+1-\eta$ for the hydrogen atom energy level.
Expanding Eq.\eqref{Eq:HAenergy} in powers of $\alpha^2$, one finds
\begin{multline}\label{Eq:E_0KG}
 E_{0\,N}=\left(m\,c^2+ \kappa  \gamma  m^3c^4 \right)-\frac{ \alpha^2 \left(m\,c^2 \right)}{2N^2}\\
+\frac{3 \alpha^4  \left(mc^2\right)}{8N^4}+\frac{ 3\alpha ^4  \left(\kappa\gamma m^3c^4\right)}{8N^4}-\frac{ \alpha ^2 \left(\kappa\gamma m^3c^4\right)}{2N^2}.
\end{multline}
The first two terms in the above expression correspond to the rest energy and 
GUP corrections to the rest energy. The third term is the Schr\"odinger term, 
with the corresponding 
relativistic correction
(which vanishes in the 
$c\rightarrow\infty$ limit),
GUP correction 
(which vanishes in the 
$\gamma\rightarrow 0$ limit)
and the GUP $+$ relativistic correction 
(which vanishes in the
$\gamma\rightarrow 0$ 
or $c\rightarrow\infty$ limit)
in the fourth, fifth and sixth terms respectively. 
However, the corrected energy levels are corrections to $E_0$, {\it i.e.} the $0^{th}$ component of the auxiliary momentum $p_0^\mu$.
To obtain the corrections to the physical energy
$E$, {\it i.e.} the $0^{th}$ component of physical momentum $p^\mu$, one needs to use the appropriate component of 
Eq.\eqref{Eq:P}, namely 
\begin{equation}
   E = E_0\left(1+\kappa\gamma p_0^\rho p_{0\,\rho}\right)
    = E_0 \left[1 - \kappa \gamma (mc)^2 \right] + {\cal O}(\gamma^4)\,,\label{Eq:PhysEnrg}
\end{equation}
to obtain 
\begin{align}\label{Eq:EnergyWithCorrections}
 E_N=&\left(m\,c^2
\right)-\frac{ \alpha^2 \left(m\,c^2 \right)}{2N^2}
+\frac{ \alpha^4  \left(mc^2\right)}{8N^4}\,.
\end{align}
As evident from the above equation, all GUP corrections, or in other words terms proportional $\gamma$, vanish. It is important to clarify that the Casimir operator remains modified. This result  has implications for field theory as shown in Chapter \ref{Ch4:QFTWML}.
Furthermore, considering the remaining two solutions of the fourth order equation \eqref{Eq:ModKG} may give rise to GUP corrections. 

\section{Schr\"odinger equation with relativistic and GUP corrections}

  In relativistic QM one can use the KG equation to obtain relativistic corrections to the kinetic energy in the Schr\"odinger equation. Such effects can be observed in the fine structure of non-relativistic systems. Therefore, one can argue that even though this chapter deals with relativistic QG effects. The fine structure of quasi relativistic systems, cannot be ignored as a promising search for low energy QG effects.
  
In the following section one can find derived RGUP corrections to the relativistically corrected kinetic terms of the Schr\"odinger equation.

One must first derive the corrected Schr\"odinger equation from Eq.\eqref{Eq:KGMassApproximation} 
\begin{equation}
    -E_0^2+c^2\vec{p}_0^2+(mc^2)^2+2\kappa\gamma m^4c^6=0.
\end{equation}
The above can be rewritten as 
\begin{equation}
    E_0=\sqrt{c^2\vec{p}_0^2+(mc^2)^2+2\kappa\gamma m^4c^6}\,,
\end{equation}
which can be further expanded to fourth order in $\vec{p}_0$ and second order in $\gamma$ by using Taylor series, and truncating the higher order terms
\begin{equation}\label{Eq:ModSEp}
 E_0=mc^2\left(1+\kappa\gamma(mc)^2\right)+\frac{\vec{p}_0^2}{2m}\left(1-\frac{1}{2}\kappa\gamma(mc)^2\right)-\frac{\vec{p}_0^4}{8m^3c^2}\left(1-3\kappa\gamma(mc)^2\right)\,.
\end{equation}
Next, using Eq.\eqref{Eq:PhysEnrg}, one can get the expression for the physical energy
\begin{equation}
    E = mc^2 + \frac{\vec{p}_0^2}{2m} \left[1 - \frac{3}{2} \kappa \gamma (mc)^2\right] 
    - \frac{\vec{p}_0^4}{8 m^3 c^2}\left[1 - 4 \kappa \gamma (mc)^2\right]\,,
\end{equation}
which consists of the rest mass, 
non-relativistic kinetic energy, relativistic and GUP corrections. 

  Next, the operator version of Eq.\eqref{Eq:ModSEp}, using $E_0 = i \hbar \frac{\partial}{\partial t_0}$ and $\vec{p}_0 = - i \hbar \vec\nabla_0$, and including a potential $V (\vec x)$, yields the modified Schr\"odinger equation with relativistic and GUP corrections
\begin{multline}\label{Eq:SchrodingerDifferentialEquation}
    i \hbar \frac{\partial}{\partial t_0} \Psi(t_0,\vec{x}_0) 
    = \left[mc^2 \left(1 + \kappa \gamma m^2 c^2 \right) \right. 
    + \frac{(- i \hbar)^2}{2 m} \left(1 - \frac{1}{2}\kappa \gamma m^2 c^2 \right) \nabla_0^2 \\
    \left. - \frac{(-i\hbar)^4}{8 m^3 c^2}\left(1 - 3 \kappa \gamma m^2 c^2 \right) \nabla_0^4 + V(\vec{x})\right] \Psi(t_0,\vec{x}_0)\,.
\end{multline}
One observes that, although the GUP-induced effective reparametrization of the mass amounted to no modifications of the Klein-Gordon equation, in this case there are GUP corrections to the kinetic term, as well as to the relativistic corrections thereof. 
It will be shown that these give rise to potentially measurable corrections in laboratory based systems.
  Omitting the rest energy term, one applies the above equation to a couple of 
problems. 

\subsection{Corrections for a particle in a box}
  Using Eq.\eqref{Eq:SchrodingerDifferentialEquation}, one can write the Schr\"odinger equation with relativistic and GUP corrections.
  
  As it is the simplest case, the applications of the results presented above to physical system begin by considering a $1+1$-dimensional. The particular example chosen is the particle in a box, with the standard potential presented below
\begin{equation}
    V(x)=\left\{
    \begin{array}{lll}
        V_0 & \text{for} & 0<x<L,\\  
        \infty & \text{for} & x \leq 0 ~ \cup ~ x\geq L.\\
    \end{array} \right.
\end{equation}
Including the potential in Eq.\eqref{Eq:ModSEp}, one
obtains the Schr\"odinger equation for one dimensional particle in a box plus small perturbations.

  Consider the wave function for the unperturbed and non-relativistic ($\gamma = 0,c\rightarrow \infty$) case
\begin{equation}
    \Psi_n(t_0,\vec{x}_0)=\sqrt{\frac{2}{L_0}}\sin\left(\frac{n\pi}{L_0}x_0\right)e^{-i\omega_n\,t_0}\,,
\end{equation}
where $L_0$ is the unmodified length od the box, and $w_n$ is the energy of the $n$-th exited state. From Eq.\eqref{Eq:X}, 
one can read-off the physical dimensions of the box to be
\begin{equation}
    L=L_0(1+\kappa\gamma (m\,c)^2+{\cal O}(\gamma^4) )\,.
\end{equation}
The corrected  spectrum of $E_0$ is
\begin{multline}
 E_{0\,n}=\int_0^L d\,x  \Psi^*\Big[\left(m\,c^2+\kappa\gamma m^3c^4-\frac{1}{4}\hbar^2\kappa\gamma m\,c^2 \nabla^2\right)\\
 -\frac{(-i\hbar)^4}{8m^3c^2}\nabla^4 +\frac{3(-i\hbar)^4}{8m}  \kappa  \gamma\nabla^4\Big]\Psi\,,
\end{multline}
which simplifies to
\begin{equation}\label{Eq:Pbox}
    E_{0n} =  
     \frac{1}{2m} \left(\frac{n \pi \hbar}{L_0}\right)^2 \left[1 - \frac{1}{2} \kappa \gamma (mc)^2\right]
    + \frac{\hbar^4}{8 m^3 c^2} \left(\frac{n \pi}{L_0}\right)^4 \left[1 - 3 \kappa \gamma (mc)^2 \right]\,,
\end{equation}
where the rest energy and a GUP correction to the rest energy terms are omitted, the first term is RGUP correction to the Schr\"odinger equation and the  and Schr\"odinger term itself, the third term is just  relativistic corrections and  RGUP corrections to the relativistic terms.

Using Eq.\eqref{Eq:PhysEnrg}, one can translate this to the following expression for the physical energy
\begin{equation}
    E_{n} =
   \frac{1}{2m} \left(\frac{n \pi \hbar}{L}\right)^2 \left[1 + \frac{3}{2} \kappa \gamma (mc)^2\right]
   - \frac{\hbar^4}{8 m^3 c^2} \left(\frac{n\pi}{L}\right)^4.
\end{equation}
The first term corresponds to the non-relativistic energy with GUP-corrections, while the last consists of relativistic corrections.

  The results of this section can be applied to experiments measuring directly the energy levels of quantum dots. Quantum dots are another name for the holes and electrons propagating in a crystal lattice. The physical edges of the crystal are in this case acting as the potential barriers. The dots are free to move inside the crystal allowing the behaviour of the dots to be modeled like particles in a box.
  
Comparing GUP corrections to the unperturbed energy term found above and equating these to the accuracy of experiments measuring the energy levels of single quantum dot \cite{Hill2002}:
\begin{equation}
    \frac{\Delta E_{n}}{E_{n}} = \frac{2}{3}\kappa\gamma m^2c^2 \sim \kappa 10^{-42} \lesssim 10^{-1}~.
\end{equation}
From this, one gets an upper bound on $\kappa$
\begin{equation}
    \kappa \lesssim 10^{41}\,.
\end{equation}

\subsection{Corrections to the linear harmonic oscillator }
  Consider Eq.\eqref{Eq:ModSEp} with the harmonic oscillator potential 
\begin{equation}
V(x)=\frac{1}{2}m\,\omega^2\,x^2\,.
\end{equation}
This gives rise to the following $E_0$
\begin{equation}
    E_0 = \frac{p_0^2}{2m} + \frac{1}{2} m \omega^2 x_0^2 [1 + 2 \kappa \gamma (m c)^2]
    - \frac{p_0^2}{4m} \left(\kappa \gamma m^2 c^2\right) - \frac{p_0^4}{8 c^2 m^3} + \frac{3p_0^4}{8m} \kappa \gamma.
\end{equation}
The annihilation and creation operators are defined as in
\cite{messiah1999quantum}
\begin{align}
 x_0 = \sqrt{\frac{\hbar}{2m\omega}}(a+a^{\dagger}), ~~~
 p_0 = -i\sqrt{\frac{\hbar m\omega}{2}}(a-a^{\dagger}).
\end{align}
Treating the GUP corrections as perturbations, one now has 
\begin{multline}
 E_{0n}^{(1)} = \bra{n}
m c^2  + \kappa\gamma m^3 c^4+
\kappa\gamma(m\,c)^2\frac{\hbar\omega}{2}(a+a^{\dagger})^2 \\
+\frac{\hbar\omega}{8}\left( \kappa \gamma m^2 c^2 \right)(a-a^{\dagger})^2\\
-\frac{\hbar^2\omega^2}{32\,m
\,c^2}\left(1 - 3 \kappa\gamma m^2c^2  \right)(a-a^{\dagger})^4\ket{n}\,,
\end{multline}
where $\ket{n}$ are the unperturbed states.
For the first order correction in perturbation theory, one obtains
\begin{multline}
    E_{0n} = 
    \hbar \omega \left(n + \frac{1}{2}\right) \left(1 + \frac{1}{2} \kappa \gamma m^2 c^2 \right)\\
    - \frac{\hbar^2 \omega^2}{32 m c^2} \left(1 - 3 \kappa \gamma m^2 c^2\right) \left[5 n (n+1) + 3\right].
\end{multline}
From this, using Eq.\eqref{Eq:PhysEnrg}, one finds the following for the physical energy
\begin{multline}
    E_{n} = 
    \hbar \omega \left(n + \frac{1}{2}\right) \left(1 - \frac{1}{2} \kappa \gamma m^2 c^2 \right)\\
    - \frac{\hbar^2 \omega^2}{32 m c^2} \left(1 - 4 \kappa \gamma m^2 c^2\right) \left[5 n (n+1) + 3\right].
\end{multline}
  Similar to the calculations for Landau levels done in \cite{Das2008},
one can use experiments to put bounds on $\kappa$ 
\begin{equation}
    \frac{ \Delta E_{n}}{E_{n}}=-\frac{3}{4}\kappa\gamma m^2c^2\sim \kappa\,10^{-44}~.
\end{equation}
Equating this to the accuracy of direct measurements of Landau levels
\cite{yin2016experimental,wildoer1997observation}, the following bound on the RGUP parameter was placed
\begin{equation}
   \kappa\,10^{-44}\leq 10^{-3}\Rightarrow\kappa\leq 10^{41}.
\end{equation}
It is worth noting that this is several orders smaller than
what was obtained for a quadratic GUP in
\cite{Das2008}. This is due to the RGUP modifications to the relativistic correction term 
to kinetic energy.

\section{\label{Section:Dirac}Dirac equation and GUP corrections}

  Starting from Eq.\eqref{Eq:ModPar}, working in the following signature $\{+,-,-,-\}$ signature, and considering the following Dirac matrices 
\begin{align}
\tau^0=\begin{pmatrix} 
\mathbf{I}&0 \\
0 & -\mathbf{I} 
\end{pmatrix},\,\,\,\,\,\tau^i=\begin{pmatrix} 
0&\sigma^i \\
\sigma^i &0
\end{pmatrix}\,,
\end{align}
%
%
one derives the following GUP-modified Dirac equation \cite{Das:2010zf}
\begin{equation}
i\hbar\tau^{\mu}\frac{\partial}{\partial x_0^{\mu}}\Psi- \left(\tau^0\sqrt{\frac{1}{4\kappa\gamma}-\sqrt{\frac{1}{\left(4\kappa\gamma\right)^2}-\frac{(mc)^2}{2\kappa\gamma}}}\right)\Psi=0\,.
 \end{equation}
Truncating to ${\cal O}(\gamma)$, the differential form of the RGUP Dirac equation was obtained as follows 
\begin{equation}\label{Eq:ModD}
   i\hbar\tau^{\mu}\partial_{\mu}\Psi-\tau^0(mc+\kappa\gamma m^3c^3)\Psi=0,
\end{equation}
the full account of the calculations can be found in  \cite{Todorinov:2018arx}.
\subsection{Hydrogen atom}
  Using the modified Dirac equation  in Eq.\eqref{Eq:ModD}, one can calculate corrections to the fine and hyperfine structure of the hydrogen atom. Again, because the differential form of Eq.\eqref{Eq:ModKGDiff} is the same as the classical case, one can use the solutions derived in the literature \cite{Citation2017}, in order to calculate the corrections for the parameters of the solutions. 
  
Using Eqs.\eqref{Eq:MinimalCoupling} and \eqref{Eq:Coulomb} for minimal coupling and Coulomb potential, from Eq.\eqref{Eq:ModD} one gets
\begin{equation}
    \left[i \tau^{\mu} \frac{\partial}{\partial x_0^{\mu}} - \frac{e}{\hbar} \tau^{\mu} A_{\mu} - \tau^0 \frac{(mc + \kappa \gamma m^3 c^3)}{\hbar}\right]\Psi = 0,
\end{equation}
with the solution in spherical coordinates $\{t,r,\theta,\phi\}$ taking the form
\begin{equation}\label{Eq:DiracWavefunction}
    \Psi(t,r,\theta,\phi)=T(t)\frac{1}{r}\left(\begin{array}{cc}
    F(r)Y_{jm}  (\theta,\phi)     \\
           i\,G(r)Y'_{jm}  (\theta,\phi)  
    \end{array}\right)\,,
\end{equation}
where $T(t)$ is the temporal part of the wave function. In the above, $F(r)$ and $G(r)$ are spherical Bessel functions, and $Y_{jm}$ and $Y'_{jm}$ are the spherical spinors. 
Following \cite{Citation2017}, one finds the energy spectrum
\begin{equation}
    E_{0N}=(mc^2+\kappa\gamma m^3c^4)\left[1+\frac{\alpha^2}{(N-2j-1)^2}\right]^{-1/2}\,,
\end{equation}
where again $\alpha=e^2/4\pi\varepsilon_0\hbar c$ is the fine structure constant, and $N$ is the principle quantum number. Expanding the above equation in a Taylor series in $\alpha$, one obtains
\begin{align}
  \nonumber E_{0N}= &(mc^2+\kappa\gamma m^3c^4)-\frac{ mc^2\alpha ^2}{2 N^2}-\frac{\kappa\gamma m^3c^4\alpha ^2}{2 N^2}\\
  &+\frac{3 mc^2 \alpha^4}{8 N^4}+\frac{\kappa\gamma m^3c^4 \alpha ^4}{8 N^4}\,,
\end{align}
As before, using 
Eq.\eqref{Eq:PhysEnrg}, one gets for the physical energy spectrum 
$E_N$ 
\begin{align}
    E_{N}= mc^2
    -\frac{ mc^2\alpha^2}{2 N^2}
  +\frac{3 mc^2 \alpha^4}{8 N^4}\,,
\end{align}
Once again, it can be observed that the RGUP modifications vanish for the physical energy.

\section{Summary}
  The chapter above presents a method of incorporating minimal length in relativistic quantum mechanics, through the use of a relativistic expression for the GUP. 
General form of the position and momentum operators are proposed to be functions of canonically conjugate auxiliary variables. 
The general functions were then expanded perturbatively, retaining correction terms up to second order in the minimum length parameter. 
The RGUP and the spacetime commutators were formulated. Furthermore, the consequences for the symmetries of spacetime were explored. Importantly, by considering RGUP and the Lorentz invariant minimum length it is shown that in the general case one needs to modify the Poincar\'e algebra. An interesting discovery was the fact that the physical and auxiliary momenta squared are Casimir invariants of the same group, leading to the conclusion that the RGUP proposed here does not have the composition law problem. That is both the the auxiliary and physical energy and momenta sum up linearly for a composite system. 

Another interesting conclusion is that one needs a non-commutative spacetime in order to incorporate frame independent minimum length.

Furthermore, a particular set of RGUP parameters was found. This set of parameters is represented by a line in parameter space $\kappa=\varepsilon$. A theory residing on the $\kappa=\varepsilon$ line in parameter space has the usual spacetime symmetries, {\it i.e.} the Poincar\'e algebra for this theory will remain unchanged.  Additionally, it was shown that even though the symmetries of spacetime are preserved, a theory on that line retains the non-commutativity of spacetime and RGUP, and thus has minimum measurable length.  

Through the use of the modified Casimir operator a RGUP-modified quantum mechanical wave equations are obtained, and applied to several examples of well known quantum mechanical systems. The analysis allowed for comparison between the proposed models and existing experimental data. Through this comparison, bounds on the numerical coefficients fixing the model were obtained. Namely, the coefficient $\kappa$ was found to be smaller than $\leq 10^{41}$. The bounds one arrives at using RGUP are still a long way from the expected ones. however, they are ten orders of magnitude better than some of the previous estimations. This result can be attributed to several things, one of which is the fact that relativistic corrections occur alongside the ones arising from GUP. The improvement on the bound can also be attributed to the increase in accuracy of the experiments used to test the model. 

\chapter{Quantum Field Theory with Minimum Length}
\label{Ch4:QFTWML}
\begin{quote}
“Physicists have come to realize that mathematics, when used with sufficient care, is a proven pathway to truth.” \\
	\begin{flushright}
	Brian Greene, The Fabric of the Cosmos: Space, Time, and the Texture of Reality 2004
	\end{flushright}
\end{quote}
 The observable effects arising from the existence of minimum measurable length are expected to be relevant at Planck  scale, {\it i.e.} $E_{\text{Pl}}\sim 10^{19}\,\text{GeV}$. According to Eq.\eqref{Eq:UniversalCorrections}, low energy remnants of the existence of minimum lengths will manifest themselves as perturbative corrections to the energy of the system. Furthermore, the relative magnitude of RGUP corrections depends on the ratio between the energy of the measured system and Planck energy 
 \begin{equation}
    \frac{E_0-E_1}{E_0}\sim\frac{E_{\text{system}}}{E_{\text{Pl}}}\sim\frac{E_{\text{system}}}{10^{19}\,\text{GeV}}\,.
 \end{equation}
 Therefore, one needs very high energy in order to maximize that ratio, thus bringing the QG effects above the noise and accuracy of the measurements. Current experiment study the behaviour of highly energetic systems by accelerating particles to speeds close to the speed of light. The accurate modeling of the resulting scattering requires one use the theoretical framework of QFT.
 
This reveals one of the advantages of RGUP and its corresponding Lorentz invariant minimum measurable length. RGUP provides a base on which one can formulate modified QFT, and therefore search for low energy effects of minimum length in the most highly energetic Earth-based experiments.

Lagrangians of an effective QFT accommodating the existence of minimum length, utilize Eqs. \eqref{Eq:XPRepresentation}, \eqref{Eq:GUPxp}, and \eqref{Eq:PoincareAlgebra} presented in  chapter \ref{Ch3:RGUP} to find the equations of motion for a scalar and spinor fields. The Lagrangians are then minimally coupled to a gauge field allowing the Feynman rules to be read from them. First order and RGUP corrections to an electron-muon scattering amplitudes are then calculated and the results are compared with experimental data from ATLAS. The results obtained  are summarized and published in \cite{BOSSO2020168319,Bosso:2020jay}.

The calculation begins with a choice of the RGUP model by fixing the $\kappa$, $\beta$, $\xi$, and $\varepsilon$ parameters in Eq.\eqref{Eq:GUPxp}.
 The particular relationship chosen between the parameters $\kappa$, $\beta$, $\xi$, and $\varepsilon$ is $\kappa=\beta=\xi=0$. 
 This choice is  made for to two reasons: the first is that the terms proportional to $\beta$ and $\xi$ introduce preferred direction in spacetime; the second is that the terms proportional to $\kappa$ make spacetime non-commutative or ``fuzzy" in addition to making it a non-Lie manifold. 
 A prescription for how to approach this problem is not currently known. 
 Therefore for the purposes of the calculations presented in this chapter, the case for the parameters  $\kappa=0$ is considered, or in other words the auxiliary and physical position operators are one and the same .
In that case one has the following representation of the physical position and momentum in terms of the auxiliary ones
 \begin{subequations}
\begin{align}\label{Eq:XPParticualrCase}
x^{\mu}&=x_0^{\mu} \\
p^{\mu}&=p_0^\mu\,(1+\gamma p_0^{\rho}p_{0\rho})\,,
\end{align} 
\end{subequations}
where the dimensionless parameter $\varepsilon$ is redefined as $\varepsilon:=\gamma_0$, and the minimum length scale $\gamma$ takes the from 
\begin{equation}
    \gamma=\frac{\gamma_0}{(M_{\text{Pl}}c)^2}\,.
\end{equation}
Additionally, in this chapter, the results will be presented in natural units, {\it i.e.} $c=\hbar=1$. With the parameters redefined as such, the position-momentum commutator which gives the RGUP has the following form
\begin{equation}
    [x^{\mu},p^{\nu}]=i\hbar\,\left(1+\gamma p^{\rho}p_{\rho}\right)\eta^{\mu\nu}+2i\,\hbar\gamma p^{\mu}p^{\nu}\,.
\end{equation}
The Poincar\'e group for this case is represented by the following algebra
\begin{subequations}\label{Eq:QFTPoincare}
\begin{align}
   [x^\mu,M^{\nu\rho}] &=  i\hbar[1 + \gamma p^{\rho} p_{\,\rho}]\left(x^{\nu}\delta^{\mu\rho}-x^{\rho}\delta^{\mu\nu}\right) + i\hbar 2 \gamma p^{\mu} M^{\nu\rho}\,\\
   [p^\mu, M^{\nu\rho}]& =  i\hbar[1 + \gamma  p^{\rho} p_{\,\rho}]\left(p^{\nu}\delta^{\mu\rho}-p^{\rho}\delta^{\mu\nu}\right)\,\\
 [M^{\mu\nu},M^{\rho\sigma}]& = i\hbar\left(1 + \gamma p^{\rho} p_{\,\rho}\right)\left(\eta^{\mu\rho}M^{\nu\sigma}-\eta^{\mu\sigma} M^{\nu\rho} 
  -\eta^{\nu\rho}M^{\mu\sigma}+\eta^{\nu\sigma}M^{\mu\rho}\right)\,.
\end{align}
\end{subequations}
It can be easily double check that the physical momentum squared is a Casimir invariant of the modified Poincar\'e algebra presented above in Eq.\eqref{Eq:QFTPoincare}. Furthermore, the Einstein dispersion relation has the following form
\begin{equation}\label{Eq:QFTDispersionRelation}
   p^\mu p_\mu=-m^2\,
\end{equation}
where $p_\mu$ is the physical momentum. The differential form of Eq.\eqref{Eq:QFTDispersionRelation}
needs to be an equation of motion for any Lagrangian describing a boson field. The equations of motion that fermion fields obey can also be easily derived.
Starting from the dispersion relation Eq.\eqref{Eq:QFTDispersionRelation} and  expressing it in terms of the auxiliary variables defined in Eq.\eqref{Eq:XPParticualrCase}, one gets
 \begin{equation}
    \label{Eq:ModKGRedefined}
    p_0^{\rho}p_{0\rho}(1+\gamma p_0^{\sigma}p_{0\sigma})^2=-m^2 \,,
\end{equation}
or in its differential form 
 \begin{equation}\label{Eq:DIfferentialKGEoM}
       \left[\partial_\mu\partial^\mu \left(1+ \gamma \partial_\nu \partial^\nu\right)^2
     + m\right] \phi = 0 \,.
 \end{equation}
As derived in the appendix \ref{app:DiracSolutions}, one can then prove that the Dirac equation has the form 
 \begin{equation}\label{Eq:DiracEoM}
    \left(\tau^\mu p_\mu-m\right)\psi =\left[\tau^{\mu}p_{0\,\mu}(1+\gamma p_{0\rho}p_0^\rho)-m\right]\psi=0\,,
 \end{equation}
 where $\tau^\mu$ are the Dirac matrices and $\psi$ is a Dirac spinor. Dirac matrices are the generators of the Clifford algebra $Cl_{(3,1)}(\mathbb{R})$ which transforms spinor representation of the Lorentz group, {\it i.e.} the Dirac or spinor fields. As part of the Lorentz group they obey the Einstein dispersion relation reflected by the Casimir operator Eq.\eqref{Eq:QFTDispersionRelation} of the Poincar\'e algebra. Therefore, one can safely conclude that if the Casimir is preserved  then the properties of the Clifford algebra $Cl_{(3,1)}(\mathbb{R})$ will remain unchanged.
This allows one to use the usual representation of its generators 
 \begin{align}\label{gamma}
\tau^0=\begin{pmatrix} 
\mathbf{I}&0 \\
0 & -\mathbf{I} 
\end{pmatrix},\,\,\,\,\,\tau^i=\begin{pmatrix} 
0&\sigma^i \\
\sigma^i &0
\end{pmatrix}\,,
\end{align}
where $\sigma^i$ are the Pauli matrices and $i\in\{1,2,3\}$. More detailed explanations of the calculations are presented in appendix \ref{app:DiracSolutions}. From the unmodified Electrodynamics one knows that the differential form of
Eq.\eqref{Eq:DiracEoM}, {\it i.e.}
\begin{equation}\label{Eq:DifferentialDiracEoM}
     \left[ i\tau^\mu\partial_\mu(1+\gamma\partial_\rho\partial^\rho)^2 -m\right]\psi=0\,,
\end{equation}
and its Dirac conjugate
are the equations of motion for the RGUP-QED Lagrangian.
\section{RGUP modified Lagrangians}\label{sec:RGUPLagrangians}
  In the previous section, the equations of motion corresponding to the scalar and spinor fields respectively have been established. Note that the equations of motions are differential equations of higher than second order. Therefore, the Lagrangians giving these equations of motion need to have higher than second derivative. The methodology of working with higher derivative Lagrangians
\begin{subequations}
\begin{align}
    \label{Eq:LagrangianGeneral}  L=&L(\phi,\dot\phi,\ddot\phi,\ldots,\overset{(n)}{\phi})\\
  \label{Eq:LagrangianDensityGeneral}  \mathcal{L}=&\mathcal{L}(\phi,\partial_{\mu_1}\phi,\partial_{\mu_1}\partial_{\mu_2}\phi,\ldots,\partial_{\mu_1}\ldots\partial_{\mu_n}\phi)
\end{align}
\end{subequations}
is given by the Ostrogradsky method \cite{Pons:1988tj,Woodard:2015zca,deUrries:1998obu}. According to the Ostrogradsky method, the Euler-Lagrange equations for theories with higher derivatives will have the form:
\begin{equation}\label{Eq:OstrogradskyEulerLagrange}
    \frac{dL}{dq} -\frac{d}{dt}\frac{dL}{d\dot{q}}+\frac{d^2}{dt^2}\frac{dL}{d\ddot{q}}+\ldots+(-1)^n\frac{d^n}{dt^n}\frac{dL}{d (d^nq/dt^n)}=0\,,
\end{equation}
which in the case of fields is
\begin{equation}
    \frac{\partial\mathcal{L}}{\partial\phi}-  \partial_\mu  \frac{\partial\mathcal{L}}{\partial(\partial_\mu\phi)}+   \partial_{\mu_1} \partial_{\mu_2}\frac{\partial\mathcal{L}}{\partial(\partial_{\mu_1} \partial_{\mu_2}\phi)}+\ldots
    +(-1)^m\partial_{\mu_1}\ldots\partial_{\mu_m}\frac{\partial\mathcal{L}}{\partial(\partial_{\mu_1} \ldots\partial_{\mu_m}\phi)}=0\,.
\end{equation}
The Ostrogradsky method allows us to reconstruct the Lagrangian describing the dynamics of scalar and spinor fields from their equations of motion. 

\subsection{\label{sec:Lag}Scalar field Lagrangian}

  In order to perform a quantization procedure for the different cases, one needs to obtain their Lagrangians. In the following subsection, from its equations of motion the author derives the Lagrangians for a free massive scalar field.  When one  obtains  the equations of motion for the scalar field one should  find Eq.\eqref{Eq:DIfferentialKGEoM}.

One begins by assuming the most general form of higher derivative Lagrangian, which has one order higher than the equations of motion
\begin{align}\label{Eq:Lagrangian}
   \nonumber \mathcal{L}=\frac{1}{2}\partial_{\mu}\phi\partial^{\mu}\phi&+\gamma \left( C_1\,\partial_\mu\partial^\mu\phi\,\partial_\nu\partial^\nu\phi+C_2\,\partial_\mu\phi\,\partial^\mu\partial_\nu\partial^\nu\phi+C_3\,\partial_\nu\partial^\nu\partial^\mu\phi\,\partial_\mu\phi\right)\\\nonumber&+\gamma^2\left(C_4\, \partial_\mu\partial^\mu\partial_\nu\phi\,\partial^\nu\partial_\rho\partial^\rho\phi+C_5\, \partial_\mu\partial^\mu\partial_\nu\partial^\nu\phi\,\partial_\rho\partial^\rho\phi+C_6\, \partial_\mu\partial^\mu\partial_\nu\partial^\nu\partial_\rho\phi\,\partial^\rho\phi\right.\\
   &\left.+C_7\, \partial_\mu\partial^\mu\phi\,\partial_\nu\partial^\nu\partial_\rho\partial^\rho\phi+C_8\, \partial_\mu\phi\,\partial^\mu\partial_\nu\partial^\nu\partial_\rho\partial^\rho\phi\right)+C_9m^2\phi^2\,,
\end{align}
Using the Euler-Lagrange  equation prescribed by the Ostrogradsky method Eq.\eqref{Eq:OstrogradskyEulerLagrange}, one obtains the following Lagrangian 
\begin{equation}\label{Eq:RealLagrangian}
    \mathcal{L}_{\phi,\mathbb{R}}=\frac{1}{2}\partial_{\mu}\phi\partial^{\mu}\phi-\frac{1}{2}m^2\phi^2+\gamma \,\partial_\nu\partial^\nu\partial^\mu\phi\,\partial_\mu\phi
    +\frac{\gamma^2}{2}\partial_\mu\phi\,\partial^\mu\partial_\nu\partial^\nu\partial_\rho\partial^\rho\phi
    \,,
\end{equation}
where $\partial_\mu = \partial / \partial x^\mu$. One can see in Appendix \ref{app:Lagrangian} how the coefficients are fixed. Additionally, it is shown that for the particular case the coefficients  are unique.
As for the Lagrangian for a complex scalar field $\phi$, we generalize Eq.\eqref{Eq:RealLagrangian} by including additional terms obtaining 
\begin{multline}\label{Eq:ComplexLagrangian}
 \mathcal{L}_{\phi,\mathbb{C}} = \frac{1}{2} \left(\partial_{\mu}\phi\right)^\dagger \partial^{\mu}\phi - \frac{1}{2} m^2 \phi^\dagger \phi + \gamma \left[\left(\partial_\nu \partial^\nu \partial^\mu \phi \right)^\dagger \partial_\mu \phi + \partial_\nu \partial^\nu \partial^\mu \phi \left(\partial_\mu \phi \right)^\dagger \right]\\
+ \frac{\gamma^2}{2} \left[\left(\partial_\mu \phi \right)^\dagger \partial^\mu \partial_\nu \partial^\nu \partial_\rho \partial^\rho \phi + \partial_\mu \phi \left(\partial^\mu \partial_\nu \partial^\nu \partial_\rho \partial^\rho \phi \right)^\dagger \right],
\end{multline}
such that hermiticity is restored, \textit{i.e.} $\mathcal{L}_{\phi,\mathbb{C}}^{\dagger}=\mathcal{L}_{\phi,\mathbb{C}}$.
Furthermore, it is worth noticing that Eq.\eqref{Eq:ComplexLagrangian} is consistent with Eq.(55) in \cite{Kober:2011dn} up to a numerical factor.
\subsection{\label{sec:QED}Spinor field Lagrangian}
  The first step in the derivation of the spinor field Lagrangian is to assume a general form of the Lagrangian, where the order of derivatives is determined by the order of the equation of motion Eq.\eqref{Eq:DifferentialDiracEoM}.
Moreover, an  assumption is made that different terms will have an arbitrary numerical coefficients multiplying every term
\begin{equation}\label{Eq:GenearalDiracLagrangian}
     \mathcal{L}_{\psi}=\bar\psi\left[ iC_1\tau^\mu\partial_\mu(1+C_2\gamma\partial_\rho\partial^\rho) -C_3 m\right]\psi\,.
 \end{equation}
Next step is to prove that it has Eq.\eqref{Eq:DifferentialDiracEoM} as an equation of motion.
Applying the Ostrogradsky method, one gets the following equations of motion for the field and its complex conjugate
\begin{align}
     &C_1i\tau^\mu\partial_\mu\psi +C_1C_2\gamma\partial_\rho\partial^\rho\psi -C_3 m\psi=0,\\
     &C_1i\tau^\mu\partial_\mu\bar\psi+C_1C_2\gamma\partial_\rho\partial^\rho\bar\psi -C_3 m\bar\psi=0\,.
\end{align}
The equations of motion obtained through the Ostrogradsky method from Eq.\eqref{Eq:GenearalDiracLagrangian} need to be identical to Eq.\eqref{Eq:DifferentialDiracEoM}, which is  obtained from the dispersion relation.  
Therefore, the Lagrangian corresponding to the QFT spinor with minimum length will be of the form 
 \begin{equation}\label{Eq:DirackLagrangian}
     \mathcal{L}_{\psi}=\bar\psi\left[ i\tau^\mu\partial_\mu(1-\gamma\,\partial_\rho\partial^\rho) -m\right]\psi\,.
 \end{equation}
The details of the calculation are shown in Appendix \ref{app:lagrangian}.

\subsection{U(1) gauge field theory}
  The gauge field Lagrangian is derived in a slightly different way. The Ostrogradsky method is not utilized in this case. The derivation is  performed by assuming that the dynamics of the gauge field, described by the electrodynamics Lagrangian, are invariant under the standard spacetime and $U(1)$ gauge symmetries.
As for the equations of motion, one can  assume that they have the same RGUP corrections as the KG equation of the form Eq.\eqref{Eq:DIfferentialKGEoM}
 \begin{equation}\label{Eq:GaugeFieldEOM}
     \partial_\mu F^{\mu\nu}= \partial_\mu\partial^\mu A^\nu +2\gamma \partial_\mu\partial^\mu\partial_\rho\partial^\rho A^\nu  
     =0\,.
 \end{equation}
 Defining the standard gauge invariant
 field strength tensor as
 \begin{equation}
     F^{\mu\nu}_0=\partial^\mu A^\nu - \partial^\nu A^\mu,
 \end{equation}
one can express the RGUP modified field strength tensor in terms of the standard one up to first order in $\gamma$, as follows:
 \begin{equation}\label{Eq:EMFieldTensor}
    F^{\mu\nu}=F^{\mu\nu}_0+2\gamma\, \partial_\rho \partial^\rho F^{\mu\nu}_0\,.
 \end{equation}
 Then Eq.\eqref{Eq:GaugeFieldEOM} can be rewritten as 
 \begin{equation}\label{Eq:FieldTensorEoM}
    \partial_\mu F^{\mu\nu}=\partial_\mu F^{\mu\nu}_0+2\gamma  \partial_\rho \partial^\rho\partial_\mu F^{\mu\nu}_0
    +\gamma^2\partial_\sigma \partial^\sigma\partial_\rho \partial^\rho\partial_\mu F^{\mu\nu}_0
    \,.
 \end{equation}
 A test of gauge invariance of $F^{\mu\nu}$ is needed and carried through by considering the following gauge transformation of the four-potential
 \begin{equation}
     A^\mu\rightarrow  A^{\prime\mu}=A^{\mu}+\partial^\mu\Lambda\,.
 \end{equation}
 Then the gauge transformed  Eq.\eqref{Eq:EMFieldTensor} can be written as follows 
 \begin{align}
\nonumber     F^{\prime\mu\nu}=&\partial^\mu A^{\prime\nu} - \partial^{\nu} A^{\prime\mu}+2\gamma\left[\partial_\rho \partial^\rho \partial^\mu A^{\prime\nu} - \partial_\rho \partial^\rho\partial^\nu\partial_\nu A^{\prime\mu}\right]\\
 \nonumber    =&\partial^\mu A^{\nu}+\partial^\mu\partial^\nu\Lambda-\partial^{\nu}A^{\mu}-\partial^{\nu}\partial^\mu\Lambda\\\nonumber
 &+2\gamma\left[\partial_\rho \partial^\rho\partial^\mu A^{\nu}+\partial_\rho \partial^\rho\partial^\mu\partial^\nu\Lambda-\partial_\rho \partial^\rho\partial^{\nu}A^{\mu}-\partial_\rho \partial^\rho\partial^{\nu}\partial^\mu\Lambda\right]\\
 =&\partial^\mu A^{\nu}-\partial^{\nu}A^{\mu}
 +2\gamma\left[\partial_\rho \partial^\rho\partial^\mu A^{\nu}-\partial_\rho \partial^\rho\partial^{\nu}A^{\mu}\right]=F^{\mu\nu}\,,
  \end{align}
where the fact that derivatives commute with each other is  used. Thus, up to first order in the RGUP parameter $\gamma$, the gauge field Lagrangian reads
 \begin{equation}\label{Eq:GaugeFieldL}
     \mathcal{L}_{A}=-\frac{1}{4}F^{\mu\nu}F_{\mu\nu}=-\frac{1}{4}F^{\mu\nu}_0F_{\mu\nu 0}-\frac{\gamma }{2}F_{\mu\nu 0} \partial_\rho \partial^\rho F^{\mu\nu}_0
     \,.
 \end{equation}
Notice that both the field tensor in Eq.\eqref{Eq:EMFieldTensor} and the gauge field Lagrangian Eq.\eqref{Eq:GaugeFieldL} are invariant under $U(1)$ gauge transformations.

\section{Feynman rules}\label{Sec:FeynmanRules}
  The scalar, spinor and gauge fields Lagrangians derived in the previous section can be used to formulate a RGUP deformed scalar and spinor electrodynamics. In the following section the lepton fields are minimally coupled to the gauge field.  The Feynman rules, consisting of the propagators and vertices for both cases are  calculated and presented. The methodology used can be found in any QFT textbook. In particular the ones used here are \cite{Srednicki2005QFT,Weinberg2005QFT,Das2008QFT}. 
\subsection{Scalar field coupled to U(1) gauge theory}
  Following standard procedure, the Feynman propagator for the scalar field with a minimum length is calculated.
Specifically, from the modified KG equation in Eq.\eqref{Eq:DIfferentialKGEoM} one has:
\begin{equation}\label{GreenFunc}
     \left[\partial_\mu\partial^\mu \left(1+ \gamma \partial_\nu \partial^\nu\right)
     + (mc)^2 \right] G(x-x') = - i \delta(x-x')\,.
\end{equation}
Expressing the Green's function $G(x - x')$ in terms of its Fourier transform
\begin{equation}\label{GreenFourier}
    G(x-x')=\int \frac{d^4p_0}{(2\pi)^4} \tilde G(p_0) e^{-i p_0\cdot(x-x')},
\end{equation}
and substituting it in Eq.\eqref{GreenFunc}, one gets
\begin{equation}
\int \frac{d^4p_0}{(2\pi)^4} \tilde G(p_0) \left[-p_0^2 (1 - \gamma p_0^2) + (mc)^2\right] e^{-i p_0\cdot(x-x')}=-i \int\frac{d^4p_0}{(2\pi)^4}e^{-i p_0\cdot(x-x')}\,.
\end{equation}
Therefore, the Fourier transform of the Feynman propagator has the form 
\begin{equation}
     \tilde G(p_0)  =\frac{-i}{-p_0^2 (1+ \gamma p_0^2) + (mc)^2}\,,
\end{equation}
while the propagator itself is
\begin{equation}\label{Eq:ScalarPropagator}
     G(x-x')=\int \frac{d^4p_0}{(2\pi)^4} \frac{-i}{-p_0^2 ( 1 + \gamma p_0^2) + (mc)^2}e^{-i p_0\cdot(x-x')}\,.
\end{equation}
The gauge field propagator can be treated in a similar manner.
In this case, the Feynman propagator has the following form
\begin{equation}\label{Eq:GaugePropagator}
G(x-x')=\int \frac{d^4q_0}{(2\pi)^4} \frac{-i}{-q_0^2+2\gamma  q_0^4}e^{-i q_0\cdot(x-x')}\,,
\end{equation}
where $q_0$ is the auxiliary four-momentum of the gauge field. This construction assumes that one can define Dirac delta functions in position space, and the minimal length arises when one tries to localize the fields. 

The complete set of Feynman rules for the system requires the calculation of the vertices for the charged and gauge fields. Starting from the Lagrangian in Eq.\eqref{Eq:ComplexLagrangian}, one introduces 
the minimal coupling. The derivatives $\partial_\mu=\frac{\partial}{\partial x_0^\mu}$ are coupled to the gauge field, forming covariant derivatives of the from
\begin{equation}
     \partial_\mu \rightarrow D_\mu=\partial_\mu -ie A_\mu\,,
\end{equation}
where $A_\mu$ is the four-potential or the gauge field. One can notice that since the gauge symmetries of the electromagnetic field are not modified, the covariant derivative keeps its usual form. 

Replacing partials in Eq.\eqref{Eq:ComplexLagrangian} by covariant derivatives,   the full action of the minimally coupled complex scalar field and the gauge field is obtained. The action reads as follows
\begin{multline}
    \int \mathcal{L}\,d^4x = \int\left[\mathcal{L}_{A} + \mathcal{L}_{\phi,\mathbb{C}}\right]\,d^4x = \int\left\{\frac{1}{2} \left(D_{\mu} \phi\right)^\dagger D^{\mu} \phi - \frac{1}{2} m^2 \phi^\dagger \phi - \frac{1}{4}F^{\mu\nu}F_{\mu\nu}\right.\\
    + \gamma \left. \left[\left(D_\nu D^\nu D^\mu \phi\right)^\dagger D_\mu \phi + D_\nu D^\nu D^\mu \phi \left(D_\mu \phi\right)^\dagger \right] \right\}
+ \frac{\gamma^2}{2} \left[\left(D_\mu \phi\right)^\dagger D^\mu D_\nu D^\nu D_\rho D^\rho \phi\right.\\\left.\left. + D_\mu \phi \left(D^\mu D_\nu D^\nu D_\rho D^\rho \phi\right)^\dagger \right]\right\}
\,d^4x\, .
\end{multline}
By expanding the covariant derivatives, the Lagrangian can be rewritten as
\begin{multline}\label{Eq:CoupledScalarLagrangian}
    \mathcal{L} = \frac{1}{2} \left(\partial_{\mu} \phi\right)^\dagger \partial^{\mu} \phi - i e A^\mu \left[\phi^\dagger \partial_\mu \phi - \phi \left(\partial_\mu \phi\right)^\dagger \right] - \frac{1}{2} m^2 \phi^\dagger \phi + e^2 A_\mu A^\mu \phi^\dagger \phi - \frac{1}{4} F^{\mu\nu}F_{\mu\nu}\\
    \gamma \left\{\left(\partial_\nu \partial^\nu \partial^\mu \phi\right)^\dagger \partial_\mu \phi + \partial_\nu \partial^\nu \partial^\mu \phi \left(\partial_\mu \phi\right)^\dagger - \frac{1}{4} F^{\mu\nu} F_{\mu\nu} \partial_\mu \partial_\nu F^{\mu\nu} \right. \\
    - i e \left\{\left(\partial_\nu \partial^\nu A^\mu\right) \left[\phi^\dagger \partial_\mu \phi - \phi \left(\partial_\mu\phi\right)^\dagger \right] + 2 \partial_\nu A^\mu \left[\left(\partial^\nu \phi\right)^\dagger \partial_\mu \phi - \partial^\nu \phi \left(\partial_\mu \phi^\dagger\right)\right] \right.\\
    \left. + A^\mu \left[\left(\partial_\nu \partial^\nu \phi\right)^\dagger \partial_\mu \phi - \partial_\nu \partial^\nu \phi \left(\partial_\mu \phi^\dagger\right)\right]\right\}
   + e^2 \left\{ A^\nu(\partial_\nu A_\mu) \left[\left(\partial^\mu \phi\right)^\dagger \phi + (\partial^\mu \phi) \phi^\dagger\right]\right.\\ + 2A^\nu A_\mu \left[\left(\partial^\mu \phi\right)^\dagger \partial_\nu \phi + \left(\partial^\mu \phi\right) \left(\partial_\nu \phi\right)^\dagger\right] \\
    + A^\nu A_\nu \left[ \left(\partial^\mu \phi\right)^\dagger \partial_\mu \phi + \left(\partial^\mu \phi\right) \left(\partial_\mu \phi\right)^\dagger\right] - 2 A^\mu A^\nu \left[\phi^\dagger\partial_\nu\partial_\mu\phi + \phi \left(\partial_\nu\partial_\mu\phi\right)^\dagger\right]\\
    \left. + 2 A^\mu (\partial_\nu \partial^\nu A_\mu) \phi^\dagger \phi + A^\mu (\partial^\nu A_\mu) \left[\phi^\dagger \partial_\nu \phi  + \phi \left(\partial_\nu \phi\right)^\dagger\right] + A^\mu A_\mu \left[\phi^\dagger \partial_\nu \partial^\nu \phi + \phi \left(\partial_\nu \partial^\nu \phi\right)^\dagger\right]\right\}\\
    + i e^3 \left\{ A_\mu A_\nu A^\nu \left[\phi \left(\partial^\mu \phi\right)^\dagger - \phi^\dagger\partial^\mu\phi\right] + 2 A^\mu A_\mu A^\nu \left[\phi^\dagger \partial_\nu \phi - \phi \left(\partial_\nu \phi\right)^\dagger\right] \right\} \\
    \left. + 2 e^4 A^\mu A_\mu A^\nu A_\nu\phi^\dagger\phi\right\} 
    + \mathcal{O}(\gamma^2)\,.
\end{multline}
The above expression contains all terms relevant to Feynman diagrams predicted by the usual scalar Quantum Electrodynamics (QED) Lagrangian with the addition of RGUP corrections.  It is easy to see that up to  6-point vertices are allowed. This can be seen by examining  Eq.\eqref{Eq:CoupledScalarLagrangian} and observing that the maximum number of lines meeting at a vertex will have two scalars and four gauge bosons.

One can check that there are 74 Feynman diagrams, with coupling constants proportional up to $e^4$ and $\gamma^3$. It is also worth mentioning that due to the presence of higher than second order derivatives the momentum conservation rules for each RGUP modified Feynman vertex will also be modified.  Additionally, one can check that for the limit $\gamma \rightarrow 0$ or infinitely small minimum length we recover the usual Lagrangian for a complex scalar fields.
\begin{table}[ht]
\begin{center}
\caption{\label{Tbl:CouplingConstantTable}Classification of the Feynman vertices arising from Eq.\eqref{Eq:CoupledScalarLagrangian} in terms of the powers of the coupling constants: $\alpha$ the fine structure constant; $\gamma$ the minimum length coefficient. }
\begin{tabular}{c | c c c c}
\toprule
  \quad  & \multicolumn{4}{c}{\text{Powers of }$\gamma$}\\  
 \hline\hline
   \multirow{6}{*}{\text{Powers of }$\alpha$}&\cellcolor{blue!25}$\alpha^{1/2}$&\cellcolor{blue!25}$\alpha^{1/2}\gamma$ &\cellcolor{red!25}$\alpha^{1/2}\gamma^2$&\cellcolor{red!25}$\alpha^{1/2}\gamma^3$ \\
  &\cellcolor{red!25}$\alpha$&\cellcolor{red!25}$\alpha\gamma$&\cellcolor{red!25}$\alpha\gamma^2$&\cellcolor{red!25}$\alpha\gamma^3$\\
   & N/A &\cellcolor{red!25}$\alpha^{3/2}\gamma$&\cellcolor{red!25}$\alpha^{3/2}\gamma^2$ &\cellcolor{red!25}$\alpha^{3/2}\gamma^3$  \\
 & N/A &\cellcolor{red!25} $\alpha^{2}\gamma$&\cellcolor{red!25}$\alpha^{2}\gamma^2$&\cellcolor{red!25}$\alpha^{2}\gamma^3$ \\
   & N/A &\cellcolor{red!25}$\alpha^{5/2}\gamma$&\cellcolor{red!25}$\alpha^{5/2}\gamma^2$&\cellcolor{red!25} $\alpha^{5/2}\gamma^3$ \\
  & N/A &\cellcolor{red!25}$\alpha^3\gamma$&\cellcolor{red!25}$\alpha^3\gamma^2$&\cellcolor{red!25}$\alpha^3\gamma^3$\\
  \bottomrule
  \end{tabular}
  \end{center}
\end{table}
 The coupling constants of the Feynman vertices which correspond to terms in Eq.\eqref{Eq:CoupledScalarLagrangian} are presented  in Table.\ref{Tbl:CouplingConstantTable}. 
However, when calculating the scattering amplitudes the focus will be on the 3-point vertices, containing up to first order in the coupling constant $e$ and the RGUP coefficient $\gamma$.
This approximation is justified by the fact that the 3-point vertices will have the largest contribution to the scattering amplitudes. The truncation up to first order in $\gamma$ is  done to isolate the RGUP corrections to only the 3-point vertices. 

\subsection{Dirac field coupled to U(1) gauge theory}\label{sec:DirackFieldFeynmanRules}
  The modified Feynman propagator is the Green's function of the modified Dirac differential operator.
Therefore, from the modified Dirac equation in Eq.\eqref{Eq:DiracEoM}, one has
 \begin{equation}\label{Eq:DiracGreenFunc}
   \left[ i\tau^\mu\partial_\mu(1-\gamma\partial_\rho\partial^\rho) -m\right] G(x-x') = - i \delta(x-x')\,.
\end{equation}
Expressing the Green's function $G(x - x')$ in terms of its Fourier transform
\begin{equation}\label{Eq:DiracGreenFourier}
    G(x-x')=\int \frac{d^4p_0}{(2\pi)^4} \tilde G(p_0) e^{-i p_0\cdot(x-x')},
\end{equation}
and substituting it in Eq.\eqref{Eq:DiracGreenFunc}, one gets
\begin{equation}
\int \frac{d^4p_0}{(2\pi)^4} \tilde G(p_0)\left[ i\tau^\mu\partial_\mu(1-\gamma\partial_\rho\partial^\rho) -m\right] e^{-i p_0\cdot(x-x')}=-i \int\frac{d^4p_0}{(2\pi)^4}e^{-i p_0\cdot(x-x')}\,.
\end{equation}
Therefore, the Fourier transform of the Feynman propagator has the form
\begin{equation}
     \tilde G(p_0) \left[\tau^{\mu}p_{0\,\mu}(1+\gamma p_{0\rho}p_0^\rho)-m\right] =-i \mathbb{I}^4\,,
\end{equation}
where $\mathbb{I}^4$ is a four dimensional unit matrix. 
Multiplying both sides by 
\begin{equation}
    \left[\tau^{\mu}p_{0\,\mu}(1-\gamma p_{0\rho}p_0^\rho)+m\right]\,,
\end{equation}
one obtains the propagator
%
%
\begin{equation}\label{Eq:DiracPropagator}
     G(x-x')=\int \frac{d^4p_0}{(2\pi)^4}\frac{-i \left[\tau^{\mu}p_{0\,\mu}(1+\gamma p_{0\rho}p_0^\rho)+m\right]}{p_0^{\mu}p_{0\,\mu}(1+\gamma p_{0\rho}p_0^\rho)^2-m^2}e^{-i p_0\cdot(x-x')}\,.
\end{equation}
Further, when the
full RGUP-QED action is considered, it reads as follows 
\begin{align}
   S= \int \mathcal{L}\,d^4x &= \int\left[\mathcal{L}_{A} + \mathcal{L}_{\psi}\right]\,d^4x \nonumber\\ &= \int\bar{\psi}\left[ i\tau^\mu D_{\mu}\psi -\gamma \bar{\psi} i\tau^\mu D_{\mu}D_{\nu}D^{\nu}\psi  - \frac{1}{4}F^{\mu\nu}F_{\mu\nu}\right]\,\,d^4x \,,
   \label{qedaction1}
\end{align}
in which the minimal coupling to the $U(1)$ gauge field has been introduced in the usual form
 \begin{equation}
     \partial_\mu \rightarrow D_\mu=\partial_\mu +ie A_\mu\,.
\end{equation}
It can be easily verified that the above, along with a local phase transformations of $\psi$ and $\bar\psi$, leaves the action
in Eq.\eqref{qedaction1} invariant.
The vertices can be read from the minimally coupled  modified Dirac field Lagrangian Eq.\eqref{Eq:DirackLagrangian} 
\begin{multline}\label{Eq:RGUPQED}
    \mathcal{L}_{\psi}=i\bar{\psi} \tau^\mu \partial_{\mu}\psi +i\gamma\bar{\psi}\tau^\mu \partial^{\rho}\partial_{\rho}\partial_{\mu}\psi-m\bar{\psi}\psi\\
    -e \left[\bar{\psi} \tau^\mu A_{\mu}\psi - 2 \gamma \bar{\psi} \tau^\mu \left(\partial_{\mu}A^{\rho}\right) \partial_\rho\psi - 2 \gamma\bar{\psi} \tau^\mu A^{\rho}\partial_{\mu}\partial_\rho\psi - \gamma\bar{\psi} \tau^\mu A_\mu\partial^{\rho}\partial_{\rho}\psi\right]\\
    - i e^2 \gamma \left[2 \bar{\psi} \tau^\mu A_\mu A^{\rho}\partial_{\rho}\psi + 2 \bar{\psi} \tau^\mu \left(\partial_\mu A^{\rho}\right)A_{\rho}\psi - \bar{\psi} \tau^\mu A^{\rho}A_{\rho}\partial_\mu\psi\right]\\
    - e^3\gamma\bar{\psi}\tau_{\mu}A^{\mu}A^{\rho}A_{\rho}\psi\,.
\end{multline}
One can see that there are up to five particle vertices.
They include always two fermions and from one to three gauge bosons.
Furthermore, one can see that the coupling constants for each vertex is a product of powers of the electronic charge $e$ and the RGUP coefficient $\gamma$.
In fact, the power of the electronic charge determines how many bosons couple to the vertex and the power of $\gamma$ is $0$ for the usual terms and $1$ for the RGUP corrections terms.

\section{RGUP Corrections to QED scattering amplitudes}\label{sec:Crosssections}
  The Lagrangians and the Feynman rules derived in the previous section deal with complex scalar fields and spinor fields. The Feynman rules for the complex scalar fields describe the interaction of charged systems that have no spin. The behaviour of electromagnetically interacting particle with half-spin is described by the Dirac field Lagrangian. In this section the author calculates the RGUP corrected amplitudes of electromagnetic scattering of a scalar electron and a scalar muon. In addition, one can find the  RGUP scattering amplitudes for real electron-muon interaction. The scattering amplitudes  for an electrom muon scattering 
\begin{figure}
\centering
\includegraphics[width=0.3\textwidth]{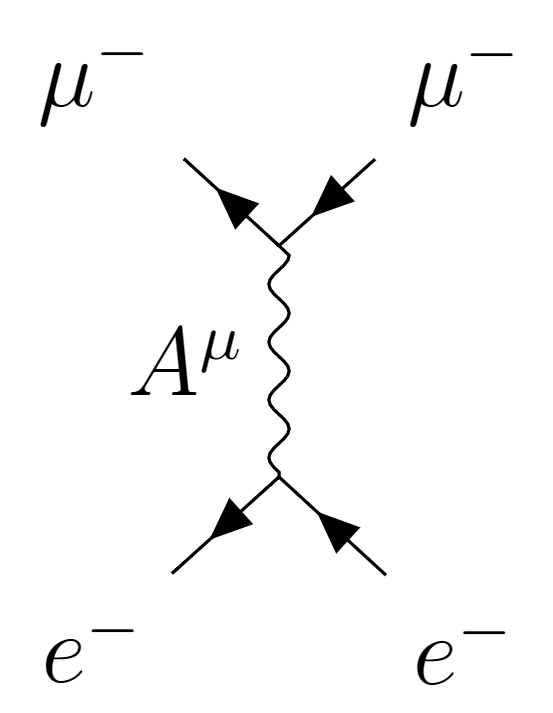}
\caption{\label{Fig:Muonscattering}The Feynman diagram of the leading order electron, muon  scattering, in $t$ channel. }
\end{figure}
and the corrections to them presented in the text below follow the methodology presented in \cite{Halzen:1984mc}.

\subsection{High energy scattering of charged scalar particles}
  The leading order of the scattering amplitude, on which the following calculation focuses, are  provided by the three-particle Feynman vertices derived from Eq.\eqref{Eq:CoupledScalarLagrangian}.

The transition amplitude for this case is given by integrating all three particle interaction terms 
\begin{equation}
    T_{fi}=-i\int  A^\mu j_\mu\,d^4x\,, \label{Eq:ScalarAmplitude}
\end{equation}
where $j_{\mu}$ is the current corresponding to the electrons and muons.
In what follows, the form of the transition amplitude shown in Eq.\eqref{Eq:ScalarAmplitude} is the same for both vertices, where the potential is provided by the same gauge boson.
One can then easily see that from Eq.\eqref{Eq:CoupledScalarLagrangian}, the terms containing one gauge field and two scalar fields will contribute to the following two terms in the transition amplitude
\begin{subequations} \label{Eq:AmplitudeSeparated}
\begin{align}
 T_{fi}^{(1)}=&-i \int e A^\mu\left[\phi_f^\dagger \partial_\mu\phi_i-\phi_f\partial_\mu\phi_i^\dagger \right] d^4x\,,\\
 T_{fi}^{(2)}=&-i \int e\gamma  A^\mu\left[4\partial_\nu\partial^\nu\phi^\dagger \partial_\mu\phi+4\partial^\nu\phi^\dagger \partial_\nu\partial_\mu\phi+\phi^\dagger \partial_\nu\partial^\nu\partial_\mu\phi-4\partial_\nu\partial^\nu\phi\partial_\mu\phi^\dagger \right. \nonumber\\ 
 &\left.-4\partial^\nu\phi\partial_\nu\partial_\mu\phi^\dagger -\phi\partial_\nu\partial^\nu\partial_\mu\phi^\dagger  \right]d^4x\,.
\end{align}
\end{subequations}
The full transition amplitude for the three particle scattering is then given by the sum of these two terms
\begin{equation}
    T_{fi}= T_{fi}^{(1)}+ T_{fi}^{(2)}
    \,.
\end{equation}
The term $T_{fi}^{(1)}$, obtainable from the unmodified scalar QED Lagrangian, corresponds to the limit $\gamma \rightarrow 0$ of the expression above.
Comparing Eqs. \eqref{Eq:ScalarAmplitude} and \eqref{Eq:AmplitudeSeparated}, one can decompose the current $j_{\mu}$ into the unmodified  term and the RGUP correction
\begin{subequations} \label{Eq:Currents}
\begin{align}\label{Eq:Currents1}
 j_{fi\,\mu}^{(1)}=&-i e\left[\phi_f^\dagger \partial_\mu\phi_i-\phi_f\partial_\mu\phi_i^\dagger \right]\,,\\
\nonumber j_{fi\,\mu}^{(2)}=&-i e\gamma  \left[4\partial_\nu\partial^\nu\phi^\dagger \partial_\mu\phi+4\partial^\nu\phi^\dagger \partial_\nu\partial_\mu\phi+\phi^\dagger \partial_\nu\partial^\nu\partial_\mu\phi-4\partial_\nu\partial^\nu\phi\partial_\mu\phi^\dagger \right.\\ &\left.-4\partial^\nu\phi\partial_\nu\partial_\mu\phi^\dagger -\phi\partial_\nu\partial^\nu\partial_\mu\phi^\dagger\right]~.
\end{align}
\end{subequations}
The following standard form for the scalar field is assumed
\begin{equation}\label{Eq:ScalarField}
    \phi(x_\mu)=N e^{-i p_\mu x^\mu}\,,
\end{equation}
where $N$ is the normalization constant and $p_\mu$ is the physical momentum of the field.
Using Eqs. \eqref{Eq:Currents}, one obtains
\begin{subequations} \label{Eq:CurrentsSUbbstituted}
\begin{align}
 j_{fi\,\mu}^{(1)}=&- e N_f N_i(p_f+p_i)_\mu e^{i(p_f-p_i)\cdot x}\,,\\
 j_{fi\,\mu}^{(2)}=&- e \gamma N_f N_i \left[-(p_f\cdot p_f)(4 p_i+-p_f)_\mu +4(p_f\cdot p_i)( p_i+p_f)_\mu\right.\nonumber\\
 &\left.-(p_i\cdot p_i)( p_i+4 p_f)_\mu\right]e^{i(p_f-p_i)\cdot x}
 \,.
\end{align}
\end{subequations}
The Feynman diagram for the scattering is obtained by connecting two 3-point vertices through the gauge boson $A^\mu$, as shown in Fig. \ref{Fig:Muonscattering}.
Thus the scattering amplitude is expressed as 
\begin{equation}
  T_{ABCD}=-i \int  \left(j_{AC\,\mu}^{(1)}+ j_{AC\,\mu}^{(2)}
  \right)\frac{ig^{\mu\nu}}{-q+2\gamma q^4}
  \left(j_{BD\,\nu}^{(1)}+ j_{BD\,\nu}^{(2)}
  \right)\,.
\end{equation}
Using Eqs. \eqref{Eq:CurrentsSUbbstituted} and doing the integration, an expression for the scattering amplitude is found
\begin{equation}\label{Eq:ScalarTransition}
     T_{ABCD}=-N_A N_B N_C N_D (2\pi)^4 \delta^{(4)}(p_D+p_C-p_A-p_B)\,\mathfrak{M}\,,
\end{equation}
where $\delta^{(4)}(p_D+p_C-p_A-p_B)$ ensures conservation of momentum, $\mathfrak{M}$ is the invariant amplitude
\begin{multline}\label{Eq:InvariantAmplitude}
    -i \mathfrak{M}=-i \frac{e^2}{-q^2(1-2\gamma q^2)}\left\{(p_A+p_C)\cdot(p_B+p_D)\left[1-4\gamma \left(p_A\cdot p_C+p_B\cdot p_D\right)\right]\right.\\+\gamma (p_A+ p_C)\cdot\left(4p_B-p_D\right)m_{\mu^-}^2-\gamma\left(p_A+ p_C\right)\cdot\left(4p_D-p_B\right)m_{\mu^-}^2\\
    +\left.\gamma (p_B+ p_D)\cdot\left(4p_A-p_C\right)m_{e^-}^2-\gamma(p_B+ p_D)\cdot\left(4p_C-p_A\right)m_{e^-}^2\right\}\,,
\end{multline}
and the $\frac{-i}{-q^2+2\gamma q^4}$ is the propagator of the gauge field $A_\mu$.The terms containing higher than linear order in the RGUP coefficient $\gamma$ are  omitted.
Computing the scattering amplitude requires fixing the normalization for the free scalar field described by Eq.\eqref{Eq:ScalarField}.
The temporal part of the current $j_\mu$ is the probability density while the spatial part is the probability current density
\begin{equation}
    j^\mu=(\rho,\vec{\mathbf{j}})\,.
    \end{equation}
As per convention, the field is normalized such that the integral of the density over a fixed volume is equal to the sum of the energies of all the scalar particles in the system.
Thus for the free scalar field one obtains
\begin{align}\label{Eq:Volume}
    \int_{V}\rho\, dV = & 2E\,, & \int_{V} \vec{j}\cdot d\vec{V} = & 2p\,.
\end{align}
From the modified KG equation we find the different components of the flux
\begin{align}\label{Eq:ProbabilityDencitiy}
  \nonumber  \rho=&-i\left[\left(\phi^\dagger \frac{\partial\phi}{\partial t}-\phi \frac{\partial\phi^\dagger}{\partial t}\right)+2\gamma \frac{\partial}{\partial t}\left(\phi^\dagger \frac{\partial^2\phi}{\partial t^2}-\phi \frac{\partial^2\phi^\dagger}{\partial t^2}\right)\right.\\
    &\left.-4\gamma \left(\frac{\partial\phi^\dagger}{\partial t} \frac{\partial^2\phi}{\partial t^2}-\frac{\partial\phi}{\partial t} \frac{\partial^2\phi^\dagger}{\partial t^2}\right)+4\gamma \left(\phi^\dagger\nabla^2 \frac{\partial\phi}{\partial t}-\phi\nabla^2 \frac{\partial^2\phi^\dagger}{\partial t^2}\right)\right]\,,\\
    \nonumber  \vec{j}=-&i\left[\left(\phi^\dagger \nabla\phi-\phi\nabla\phi^\dagger\right)+4\gamma \left(\frac{\partial\phi^\dagger}{\partial t}\nabla \frac{\partial\phi}{\partial t}-\frac{\partial\phi}{\partial t}\nabla \frac{\partial^\dagger\phi}{\partial t}\right)\right.\\
    &\left.-2\gamma \nabla\left(\phi^\dagger\nabla^2\phi-\phi\nabla^2\phi^\dagger\right)+4\gamma \left(\nabla\phi^\dagger\nabla^2 \phi-\nabla\phi\nabla^2 \phi\right)\right]\label{Eq:MomentumDensity}\,.
\end{align}
Using Eq.\eqref{Eq:ScalarField}, Eq.\eqref{Eq:MomentumDensity}, and  substituting in   Eq.\eqref{Eq:Volume} one arrives at the following normalization constant
\begin{equation}\label{Eq:Volumescale}
    N=\frac{1}{\sqrt{V\left(1-4\gamma (E^2+|\vec{p}|^2)\right)}}\,.
\end{equation}
Note that RGUP affects the normalization constant for the fields,  in addition to the corrections to the invariant amplitude $\mathfrak{M}$. An interesting thing to note is that despite working in the $x=x_0$ approximation, the fields $\phi$ behave as if they inhabit a volume scaled in a similar way to the position operator in \cite{Todorinov:2018arx}.

The scattering transition rate per unit volume is then given by 
\begin{equation}
    W_{fi}=\frac{|T_{ABCD}|^2}{\tau V}\,,
\end{equation}
where $\tau$ is the time of interaction and $T_{ABCD}$ is the scattering transition amplitude.
Substituting Eq.\eqref{Eq:CurrentsSUbbstituted} with Eq.\eqref{Eq:Volumescale} into the expression above results in 
\begin{equation}
    W_{fi}=(2\pi)^4\frac{\delta^{(4)}(p_D+p_C-p_A-p_B)|\mathfrak{M}|^2}{\mu_A \mu_B \mu_C \mu_D \,V^4}\,,
\end{equation}
where
\begin{equation}
 \mu_A=1-4\gamma (E_A^2+|\vec{p}_A|^2)\,, \label{Eq:Correction}
\end{equation}
and $\mu_B, \mu_C, \mu_D$ are similarly defined. 
Dividing the transition amplitude by the initial flux, and subsequently multiplying it by the number of final states in the volume, we get the cross section of the scattering process
\begin{equation}
    d\sigma=\frac{V^4}{|v_A|^2 2E_A 2E_B}\frac{\delta^{(4)}(p_D+p_C-p_A-p_B)|\mathfrak{M}|^2}{\mu_A \mu_B \mu_C \mu_D \,V^4}\frac{(2\pi)^4}{(2\pi)^6}\frac{d^3p_C}{2E_C}\frac{d^3p_D}{2E_D}\,,
\end{equation}
where $v_A=p_A/E_A$. One can rewrite the cross section  as 
\begin{equation}\label{Eq:DifferentialCrossSection}
    d\sigma=\frac{|\mathfrak{M}|^2}{F\,\mu_A \mu_B \mu_C \mu_D}dQ\,,
\end{equation}
where $F$ is the initial flux 
\begin{equation}\label{Eq:InitialFlux}
    F=|v_A|^2\, 2E_A \, 2E_B = 4\left(\left(p_A\cdot p_B\right)^2-m_A^2m_B^2\right)^{1/2}\,,
\end{equation}
and the Lorentz invariant phase space factor is 
\begin{equation}\label{Eq:AngleAndPhaseSpaceVolume}
    dQ=\delta^{(4)}(p_D+p_C-p_A-p_B)\frac{(2\pi)^4}{(2\pi)^6}\frac{d^3p_C}{2E_C}\frac{d^3p_D}{2E_D}.
\end{equation}

\subsubsection{Scalar electron-muon scattering}

The QG effects which this thesis is trying to model occur at energies close to the Planck energy. At such high energies the kinetic energy of the particles are much greater than the rest mass $|\vec{p}|\gg mc$.  Therefore, in order of simplifying the calculations the scattering is assumed to happen at ultra-relativistic energy scale. 

In the case of an electron-muon scattering, the conservation of momentum reads
\begin{equation}\label{Eq:CenterOfMassFrame}
    p_A+p_C=p_B+p_D=0\,.
\end{equation}
Notice that in this particular reference frame all the information is in the magnitudes of the initial and final momenta, $p_i$ and $p_f$, respectively, with
\begin{align}\label{Eq:CenterOfMassFrameAlt}
    |\vec{p}_A| = |\vec{p}_B| = & |\vec{p}_i|\,, & |\vec{p}_C| = |\vec{p}_D| = &|\vec{p}_f|\,.
\end{align}
In addition, for high energies the following approximation can be considered
\begin{equation}\label{Eq:UltrarelativisticApproximation}
    E^2\approx |\vec{p}|^2\,.
\end{equation}
Thus, for the correction terms $\mu$ in Eq.\eqref{Eq:Correction} one has 
\begin{align}
    \mu_A = \mu_B = & \mu_i, & \mu_C = \mu_D = & \mu_f\,.
\end{align}
Substituting them in Eq.\eqref{Eq:DifferentialCrossSection} gives for the differential cross section
\begin{equation}\label{Eq:CenterOfMassScatteringAmplitude}
d\sigma=\frac{|\mathfrak{M}|^2}{F\mu_i^2 \mu_f^2}dQ\,.
\end{equation}
Imposing conditions Eqs.(\ref{Eq:CenterOfMassFrame},\ref{Eq:CenterOfMassFrameAlt},\ref{Eq:UltrarelativisticApproximation}) on Eq.\eqref{Eq:AngleAndPhaseSpaceVolume}, the Lorentz invariant phase space factor $dQ$ can be expressed in terms of the solid angle $d\Omega$ as follows,
\begin{equation}\label{Eq:Formfactor}
    dQ=\frac{1}{4\pi^2}\frac{|\vec{p}_f|}{4\sqrt{s}}d\Omega\,,
\end{equation}
where $s=(E_A+E_B)^2$ is the Mandelstam variable.

Substituting Eq.\eqref{Eq:Formfactor} and Eq.\eqref{Eq:InitialFlux} in Eq.\eqref{Eq:CenterOfMassScatteringAmplitude} one gets an expression for the differential cross section in the center of mass frame
\begin{equation}
    \left.\frac{d\sigma}{d\Omega}\right\vert_{CM}=\frac{1}{64\pi^2\,s\,\mu_i^2 \mu_f^2}\frac{|\vec{p}_f|}{|\vec{p}_i|}|\mathfrak{M}|^2\,.
\end{equation}
Imposing conditions Eqs.(\ref{Eq:CenterOfMassFrame},\ref{Eq:CenterOfMassFrameAlt},\ref{Eq:UltrarelativisticApproximation}) on Eq.\eqref{Eq:InvariantAmplitude}  and expanding the scalar product, the invariant amplitude for high energy collision in the center of mass frame $\mathfrak{M}$ up to first order in $\gamma$ reads
\begin{equation}
    \mathfrak{M}=4\pi  \alpha \frac{3+\cos\theta}{1-\cos\theta}\left[1+8\gamma E^2(1-\cos\theta)\right]\,,
\end{equation}
where $\alpha$ is the fine structure constant.
\begin{figure}
 \centering
 \includegraphics[width=0.8\textwidth]{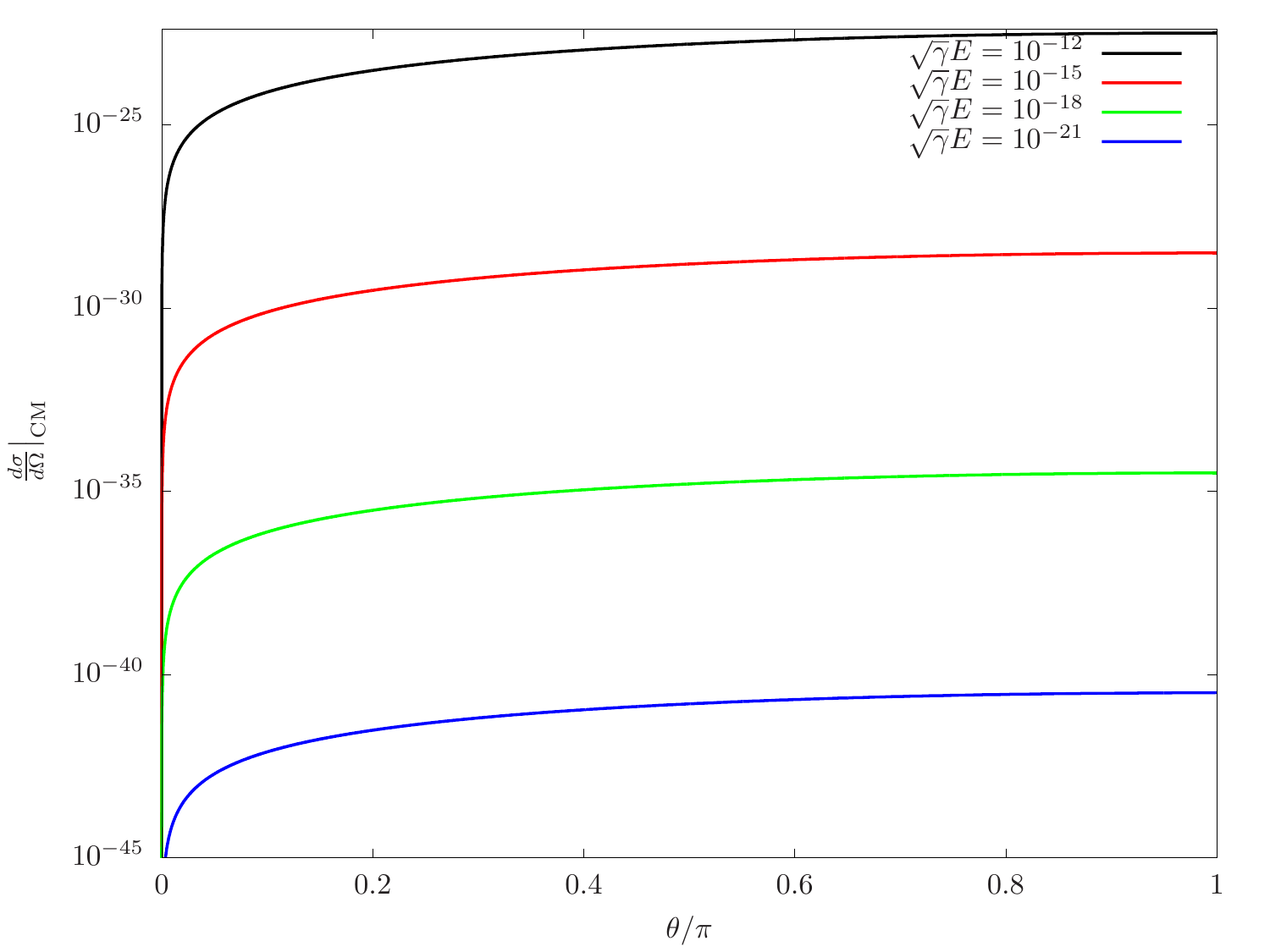}
\caption{Corrections to the differential cross section for different values of $\sqrt{\gamma} E$ with respect to the scattering angle $\theta$.
For the purposes of the plot, we consider the RGUP minimum length to be equal to the Planck length.}\label{fig2}
\end{figure}
Thus, the full expression for the differential cross section is 
\begin{equation}\label{Eq:FinalScalarDIferentialCrossSection}
    \left.\frac{d\sigma}{d\Omega}\right\vert_{CM} = \frac{1}{4\,s}\alpha^2\left(\frac{3+\cos\theta}{1-\cos\theta}\right)^2\left[1+16\gamma E^2(1-\cos\theta)\right]\,.
\end{equation}
Notice that in the limit $\gamma \rightarrow 0$, we get the standard result, namely
\begin{equation}
    \left.\frac{d\sigma}{d\Omega}\right\vert_{CM}^0 = \frac{1}{4\,s}\alpha^2\left(\frac{3+\cos\theta}{1-\cos\theta}\right)^2\,.
\end{equation}
Therefore, the magnitude of the correction is
\begin{equation}
    \frac{\left.d\sigma/d\Omega\right\vert_{CM} - \left.d\sigma/d\Omega\right\vert_{CM}^0}{\left.d\sigma/d\Omega\right\vert_{CM}^0} = 16\gamma E^2 (1 - \cos\theta).
\end{equation}
Finally, an important observation is that the correction is the largest for back scattering
($\theta=\pi$).
On Fig.\ref{fig2} one can see presented the RGUP correction term of Eq.\eqref{Eq:FinalScalarDIferentialCrossSection} for several different energies. It is  assumed that the RGUP corrections will be relevant at Planck energies, {\it i.e.} the RGUP minimum measurable length is equal to the Planck length. The $\sqrt{\gamma}  E=10^{-16}$ curve represents the magnitude or the corrections for energies used in current particle physics experiments.  The curves $\sqrt{\gamma} E=10^{-14}$ and $\sqrt{\gamma}  E=10^{-12}$ correspond respectively to hundred and ten thousand times higher energies. One can easily see that even a simplified model such as this will give corrections to the cross section of the electromagnetic scattering. Moreover, increasing the energy ten times will lead to a hundred times greater magnitude of the RGUP corrections. Another feature worth mentioning is the fact that with the lowering of the energy the correction term is very quickly suppressed, and the modified theory recovers all previous results.

 \subsection{\label{sec:scatteringcrosssecton}High energy electron-muon scattering}
  The calculated RGUP corrections to the scalar electron-muon scattering, are an effective toy model and a proof of concept. However, experiments such as the Stern–Gerlach experiment \cite{gerlach1922experimentelle} show that electrons have spin $1/2$. Furthermore, particles with spin $1/2$ such as the electron and muon obey the Pauli exclusion principle. An accurate model of the electromagnetic electron-muon scattering  needs to take that into account. In this subsection the reader can find the RGUP corrected amplitude of a spinor electron-muon scattering. The inclusion of spin requires the electron and muon wavefunctions to be solutions of the modified Dirac equation Eq.\eqref{Eq:DifferentialDiracEoM}, which introduces interesting challenges.
The calculations of the QG corrections to the QED scattering amplitude and cross section will again follow the procedure outlined in 
\cite{Halzen:1984mc}.

First, the transition amplitude can be read off
from the three particle interaction terms in the Lagrangian for the Dirac field Eq.\eqref{Eq:RGUPQED}. Again, one has
\begin{equation}
    T_{fi}=-i\int  A^\mu j_\mu\,d^4x\,, 
\end{equation}
where this time  $j_{\mu}$ is the current corresponding to the Dirac electrons and muons. For ease of 
calculations, one splits the transition amplitude for the three particle Feynman vertex into the usual term $ T_{fi}^{(1)}$ and two correction terms  $ T_{fi}^{(2)}$ and $ T_{fi}^{(3)}$ as follows
 \begin{subequations} \label{Eq:DiracAmplitudeSeparated}
\begin{align}
 T_{fi}^{(1)}&=i\int e\bar{\psi}_f \tau^\mu A_{\mu}\psi_i\, d^4 x\,,\\
 T_{fi}^{(2)}&=i\int2e\gamma\partial_{\rho}\bar{\psi}_f \tau^\rho A^{\mu}\partial_\mu\psi_i\, d^4 x\,,\\
 T_{fi}^{(3)}&=-i\int e\gamma\bar{\psi} \tau^\mu A_\mu\partial^{\rho}\partial_{\rho}\psi\, d^4 x\,.
\end{align}
\end{subequations}
From Eq.\eqref{Eq:DiracAmplitudeSeparated} presented above, the current and its corrections terms can be isolated. They take the following form
\begin{subequations} \label{Eq:CurrentsPsi}
\begin{align}
 j_{fi\,\mu}^{(1)}&=-e\bar{\psi}_f \tau^\mu\psi_i\,,\\
 j_{fi\,\mu}^{(2)}&= -2e\gamma\partial_{\mu}\bar{\psi}_f \tau^\mu \partial_\rho\psi_i\,,\\
 j_{fi\,\mu}^{(3)}&= e\gamma\bar{\psi}_f \tau^\mu \partial^{\rho}\partial_{\rho}\psi_i\,.
\end{align}
 \end{subequations}
In terms of the physical momentum $p_\mu$, the Dirac equation will have the form of Eq.\eqref{Eq:DifferentialDiracEoM}. One can recall that the modified Poincar\'e group has the  squared physical momentum as a Casimir invariant. This means that the  solutions of Eq.\eqref{Eq:DifferentialDiracEoM} need to obey the usual dispersion relation  and will be of the form shown in the appendix \ref{app:DiracSolutions}
\begin{equation}
 \psi(x)=u^{(s)}\left(p\right)e^{-ip\cdot x}\,,
 \end{equation}
 where $u^{(s)}\left(p\right)$ are four component complex spinors 
\begin{subequations}\label{freefieldsolutions}
\begin{align}
    u^{(s)}\left(p\right) = & N\begin{pmatrix}
        \chi^{(s)}\\
        \frac{\sigma \cdot p}{E+m}\chi^{(s)}
    \end{pmatrix}\,, & E > & 0\,,\\\nonumber\\
    u^{(s+2)}\left(p\right) = & N\begin{pmatrix}
        \frac{-\sigma \cdot p}{E+m}\chi^{(s+2)}\\
        \chi^{(s+2)}
    \end{pmatrix}\,, & E < & 0\,,
\end{align}
\end{subequations}
where $p$ is the physical four-momentum, $E$ is the zeroth component of that vector, $s\in\{1,2\}$, and $\chi^{(s)}$ are 
\begin{equation}
   \chi^{(1)}=\begin{pmatrix}
1\\
0
\end{pmatrix}, \qquad \qquad \chi^{(2)}=\begin{pmatrix}
0\\
1
\end{pmatrix}\,.
\end{equation}
Substituting the form of the field into Eq.\eqref{Eq:CurrentsPsi}, one gets
\begin{subequations} 
\label{Eq:CurrentsU}
\begin{align}
 j_{fi\,\mu}^{(1)}&=-e\bar{u}_f \tau^\mu u_i\,e^{i\left(p_f-p_i\right)\cdot x}\,,\\
 j_{fi\,\mu}^{(2)}&=-2e\gamma\bar{u}_f p_{f\,\rho}\tau^\rho p_{i\,\mu} u_i\,e^{i\left(p_f-p_i\right)\cdot x}\,,\\
 j_{fi\,\mu}^{(3)}&=e\gamma\bar{u}_f \tau^\mu p_{i\rho}p^{\rho}_i u_i\,e^{i\left(p_f-p_i\right)\cdot x}\,.
\end{align}
\end{subequations}
Using the currents, one can calculate the invariant amplitude for an electron-muon scattering presented in Fig. \ref{Fig:Muonscattering}. 

One can express the square of the invariant amplitude as
\begin{equation}\label{Eq:SpinorInvariantAmplitude}
    \left|\mathfrak{M}\right|^2=\frac{e^4}{-q^2(1-2\gamma q^2)}L_e^{\mu\nu}L^{muon}_{\mu\nu}\,,
\end{equation}
where the two fermions are scattered via the exchange of a gauge boson $A^\mu$ .
The tensors  $L_e^{\mu\nu}$ and $L^{muon}_{\mu\nu}$ can be expressed from Eqs.\eqref{Eq:CurrentsU} as follows 
\begin{equation}\label{Eq:TensorL}
    L_e^{\mu\nu}=\frac{1}{2}\sum_{spins}\left[j_{fi\,\mu}^{(1)} +j_{fi\,\mu}^{(2)} +j_{fi\,\mu}^{(3)}\right]\left[ j_{fi\,\nu}^{(1)} +j_{fi\,\nu}^{(2)} +j_{fi\,\nu}^{(3)}\right]^*\,,
\end{equation}
which, using Eq.\eqref{Eq:CurrentsU}, can be expressed as 
\begin{multline}\label{Eq:TensorAmpl}
    L_e^{\mu\nu}=\frac{1}{2}\sum_{spins}\left[-\bar{u}_f \tau^\mu u_i-2\gamma\bar{u}_f k_{f\,\rho}\tau^\rho k_i^\mu u_i+\gamma\bar{u}_f \tau^\mu k_{i\rho}k^{\rho}_i u_i\right]\\
    \times\left[-\bar{u}_f \tau^\nu u_i-2\gamma\bar{u}_f k_{f\,\rho}\tau^\rho k_i^\nu u_i+\gamma\bar{u}_f \tau^\nu k_{i\rho}k^{\rho}_i u_i\right]^*\,.
\end{multline}
The expression for $L^{muon}_{\mu\nu}$ is obtained analogously. 
Further expanding Eqs.\eqref{Eq:TensorL} and \eqref{Eq:TensorAmpl} and truncating to first order in $\gamma$,
one obtains
\begin{subequations}
\begin{multline}
     L_e^{\mu\nu}=\frac{1}{2}\sum_{spins}\left[(1-2\gamma k_\rho k^{\rho})\bar{u}(k') \tau^\mu u(k)\bar{u}(k) \tau^\nu u(k')\right.\\+2\gamma\bar{u}(k') \tau^\mu u(k)\bar{u}(k) k_{\rho}\tau^\rho k'^{\nu} u(k')\\\left.+2\gamma\bar{u}(k')k'_{\rho}\tau^\rho k^{\mu} u(k)\bar{u}(k) \tau^\nu u(k')\right]\,,
\end{multline}
 \begin{multline}
     L^{\text{muon}}_{\mu\nu}=\frac{1}{2}\sum_{spins}\left[(1-2\gamma p_\rho p^{\rho})\bar{u}(p') \tau_\mu u(p)\bar{u}(p) \tau_\nu u(p')\right.\\+2\gamma\bar{u}(p') \tau_\mu u(p)\bar{u}(p) p_{\rho}\tau^\rho p'_{\nu} u(p')\\\left.+2\gamma\bar{u}(p')p_{\rho}\tau^\rho p'_{\mu} u(p)\bar{u}(p) \tau_\nu u(p')\right]\,.
\end{multline}
\end{subequations}
Summing over the spins and taking the trace over the Dirac matrices, one obtains 
 \begin{subequations}\label{Eq:TracedTensors}
\begin{align}
\nonumber L_e^{\mu\nu}=2\left(1-2\gamma k\cdot k\right)&\left[k'^\mu k^\nu+k'^\nu k^\mu-\left(k'\cdot k-m_e^2\right)g^{\mu\nu}\right]\\&\qquad\qquad+8\gamma\left[\left(k\cdot k\right)k'^\mu k'^\nu+m_e^2 k'^{\mu}k^{\nu}\right]\,,\\
\nonumber   L^{\text{muon}}_{\mu\nu}=2\left(1-2\gamma p\cdot p\right)&\left[p'_\mu p_\nu+p'_\nu p_\mu-\left(p'\cdot p-m_{\text{muon}}^2\right)g_{\mu\nu}\right]\\&\qquad\qquad+8\gamma\left[\left(p\cdot p\right)p'_\mu p'_\nu+ m_{\text{muon}}^2 p'_{\mu}p_{\nu}\right]\,.
\end{align}
 \end{subequations}
Substituting the above Eqs.\eqref{Eq:TracedTensors} in Eq.\eqref{Eq:SpinorInvariantAmplitude}, one obtains the (squared) invariant amplitude including corrections up to linear order in $\gamma$
\begin{multline}
     \left\vert\mathfrak{M}\right\vert^2=  \frac{e^4}{\left(k-k'\right)^4}\\
     \left\{4\left[k'^\mu k^\nu+k'^\nu k^\mu-\left(k'\cdot k-m_e^2\right)g^{\mu\nu}\right]\left[p'_\mu p_\nu+p'_\nu p_\mu-\left(p'\cdot p-m_{\text{muon}}^2\right)g_{\mu\nu}\right]\right.\\\left.-8\gamma m_e^2\left[k'^\mu k^\nu+k'^\nu k^\mu-\left(k'\cdot k-m_e^2\right)g^{\mu\nu}\right]\left[p'_\mu p_\nu+p'_\nu p_\mu-\left(p'\cdot p-m_{\text{muon}}^2\right)g_{\mu\nu}\right]\right.\\\left.-8\gamma m_{\text{muon}}^2\left[k'^\mu k^\nu+k'^\nu k^\mu-\left(k'\cdot k-m_e^2\right)g^{\mu\nu}\right]\left[p'_\mu p_\nu+p'_\nu p_\mu-\left(p'\cdot p-m_{\text{muon}}^2\right)g_{\mu\nu}\right]\right.\\\left.+16\gamma m_e^2\left( k'^\mu k'^\nu+ k'^{\mu}k^{\nu}\right)\left[p'_\mu p_\nu+p'_\nu p_\mu-\left(p'\cdot p-m_{\text{muon}}^2\right)g_{\mu\nu}\right]\right.\\\left.+16\gamma m_{\text{muon}}^2 \left(p'_\mu p'_\nu+  p'_{\mu}p_{\nu}\right)\left[k'^\mu k^\nu+k'^\nu k^\mu-\left(k'\cdot k-m_e^2\right)g^{\mu\nu}\right]\right\}.
\end{multline}
The above expression is considered in the center of mass frame. Furthermore, the ultra-relativistic approximation is applied to the leading order. In other words the magnitude of the 3-momentum is considered to be much greater than the rest mass, that is $\vec{p}^{\,2}\gg m^2$ or $E^2\approx\vec{p}^{\,2}$.
Thus, terms proportional to the mass in the leading order can be ignored.
However, the terms proportional to the mass in the corrections must be kept. In the ultra-relativistic approximation the square of the invariant amplitude reads
\begin{multline}
       \left|\mathfrak{M}\right|^2= \frac{8\,e^4}{\left(k-k'\right)^4}\left\{ \left(k'\cdot p'\right)\left(k\cdot p\right)+\left(k'\cdot p\right)\left(k\cdot p'\right) \right.\\-2\gamma\left(m_e^2+m_{\text{muon}}^2\right)\left[\left(k'\cdot p'\right)\left(k\cdot p\right)+\left(k'\cdot p\right)\left(k\cdot p'\right)
       \right]\\
    + 2 \gamma m_e^2 \left[\left(k'\cdot p'\right) \left(2 k'\cdot p + k\cdot p\right)
    + \left(k'\cdot k\right) \left(k\cdot p' - k'\cdot p\right)
       \right]
       \\
    \left. + 2 \gamma m_{\text{muon}}^2 \left[ \left(k'\cdot p'\right) \left(2 p'\cdot k + k\cdot p\right)
    + \left(k'\cdot k\right) \left(k\cdot p' - k'\cdot p\right) 
       \right]\right\}\,.
\end{multline}
Furthermore, in terms of the Mandelstam variables in the $s$ channel
\begin{subequations}
\begin{align}
    s&=\left(k+k'\right)^2=\left(p+p'\right)^2\approx2k\cdot k'\approx 2p\cdot p'\,,\\
    t&=\left(k-p\right)^2=\left(p'-k'\right)^2\approx -2k\cdot p\approx -2k'\cdot p'\,,\\
    u&=\left(k-p'\right)^2=\left(p-k'\right)\approx -2k\cdot p'\approx -2p\cdot k'\,,
\end{align}
\end{subequations}
the differential cross section reads
\begin{equation}
 \left.\frac{d\sigma}{d\Omega}\right\vert_{CM}=\frac{1}{64\pi^2\,s}\left\vert\mathfrak{M}\right\vert^2=\frac{2e^4}{64\pi^2\,s}\left[\frac{t^2+u^2}{s^2}+ \frac{1}{2}\gamma(m_e^2+m_{\text{muon}}^2)\frac{tu-u^2}{s^2} \right]\,,
 \end{equation}
which in terms of the energy and the scattering angle becomes
\begin{equation}
\left. \frac{d\sigma}{d\Omega} \right\vert_{CM} = \frac{\alpha^2}{4\,s} \left[ \frac{1}{2} \left(1 + \cos^2\theta\right) + \frac{1}{4} \gamma (m_e^2 + m_{\text{muon}}^2) \left(\cos\theta+\cos^2\theta\right)\right]\,.
 \end{equation}
Thus, one obtains the full cross-section by integrating over the full solid angle
 \begin{multline}
    \oint \left. \frac{d\sigma}{d\Omega} \right\vert_{CM}d\Omega \\
    = \int_0^{2\pi} \int_0^{\pi} \frac{\alpha^2}{4\,s} \left[ \frac{1}{2} \left(1 + \cos^2\theta\right) + \frac{1}{4} \gamma (m_e^2 + m_{\text{muon}}^2) \left(\cos\theta+\cos^2\theta\right)\right]\sin\theta d\theta d\phi\,,
 \end{multline}
 which gives
 \begin{equation}\label{Eq:TotalCrossSection}
     \sigma\left(e^+e^-\rightarrow \mu^+\mu^-\right)=\frac{4\pi\alpha^2}{3\,s}\left[1+\frac{\gamma}{8}(m_e^2+m_{\text{muon}}^2)\right].
 \end{equation}
Finally, one can see that the total cross section is modified 
and the magnitude of the  correction depends only on the rest masses and fundamental constants.  
In fact, if one assumes that QG effects are relevant at Planckian energies, 
or equivalently if one assumes $\gamma_0=1$, the corrections for proton-proton scattering are about 
$\gamma m_p^2/4 \sim 10^{-38}$. 
The dependence of the corrections on the rest mass of the particles involved suggests that minimum length effects on scattering amplitudes may be measurable for electromagnetic scattering of heavier systems, such as scattering of heavy ions. The form of the RGUP corrections to the total cross section presented in Eq.\eqref{Eq:TotalCrossSection} will hold true when considering scattering of heavy ions with total spin of the nucleus equal to  $1/2$. For example for the Xe-Xe scattering observed in the ATLAS experiment, the order of magnitude of the corrections is $\gamma\, m_{\text{Xe}}^2 /4\sim 10^{-34}$ \cite{Balek:2019nqk}.

\section{Summary}
  In short, chapter \ref{Ch4:QFTWML} presents the methodology for the construction of an effective QFT with minimum measurable length. The author calculates the resulting scattering amplitudes and compares them to the results of current experiments. The introduction to the chapter motivates the need for RGUP modified QFT and fixes the particular RGUP model used in the calculations in the chapter. Worthy of note is the fact that the RGUP chosen does not have non-commutativity of spacetime. The decision to not consider this effect of minimum length is  due to the lack of consistent methodology to address the fact that the spacetime presented in Eq.\eqref{Eq:GUPxx} is non-commutative and it cannot be described y an algebra. This additional assumptions erode the generality of the model presented. However, the results obtained in this thesis would not have been possible without it. 
  
The square of the physical momentum $p_\mu p^\mu$ is still a Casimir invariant of the Poincar\'e group of the chosen representation. From the Einstein dispersion relation the author to reads of a modified differential form of the KG  and Dirac equations. One then utilizes the Ostrogradsky method to recover the scalar and spinor field Lagrangians  from their equations of motion, {\it i.e.} the modified KG and Dirac equations. The calculations and results are outlined in  sections \ref{sec:RGUPLagrangians}. Furthermore, as in  the usual case, there is a unique up to a total derivative  correspondence between a Lagrangian and its equation of motion is unique. The resulting Lagrangians contain higher than second derivative, this might cause problems due to the Ostrogradsky instability. This problem is explored in appendix \ref{app:Ostr} where the author shows that for systems with energies smaller than Planck energy $E<E_{\text{Pl}}$ all the Lagrangians are well defined and do not contain malicious ghosts. The Ostrogradsky instability is pushed back and shows up when the energy of the system approaches the Planck energy.

The modified gauge field Lagrangian is  constructed by a slightly different method. To accomplish this, a condition is  imposed on the gauge fields. The modified electromagnetic or gauge field must have the same symmetries as the unmodified case. In other words the Lagrangian must be $U(1)$ gauge invariant. The gauge field is a boson field and has spin equal to one. Therefore the gauge field is  required to obey a KG-like equation of motion. The first condition ensures the phase symmetry of the photon; and the second ensures that the gauge field transforms under the Poincar\'e group and therefore obeys the spacetime symmetries of SR. Then, a modified version of the $F_{\mu\nu}$
field strength tensor and by extension the Lagrangian for the gauge field is  obtained and is presented in Eq.\eqref{Eq:GaugeFieldL}.

In Section \ref{Sec:FeynmanRules}, the scalar and spinor field Lagrangians are  coupled to the gauge field through the standard use of minimal coupling. The propagators and vertices for both of the cases are  then calculated from the coupled Lagrangian. A result worth of note is the fact that there are up to six particle interactions allowed for the scalar field, and up to five for the Dirac field. Additionally in all the vertices only two of the participation particles are the matter fields. Therefore, an annihilation of an electron and antielectron has a small probability of giving off more than two photons.  

In the following section \ref{sec:scatteringcrosssecton}, the transition amplitudes and their RGUP corrections are  calculated for the three-point vertices. This allowed for the calculations of the RGUP corrected scattering amplitudes for electron-muon scattering for both cases yielding a number of interesting results. First, it is  shown that the RGUP corrections rise in magnitude with the energy of the system squared, which gives hope that with the improvement of current experiments one may be able to observe the effects. Furthermore, the magnitude of the corrections is shown to be dependent on the angle and is the highest for the backscatter. Finally, the results are  compared to the Xe-Xe scattering observed in ATLAS and the bounds on the minimum length scale are  put to be  $\gamma_0\leq 10^{34}$, which is seven orders of magnitude better than the bounds shown in chapter \ref{Ch3:RGUP}.

\chapter{Conclusion}\label{Ch5:Conclusion}

\begin{quote}
	“I would rather have questions that can't be answered than answers that can't be questioned.”
	
	\begin{flushright}
		 ― Richard Feynman
	\end{flushright}
\end{quote}
 As evident from Chapter \ref{Ch1:Introduction} of this thesis, QM and GR are comprehensive theories, which have proven themselves in countless experiments and discoveries. However, it is also clear that neither QM or GR is a complete and all encompassing theory. Evidence for this claim is the fact that both fail to explain phenomena belonging to the sphere of the other.  
 
For example GR in most cases fails to account for the quantum properties of matter, while the Standard Model of elementary particles excludes the gravitational force altogether. These facts allow one to easily conclude that a new theory is needed. Such a theory must encompass all four interactions and  matter on both classical and quantum level. It is widely believed that the main obstacle in the path of the formulation of the unified theory is the quantization of the gravitational field. This task has resulted in a collection of theories known as Quantum Gravity theories.

The most promising members of that family are LQG and  ST are all reviewed in Chapter \ref{Ch1:Introduction}. Needless to say the theories presented by the author are not the only ones, noticeable examples not mentioned are noncommutative gravity \cite{Harikumar:2006xf} and conformal gravity \cite{Mannheim:2007ki}. Unfortunately, none of these theories have provided any experimentally supported predictions. Although they are extremely well constructed mathematically and have robust collections of ideas, this lack of experimental data to support their results takes them into the realm of conjectures and hypotheses rather than full fledged theories. The problems associated with QG make the field all that more interesting to work on. 

For all its problems QG has one robust prediction. That is the fact that all theories attempting to quantize gravity predict the existence of a minimum measurable length or minimum uncertainty in position. How the three theories mentioned above arrive at the conclusion that such length exists is also described in the introduction of this thesis. A feature worthy of attention is the fact that regardless of initial assumptions and postulates, one always arrives at some form of minimum measurable length. The diversity of the initial assumptions and postulates giving minimum uncertainty in position stands as an evidence for its robustness of this result. 

Thus the field of QG Phenomenology enters the scene here. The field of QGP works to reconcile the QG theories such as ST and LQG with currently existing experiments. The main goal of QGP is to find evidence of low energy remnants of QG effects in already existing experiments and data. Such experimental evidence will provide much needed experimental insight into the quantum nature of gravity, and by extension the properties of spacetime itself.  
A more general overview of approaches to QGP are outlined in chapter \ref{Ch2:QuantmGravityPhenomenology}. Chapter two also provides a motivation for the choice of topic for this thesis. The phenomenological approach to QG presente in this thesis is the extension of the Heisenberg Uncertainty Principle to the so called
Generalized Uncertainty Principle. Such a generalization is made to accommodate the existence of minimum measurable length, suggested by QG theories. However,  most of the work done in this area considers non-relativistic systems and theories. This raises the following problems. First, the minimum length obtained in such a way is not Lorentz invariant or in other words is frame-dependant. Which means that the minimum length one may be able to measure would depends on the frame of reference of the observer. This raises the question: in which frame of reference is this a true minimum length? This violates the Equivalence principle and effectively re-introduces the concept of aether! Additionally, non-relativistic GUP gives rise to a modified dispersion relation, containing higher than second order of the momentum operator. As discussed in appendix \ref{app:CLP}, modified dispersion relation of this type leads to non-linear composition of momentum and energy when considering a classically bound system. This non-linearity is known as the Composition Law Problem. One needs to address these problems before attempting to apply GUP to relativistic systems. As for the question as to why one needs to apply GUP at high energies, the answer is quite simply that the minimum length proposed by the theories of QG is usually proportional to the Planck length $l_{\text{Pl}}$. This translates  in terms of energy to $E_{\text{Pl}}=M_{\text{Pl}}c^2\sim 10^{19}\text{GeV}$. In order to have a best chance of detecting low energy remnants of QG effects one must maximize the following ratio 
\begin{equation}
   \eta= \frac{E}{E_{\text{Pl}}}\,,
\end{equation}
where $E$ is the energy scale of the experiment. Therefore, more energetic experiments provide greater chance of detection of residual QG phenomena, simply due to the fact that the corrections themselves have greater magnitude.

\section{Relativistic Generalized Uncertainty Principle} 
    The relativistic extension of GUP, developed by the author of this thesis, is presented in Chapter \ref{Ch3:RGUP}. 
The chapter is inspired by the work of Quesne et. al. \cite{Quesne2006} and is based on the calculations presented and published in \cite{Todorinov:2018arx}.

As evident from the usual non-relativistic form of GUP Eq.\eqref{Eq:GeneralGUP}, the accommodation of minimum uncertainty in position requires the position $x_\mu$ and momentum $p_\mu$ operators to no longer be canonically conjugate. However, a pair of canonically conjugate auxiliary variables can be found from assuming that the physical position $x_\mu$ and momentum $p_\mu$ can be expressed as functions of the auxiliary variables. Since the main focus of this work is phenomenological study of QG, one can use the Taylor expansion to perturbatively expand both functions and truncate to second order in the minimum length parameter $\gamma$. Then the the following requirements are imposed on spacetime and momentum space: homogeneity and isotropy on spacetime; and commutativity on momentum space. This allows the position and momentum operators to be written as 
\begin{subequations}
\begin{align}
x^{\mu}&=x_0^{\mu}-\kappa\gamma p_0^{\rho}p_{0\rho}x_0^{\mu}+\beta\gamma p_0^{\mu}p_0^{\rho}x_{0\rho}+\xi\hbar\gamma p_0^{\mu}, \\
p^{\mu}&=p_0^\mu\,(1+\varepsilon\gamma p_0^{\rho}p_{0\rho})\,,
\end{align} 
\end{subequations}
which makes the RGUP commutator take the form presented in Eq.\eqref{Eq:GUPxp}:
\begin{equation}
[x^{\mu},p^{\nu}]=i\hbar\,\left(1+(\varepsilon-\kappa)\gamma p^{\rho}p_{\rho}\right)\eta^{\mu\nu}+i\,\hbar(\beta+2\varepsilon)\gamma p^{\mu}p^{\nu}\,. 
\end{equation}
The minimum measurable length provided by the RGUP is frame independent or Lorentz invariant. One can observe an interesting result, namely that at high energies spacetime is non-commutative. In particular the physical position operators give commutator Eq.\eqref{Eq:GUPxx}
\begin{equation}
    [x^{\mu},x^{\nu}]=
i\hbar\gamma\frac{-2\kappa+\beta}{1+(\varepsilon-\kappa)\gamma p^{\rho}p_{\rho}}\left(x^{\mu}p^{\nu}-x^{\nu}p^{\mu}\right)\,.
\end{equation}
 More important is the fact that  the commutator presented in Eq.\eqref{Eq:GUPxx} does not meet the requirements for algebraic closure.  A closed algebra is the one for which the commutator between two generators is either zero for abelian groups, or belongs to the algebra itself. An example of such algebras are the algebras corresponding to Lie groups or in other words describing infinitely smooth and continuous manifolds. This suggests that spacetime is no longer a Riemannian manifold. 

This fundamental change of the properties of spacetime called for an examination of its symmetries. The approach chosen is the study of the Poincar\'e group and in particular its corresponding algebra. The generators of the Poincar\'e group are the translation operators represented by the physical momentum $p_\mu$ and the RGUP modified Lorentz generators 
\begin{equation}
    M^{\mu\nu} = p^{\mu}x^{\nu}-p^{\nu}x^{\mu}
    = \left[1+(\varepsilon-\kappa)\gamma p^{\rho}p_{\rho}\right]\tilde{M}^{\mu\nu}\,,
\end{equation}
where $\tilde{M}^{\mu\nu}$ are the unmodified ones. Their commutators form the algebra itself, the modified algebra found is presented in Eq.\eqref{Eq:PoincareAlgebra}:
\begin{subequations}
\begin{align}
  [x^\mu,M^{\nu\rho}] &=  i\hbar[1 + (\varepsilon - \kappa) \gamma p^{\rho} p_{\,\rho}]\left(x^{\nu}\delta^{\mu\rho}-x^{\rho}\delta^{\mu\nu}\right) + i\hbar 2 (\varepsilon - \kappa) \gamma p^{\mu} M^{\nu\rho}\,,\\
   [p^\mu, M^{\nu\rho}]& =  i\hbar[1 + (\varepsilon - \kappa) \gamma p^{\rho} p_{\,\rho}]\left(p^{\nu}\delta^{\mu\rho}-p^{\rho}\delta^{\mu\nu}\right)\,,\\
 [M^{\mu\nu},M^{\rho\sigma}]& = i\hbar\left(1 + (\varepsilon - \kappa) \gamma p^{\rho} p_{\,\rho}\right)\left(\eta^{\mu\rho}M^{\nu\sigma}-\eta^{\mu\sigma} M^{\nu\rho} 
  -\eta^{\nu\rho}M^{\mu\sigma}+\eta^{\nu\sigma}M^{\mu\rho}\right)\,.
\end{align}
\end{subequations}
Writing the algebra in this way reveals a relationship between the parameters used to fix the model, namely $\varepsilon = \kappa$, for which the Poincar\'e algebra remains unmodified. This relation is represented by a line in parameter space. The RGUP commutator Eq.\eqref{Eq:RGUPk=e} and the spacetime algebra Eq.\eqref{Eq:STNCk=e} on this line take the following form 
\begin{align}
[x^{\mu},p^{\nu}]&=i\hbar\,\left(\eta^{\mu\nu}+2\kappa\gamma p^{\mu}p^{\nu}\right)\,, \\
[x^{\mu},x^{\nu}]&=-
2i\hbar\kappa\gamma\left(x^{\mu}p^{\nu}-x^{\nu}p^{\mu}\right)\,.
\end{align}
As opposed to a noncommutative Landau atom \cite{mamat2016landau}, the position-position commutator for the RGUP presented her is not a constant and it depends on the momentum. 

The conclusion can be drawn then that the existence of a Lorentz invariant minimum measurable length does not require a modification of the spacetime symmetries. The only requirement is that spacetime is a non-commutative manifold.

Another important observation is made when one calculates the four Casimir operators in the modified Poincar\'e algebra which commutate with every other operator in the set. These operators are the physical and auxiliary momenta squared $p_\mu p^\mu$ and $p_{0\mu} p^{0\,\mu}$. In addition to the squares of the physical and auxiliary  Pauli–Lubanski pseudovector.  
The physical momentum squared corresponds to the Einstein dispersion relation
\begin{equation}
    p_\mu p^\mu=-(mc)^2\,.
\end{equation}
Due to the quadratic nature of the Einstein dispersion relations of the RGUP modified Poincar\'e algebra, the composition law problem is avoided, and therefore energies and momenta sum up linearly as they should.

Phenomenological studies, however, suggest the calculation of RGUP corrections to the energy levels suitable physical systems. The Einstein dispersion relation is used to derive a RGUP-modified quantum mechanical wave equation, after which, RGUP corrections to the energies levels of several different well known relativistic QM systems are obtained. Comparison is drawn between the relative magnitude of the corrections, and the accuracy of current physical experiments. 
I applied them to
several existing experiments and estimated the upper bounds on the GUP parameter. A bound on the RGUP parameter is then found, namely
\begin{equation}
\kappa\leq
10^{41}.
\end{equation} 
While this result might not look physical, it is ten orders of magnitude lower than previous bounds on the minimum length scale obtained using similar experiments. Furthermore, the RGUP proposed  in chapter \ref{Ch3:RGUP} is a crucial step in the development of QFT with minimum measurable length. RGUP corrections to scattering amplitudes usually calculated by QFT will allow the use of high energy experiments in phenomenological studies of minimum length, potentially narrowing the parameter space even further.   

\section{Quantum Field Theory with minimum measurable length}
    There are previous attempts made to formulate QFT with minimum length.  However, very few of them employ any form of relativistic GUP or Lorentz invariant minimum length \cite{Szabo:2006wx,Chaichian:2004za,snyder1947quantized,Quesne:2006is,Faizal:2017map,Faizal:2014dua,Pramanik:2014zfa,Deriglazov:2014yta,Pramanik:2013zy,Husain:2013zda,Kober:2010sj,Hossenfelder:2006cw,Capozziello:1999wx,zakrzewski1994quantum,Kober:2011dn}. Furthermore, all of them posses a modified dispersion relation and therefore have the composition law problem. 
    
A paper that uses RGUP similar in form to the one derived in \cite{Todorinov:2018arx} is \cite{Kober:2011dn}. Although similar, the calculations presented in  chapter \ref{Ch4:QFTWML}
make fewer assumptions and take the results further by calculating transition amplitudes and RGUP corrected differential and total cross-sections for charged scalar and vector fields. At the time of writing this thesis, \cite{Todorinov:2018arx} is one of very few results in this area.

In chapter \ref{Ch4:QFTWML}, an effective QFT is derived from a small set of assumptions, namely that the physical position $x_\mu$ and momentum $p_\mu$ have the following form in terms of the auxiliary ones
 \begin{subequations}
\begin{align}
x^{\mu}&=x_0^{\mu}\,, \\
p^{\mu}&=p_0^\mu\,(1+\gamma p_0^{\rho}p_{0\rho})\,.
\end{align} 
\end{subequations}
The Poincar\'e group for the particular choice of RGUP is calculated. It was shown that the momentum squared $p_\mu p^\mu$ is still a Casimir invariant, thus avoiding the composition law problem. Utilizing the line of reasoning showed in chapter \ref{Ch3:RGUP}  RGUP modified form of the Klein-Gordon and Dirac equations are derived. 

In the usual formulation of QFT, 
the standard Klein-Gordon and Dirac equations are equations of motion of the scalar and vector fields respectively. They are obtained from the Lagrangian or the Lagrangian density using the Euler-Lagrange equations
\begin{subequations}
\begin{align}
   \frac{\partial L}{\partial q_{i}}(t,{\boldsymbol{q}}(t),{\dot {\boldsymbol {q}}}(t))-{\frac {\mathrm {d} }{\mathrm {d} t}}{\frac {\partial L}{\partial {\dot {q}}_{i}}}(t,{\boldsymbol {q}}(t),{\dot {\boldsymbol {q}}}(t))=0\,,\\
    \frac{\partial{\mathcal{L}}}{\partial\varphi}\partial _{\mu}\left({\frac {\partial {\mathcal {L}}}{\partial(\partial _{\mu }\varphi)}}\right)=0\,.
\end{align}
\end{subequations}
However, as evident from Eqs.\eqref{Eq:ModKGDiff} and \eqref{Eq:DifferentialDiracEoM}, the equations of motion of a potential field theory with minimum length contain higher than second derivatives. Therefore the action and subsequently the Lagrangian and its density need to have higher derivative terms. The methodology for working with higher derivative field theories is given by the Ostrogradky method \cite{Pons:1988tj,Woodard:2015zca,deUrries:1998obu}. The Euler-Lagrange equations are generalized to work with higher derivative Lagrangians and have the form
\begin{subequations}
\begin{align}
    &\frac{dL}{dq} -\frac{d}{dt}\frac{dL}{d\dot{q}}+\frac{d^2}{dt^2}\frac{dL}{d\ddot{q}}+\ldots+(-1)^n\frac{d^n}{dt^n}\frac{dL}{d (d^nq/dt^n)}=0\,,\\
   & \frac{\partial\mathcal{L}}{\partial\phi}-  \partial_\mu  \frac{\partial\mathcal{L}}{\partial(\partial_\mu\phi)}+   \partial_{\mu_1} \partial_{\mu_2}\frac{\partial\mathcal{L}}{\partial(\partial_{\mu_1} \partial_{\mu_2}\phi)}+\ldots
    +(-1)^m\partial_{\mu_1}\ldots\partial_{\mu_m}\frac{\partial\mathcal{L}}{\partial(\partial_{\mu_1} \ldots\partial_{\mu_m}\phi)}=0\,.
\end{align}
\end{subequations}
Through the use of the Ostrogradsky method, the RGUP modified Lagrangians are derived. The Lagrangians describing the dynamics of both the scalar and spinor fields  
are found to be 
\begin{multline}
    \mathcal{L}_{\phi,\mathbb{C}} = \frac{1}{2} \left(\partial_{\mu}\phi\right)^\dagger \partial^{\mu}\phi - \frac{1}{2} m^2 \phi^\dagger \phi + \gamma \left[\left(\partial_\nu \partial^\nu \partial^\mu \phi \right)^\dagger \partial_\mu \phi + \partial_\nu \partial^\nu \partial^\mu \phi \left(\partial_\mu \phi \right)^\dagger \right]\\
+ \frac{\gamma^2}{2} \left[\left(\partial_\mu \phi \right)^\dagger \partial^\mu \partial_\nu \partial^\nu \partial_\rho \partial^\rho \phi + \partial_\mu \phi \left(\partial^\mu \partial_\nu \partial^\nu \partial_\rho \partial^\rho \phi \right)^\dagger \right],
\end{multline}
\begin{equation}
  \mathcal{L}_{\psi}=\bar\psi\left[ i\tau^\mu\partial_\mu(1-\gamma\,\partial_\rho\partial^\rho) -m\right]\psi\,.
\end{equation}
The gauge field Lagrangian is obtained by slightly different means. A requirement is placed on the gauge field, namely the gauge $U(1)$ symmetry must be preserved. With this requirement in place, one can show that the representation of the RGUP modified Poincar\'e algebra, similarly to the unmodified one, is locally isomorphic to  $SU(2)\otimes SU(2)$. According to this isomophism, there exists an irreducible representation of spin $\left(\frac{1}{2},\frac{1}{2}\right)$ under which the gauge field is classified. Therefore, the equation of motion for the gauge field must be modified similarly to Eq.\eqref{Eq:ModKGDiff}, and has the form presented in Eq.\eqref{Eq:GaugeFieldEOM} in Lorentz gauge is
\begin{equation}
 \partial_\mu F^{\mu\nu}= \partial_\mu\partial^\mu A^\nu +2\gamma \partial_\mu\partial^\mu\partial_\rho\partial^\rho A^\nu  
     =0\,.
\end{equation}
Thus, a minimum length modified field strength tensor is obtained  in Eq.\eqref{Eq:EMFieldTensor} and has the equation of motion seen in Eq.\eqref{Eq:FieldTensorEoM} as 
\begin{equation}
 \partial_\mu F^{\mu\nu}=\partial_\mu F^{\mu\nu}_0+2\gamma  \partial_\rho \partial^\rho\partial_\mu F^{\mu\nu}_0
    +\gamma^2\partial_\sigma \partial^\sigma\partial_\rho \partial^\rho\partial_\mu F^{\mu\nu}_0\,
    \,.
\end{equation}
The gauge field Lagrangian then has the usual form as in Yang-Mills theories, with a RGUP modified field tensor. One important result is the fact that if one stays away from the Planck regime, {\it i.e.} $E<E_{\text{Pl}}$,  the effective Lagrangians found in this work do not experience the Ostrogradsky instability, and therefore do not contain malicious ghosts. This is discussed in Appendix \ref{app:Ostr}.

The Feynman rules for both the scalar and spinor fields are derived. Since the physical momentum is the operator of translations for our theory, the Feynman propagators need to be Green's functions of the modified differential forms of the Klein-Gordon Eq.\eqref{Eq:ModKGDiff} and Dirac Eq.\eqref{Eq:DifferentialDiracEoM} equations. The resulting propagators for the complex scalar, spinor, and gauge fields are presented in Eqs.\eqref{Eq:ScalarPropagator}, \eqref{Eq:DiracPropagator}, and \eqref{Eq:GaugePropagator}. The rules and the coupling constants for the vertices are calculated by coupling the matter fields to the gauge ones through standard minimal coupling and expanding and reading off the vertices from the different terms in the coupled Lagrangian density. An interesting although expected result is the fact that higher vertices are allowed. As one can see for the complex scalar field, there are up to six-point vertices,  which contain two matter fields and up to four gauge bosons in the interaction. The coupling constants are presented in Table \ref{Tbl:CouplingConstantTable}. Similarly for the Dirac field, up to five-point vertices are allowed, containing two spinors fields and up to three gauge bosons in a single interaction vertex. Additionally, due to the modifications to the propagator of the gauge field there is a small self interaction between the photons.

Then, following the standard procedure outlined in \cite{Halzen:1984mc},
the transition amplitudes, differential and total cross-sections are calculated for ultra-relativistic scalar electron-muon scattering and more importantly electron-muon scattering. The results are presented in Eqs.\eqref{Eq:FinalScalarDIferentialCrossSection} and \eqref{Eq:TotalCrossSection}, where one can see the minimum length effects on the cross-sections of these processes. Interestingly, one can see from Eq.\eqref{Eq:TotalCrossSection} in the ultra-relativistic limit, that the only relevant correction  depends on the rest mass and the fundamental constants of nature. The fact that the RGUP corrections to the total cross-section are proportional to the rest mass suggests that the minimum length effects on scattering amplitudes might be measurable using heavy ions scattering experiments such as ATLAS \cite{Balek:2019nqk}. Through the use of Xe-Xe scattering upper bounds on the RGUP parameter are set, i.e. $\gamma_0<10^{34}$. The bounds obtained from the cross-sections are seven orders of magnitude lower than the one found exploring relativistic QM. It showns that further study is needed. 

In conclusion a gauge invariant QG modified abelian effective QFT  describing both fermionic and bosonic fields is obtained. This effective field theory accommodates the existence of Lorentz-invariant minimum measurable length. Moreover, its predictions may be measurable in current or future high-energy experiments.  
This allows the derived RGUP modified cross-sections to be tested against real data. 

\section{Future Research}
    The results presented in the thesis so far provide a robust basis for a multitude of interesting research topics. The most promising and interesting of these are described below.  

\subsection{Applications to Classical Electrodynamics}
    
The Hamiltonian of RGUP modified  Electrodynamics 
 is derived using the Ostrogradsky formalism in appendix \ref{app:Ostr}. 
In addition, the modified field tensor presented in Eq.\eqref{Eq:FieldTensorEoM} can be used for the derivation of the modified expression of Maxwell's field equations. 
It is reasonable to expect that the RGUP modified Maxwell's field equations will be higher derivative differential equations of the form
\begin{align}
\nabla\left(1+f(\gamma,\Box)\right) \cdot \mathbf{E} &=0\,\\
\nabla\left(1+f(\gamma,\Box)\right) \times \mathbf {E} &=-{\frac {\partial  }{\partial t}}\left(1+f(\gamma,\Box)\right)\mathbf{B},\\
\nabla\left(1+f(\gamma,\Box)\right) \cdot \mathbf {B} &=0\,\\
\nabla\left(1+f(\gamma,\Box)\right) \times \mathbf{B} &=\mu_{0}\varepsilon _{0}{\frac {\partial  }{\partial t}}\left(1+f(\gamma,\Box)\right)\mathbf {E},
\end{align}
thus having additional solutions proportional to $\gamma=\gamma_0/M_{\text{Pl}}$.  
As one knows, Maxwell's equations are the foundation used for the theoretical description of every electromagnetically interacting system, as well as the properties of light. Therefore, this will allow us to calculate RGUP corrections and effects for classical systems, such as diffraction, interference, non-linear optics, femtosecond optics, light vertices, and many others.

\subsection{Photon self-interaction and Cosmology}
    Additionally, from the propagator shown in Eq.\eqref{Eq:GaugePropagator}, one can conclude that there is a small self-interaction terms between the RGUP modified photons. An interesting study is the application of this result to effects dealing with propagation of light, for example the Faraday rotation. Research and results in this direction will allow for a search of Quantum Gravity effects using experiments measuring the polarization map of Cosmic Microwave Background radiation, in which one can find structures which are not well described by modern physical theories. 
    
One preliminary result in that direction can be made using an analogy with magnets. Magnets have domains in them due to the fact that the atoms in the crystalline lattice interact magnetically. Similarly, even a small self-interaction between the CMB photons amplified by the  distance they have traveled can produce similar structures. Analogously, if measurements of the polarization of light received by Supernovae reveals that there is an anisotropy in its distribution, then a conclusion for the magnitude of the self-interaction of photons, can be drawn. In this particular case, there are two separate mechanisms amplifying QG effects: distance traveled; and the high energy of the photons. The first step in this avenue of research would be to use the RGUP to estimate modified Faraday rotation to calculate the effective radius of the self-interaction between photons, followed by a simulation of  CMB photons starting with a uniform polarization distribution, propagating through the Universe. The final step would be to  observe the resulting structures and compare their sizes with existing experimental data collected from experiments such as BICEP2/Keck Array, BICEP1, QUIET, and  CAPMAP.

\subsection{Higgs mechanism and boson mass}
    
As mentioned before, gauge invariant Quantum Gravity modified abelian  effective Quantum Field Theory is formulated in this thesis. 
However, this is the simplest case of gauge theory, \textit{i.e.} $U(1)$ gauge field. To have a complete modified expression of the SM one has to go through similar line of calculations  generalized to non-abelian gauge theories such as $SU(2)$ and $SU(3)$, representing the Weak and Strong interaction respectively. An effective Lagrangian for a massive abelian scalar field is presented here providing a possible insight in the RGUP modified dynamics of the Higgs boson. However, the generalization of the formalism to $SU(2)$ and $SU(3)$ gauge symmetries will allow the calculation of QG corrections to the Higgs mechanism and particularly the masses of all the particle in the SM. This research will be conducted through modifying the already existing quantization procedures to accommodate the existence of minimum measurable length.

Additionally, it would be interesting to compare and contrast the relation GR-QG with that between Fermi’s theory of the weak interaction and the weak interactions in the SM. The Fermi coupling constant is dimensionful, signalling (historically) that something should happen around 100 GeV in energy. Now  it is known that  this is roughly the mass of the W and Z bosons, exchanged in the SM description of the weak force. The dimensionful Fermi coupling constant is an approximation of the dimensionless SM weak coupling constant times a W or Z propagator.

\subsection{Applications to gravitational waves} 
     
Historically, symmetries and properties of spacetime have been explored from Galilean Relativity to Special Relativity to General Relativity. We have extended GUP from the Galilean Relativity to Special Relativity, \textit{i.e.} from space to space-time, moreover keeping all its symmetries intact. Therefore, the next natural step is to take the formulation of RGUP on a curved background. This will allow for QG effects to be tested for high curvature regimes, possibly leading to corrections for effects such as  Hawking radiation. 
Furthermore, the Lagrangians are obtained from the Equations of motion using the methodology of classical field theory, one can apply the same line of reasoning to  phenomena described by gravitational waves in GR. Tt is reasonable to expect a modification of the dispersion relation of gravity waves. Since the methodology used by LIGO and similar experiments for the detection of Gravity Waves is heavily reliant on the physical models used for the modeling of black hole merger, one can use the Signal to Noise ratio of the usual model compared to the QG modified model to put bounds on the parameters determining the scale of the minimum length. The great accuracy of experiments such as LIGO promise to greatly improve our current limits on the scales of QG effects \cite{Dapor:2020jvc,Mirshekari:2011yq,BOSSO2018498}.

\subsection{Topological defects}
    The Lagrangians obtained here can be used in cosmology. A research in this direction is to be conducted through the  search for topological defects in the ground state of our theory. This will be conducted through the use of the Bogomol'nyi-Prasad-Sommerfeld (BPS) bound to recover the energy functional of the theory. This will be followed by the search for its global minimum, after which one will be able to draw conclusions for the topology of that ground state. Particularly interesting are the  time independent defects in the ground state, such as domain walls, solitons, and vertices. These defects are  interesting  due to the fact that they are stable in time. Therefore, their existence can have low energy remnants observable in the large structures in the Universe \cite{tedesco2011fine,wang2016solutions,Faizal:2017map}. 

\section{Closing remarks}
    
There are two fundamental theoretical frameworks describing nearly every physical phenomenon in the universe: Quantum Mechanics and General Relativity. These theories began by modifying the status quo and end up fundamentally changing the way the Universe is perceived.
An example of this is the end of Classical Mechanics and the rise of Quantum Physics, marked by the discovery of the Heisenberg's Uncertainty Principle. The fall of Newtonian gravity and the rise of General Relativity is sparked by introducing a finite maximum velocity {\it i.e.} the speed of light $c$. The state of knowledge of the physical world is on the verge of a similar situation now. 

Theories of QG strongly suggest that the properties of spacetime and matter as observed by current experiments do not provide the full picture. Among the problems and disagreements between the different theories of QG, one result shines through and is uniformly agreed upon. This is the existence of minimum measurable length. One now sees the parallels between the leap of understanding of the Universe, which occurred at the fall of Galilean relativity and today.

An important element missing in the picture is experimental evidence, which supports any of the theories tackling the very difficult task of QG. As physics experiments get bigger and bigger, the field should turn its attention to Quantum Gravity Phenomenology. 
QGP has the difficult task to find and test already existing data and effects originating in much higher energy scale, or in other words smaller length scales. 

One idea that may close the gap between the scale at which QG effects are dominant and the current experiments is to consider QG effects on relativistic systems. In this work, frame independent minimum length is presented and its implications are shown for relativistic systems. The results presented here open the possibility of using higher energy experiments such as ATLAS to search for QG corrections. The high energy of these experiments drastically improves the probability of detection.  
Additionally and possibly more interesting is the fact that the results presented here strongly suggest that the infinitely smooth and continuous spacetime is not the full picture, and that the properties of spacetime need to be reconsidered if a theory of quantum gravity is to be found.

\bibliography{thesis} 

\begin{thebibliography}{167}

\bibitem{Plank1901}
M.~Planck.
\newblock \"Ueber das gesetz der energieverteilung im normalspectrum.
\newblock {\em Annalen der Physik}, 309(3):553--563, 1901.

\bibitem{rayleigh1900vi}
L.~Rayleigh.
\newblock Vi. the law of partition of kinetic energy.
\newblock {\em The London, Edinburgh, and Dublin Philosophical Magazine and
  Journal of Science}, 49(296):98--118, 1900.

\bibitem{EinsteinPhoton1905}
A.~Einstein.
\newblock \"Ueber einen die erzeugung und verwandlung des lichtes betreffenden
  heuristischen gesichtspunkt.
\newblock {\em Annalen der Physik}, 322(6):132--148, 1905.

\bibitem{Bohr1913a}
N.~Bohr.
\newblock I. on the constitution of atoms and molecules.
\newblock {\em The London, Edinburgh, and Dublin Philosophical Magazine and
  Journal of Science}, 26(151):1--25, 1913.

\bibitem{Bohr1913b}
N.~Bohr.
\newblock Xxxvii. on the constitution of atoms and molecules.
\newblock {\em The London, Edinburgh, and Dublin Philosophical Magazine and
  Journal of Science}, 26(153):476--502, 1913.

\bibitem{Bohr1913c}
N.~Bohr.
\newblock Lxxiii. on the constitution of atoms and molecules.
\newblock {\em The London, Edinburgh, and Dublin Philosophical Magazine and
  Journal of Science}, 26(155):857--875, 1913.

\bibitem{Broglie1925}
L.~De~Broglie.
\newblock Recherches sur la th\'eorie des quanta.
\newblock {\em Ann. Phys.}, 10(3):22--128, 1925.

\bibitem{Dirac1928}
P.~A.~M. Dirac and R.~H. Fowler.
\newblock The quantum theory of the electron.
\newblock {\em Proceedings of the Royal Society of London. Series A, Containing
  Papers of a Mathematical and Physical Character}, 117(778):610--624, 1928.

\bibitem{Dirac1930}
P.~A.~. Dirac and R.~H. Fowler.
\newblock A theory of electrons and protons.
\newblock {\em Proceedings of the Royal Society of London. Series A, Containing
  Papers of a Mathematical and Physical Character}, 126(801):360--365, 1930.

\bibitem{Schrodinger1926}
E.~Schr\"odinger.
\newblock An undulatory theory of the mechanics of atoms and molecules.
\newblock {\em Phys. Rev.}, 28:1049--1070, Dec 1926.

\bibitem{Heisenberg1927}
W.~Heisenberg.
\newblock {\"{U}ber den anschaulichen Inhalt der quantentheoretischen Kinematik
  und Mechanik}.
\newblock {\em Zeitschrift f\"{u}r Physik}, 33:879--893, 1925.

\bibitem{OReilly2002}
E.~P. O'Reilly.
\newblock {\em Quantum theory of solids}.
\newblock Taylor \& Francis, London;New York;, 2002.

\bibitem{OConnel2005}
J.~P. O'Connell, J.~M. Haile, and Inc NetLibrary.
\newblock {\em Thermodynamics: fundamentals and applications}.
\newblock Cambridge University Press, Cambridge [England], 2005.

\bibitem{Garrison2008}
J.~C. Garrison and R.~Y. Chiao.
\newblock {\em Quantum optics}.
\newblock Oxford University Press, Oxford;Toronto;, 2008.

\bibitem{Lewars2011}
E.~Lewars.
\newblock {\em Computational chemistry: introduction to the theory and
  applications of molecular and quantum mechanics}.
\newblock Springer, London;New York;Dordrecht [Netherlands];, 2nd edition,
  2011.

\bibitem{Brout1964}
F.~Englert and R.~Brout.
\newblock Broken symmetry and the mass of gauge vector mesons.
\newblock {\em Phys. Rev. Lett.}, 13:321--323, Aug 1964.

\bibitem{Higgs1964}
P.~W. Higgs.
\newblock Broken symmetries and the masses of gauge bosons.
\newblock {\em Phys. Rev. Lett.}, 13:508--509, Oct 1964.

\bibitem{Kibble1964}
G.~S. Guralnik, C.~R. Hagen, and T.~W.~B. Kibble.
\newblock Global conservation laws and massless particles.
\newblock {\em Phys. Rev. Lett.}, 13:585--587, Nov 1964.

\bibitem{CMSLHSHiggs2012}
S.~Chatrchyan, V.~Khachatryan, A.M. Sirunyan, A.~Tumasyan, W.~Adam, E.~Aguilo,
  T.~Bergauer, M.~Dragicevic, J.~ErÃ¶, C.~Fabjan, and Et. al.
\newblock {Observation of a new boson at a mass of 125 GeV with the CMS
  experiment at the LHC}.
\newblock {\em Phys. Lett. B}, 716(1):30 -- 61, 2012.

\bibitem{Einstein1911GR}
A.~Einstein.
\newblock Ãœber den einfluÃŸ der schwerkraft auf die ausbreitung des lichtes.
\newblock {\em Annalen der Physik}, 340(10):898--908, 1911.

\bibitem{Einstein1913GR}
A.~Einstein and M.~Grossmann.
\newblock Entwurf einer verallgemeinerten relativitÃ¤tstheorie und eine theorie
  der gravitation. i. physikalischer teil von a. einstein ii. mathematischer
  teil von m. grossmann.
\newblock {\em Zeitschrift fÃ¼r Mathematik und Physik}, 62:225â€“244,
  245â€“261, 1913.

\bibitem{Einstein1905SR}
A.~Einstein.
\newblock Zur elektrodynamik bewegter kÃ¶rper.
\newblock {\em Annalen der Physik}, 322(10):891--921, 1905.

\bibitem{chew1961principle}
G.~F. Chew and S.~C. Frautschi.
\newblock Principle of equivalence for all strongly interacting particles
  within the s-matrix framework.
\newblock {\em Physical Review Letters}, 7(10):394, 1961.

\bibitem{regge1959introduction}
T.~Regge.
\newblock Introduction to complex orbital momenta.
\newblock {\em Il Nuovo Cimento (1955-1965)}, 14(5):951--976, 1959.

\bibitem{Susskind1969}
L.~Susskind.
\newblock Harmonic-oscillator analogy for the veneziano model.
\newblock {\em Phys. Rev. Lett.}, 23:545--547, Sep 1969.

\bibitem{Susskind1970}
L.~Susskind.
\newblock Structure of hadrons implied by duality.
\newblock {\em Phys. Rev. D}, 1:1182--1186, Feb 1970.

\bibitem{Nambu:1997wf}
Y.~Nambu.
\newblock {Quark model and the factorization of the Veneziano amplitude}.
\newblock In {\em {International Conference on Symmetries and Quark Models,
  Wayne State U., Detroit}}, pages 269--278, 1997.

\bibitem{dolen1968finite}
R.~Dolen, D.~Horn, and Ch. Schmid.
\newblock Finite-energy sum rules and their application to $\pi$ n charge
  exchange.
\newblock {\em Physical Review}, 166(5):1768, 1968.

\bibitem{veneziano1968construction}
G.~Veneziano.
\newblock Construction of a crossing-simmetric, regge-behaved amplitude for
  linearly rising trajectories.
\newblock {\em Il Nuovo Cimento A (1965-1970)}, 57(1):190--197, 1968.

\bibitem{lovelace1971pomeron}
C.~Lovelace.
\newblock Pomeron form factors and dual regge cuts.
\newblock {\em Physics Letters B}, 34(6):500--506, 1971.

\bibitem{nambu1995quark}
Y.~Nambu.
\newblock Quark model and the factorization of the veneziano amplitude.
\newblock {\em Broken symmetry: selected papers of Y. Nambu}, 13:258--267,
  1995.

\bibitem{nielsen1969almost}
H.B. Nielsen.
\newblock An almost physical interpretation of the dual n-point function.
\newblock {\em Nordita Report, unpublished}, 1969.

\bibitem{susskind1970structure}
L.~Susskind.
\newblock Structure of hadrons implied by duality.
\newblock {\em Physical Review D}, 1(4):1182, 1970.

\bibitem{ramond1971dual}
P.~Ramond.
\newblock Dual theory for free fermions.
\newblock {\em Physical Review D}, 3(10):2415, 1971.

\bibitem{neveu1971tachyon}
A.~Neveu and J.~H. Schwarz.
\newblock Tachyon-free dual model with a positive-intercept trajectory.
\newblock {\em Physics Letters B}, 34(6):517--518, 1971.

\bibitem{Banks:1996vh}
T.~Banks, W.~Fischler, S.H. Shenker, and L.~Susskind.
\newblock {M theory as a matrix model: A Conjecture}.
\newblock {\em Phys. Rev. D}, 55:5112--5128, 1997.

\bibitem{SCHWARZ1982}
J.~H. Schwarz.
\newblock Superstring theory.
\newblock {\em Physics Reports}, 89(3):223 -- 322, 1982.

\bibitem{SCHERK1974}
J.~Scherk and J.~H. Schwarz.
\newblock Dual models for non-hadrons.
\newblock {\em Nuclear Physics B}, 81(1):118 -- 144, 1974.

\bibitem{Zwiebach2004}
B.~Zwiebach.
\newblock {\em A First Course in String Theory}.
\newblock Cambridge University Press, 2004.

\bibitem{Polyakov1981}
A.M. Polyakov.
\newblock Quantum geometry of bosonic strings.
\newblock {\em Physics Letters B}, 103(3):207 -- 210, 1981.

\bibitem{Martin1998}
S.~P. Martin.
\newblock {\em A supersymmetry primer}, pages 1--98.
\newblock World Scientific, 1998.

\bibitem{DeWitt:1967yk}
B.~S. DeWitt.
\newblock {Quantum Theory of Gravity. 1. The Canonical Theory}.
\newblock {\em Phys. Rev.}, 160:1113--1148, 1967.

\bibitem{Arnowitt1959}
R.~Arnowitt, S.~Deser, and C.~W. Misner.
\newblock Dynamical structure and definition of energy in general relativity.
\newblock {\em Phys. Rev.}, 116:1322--1330, Dec 1959.

\bibitem{SEN1982}
A.~Sen.
\newblock Gravity as a spin system.
\newblock {\em Physics Letters B}, 119(1):89 -- 91, 1982.

\bibitem{Penrose1971}
R.~Penrose.
\newblock Angular momentum: an approach to combinatorial space-time.
\newblock {\em Quantum theory and beyond}, pages 151--180, 1971.

\bibitem{Ashtekar1986new}
A.~Ashtekar.
\newblock New variables for classical and quantum gravity.
\newblock {\em Physical Review Letters}, 57(18):2244, 1986.

\bibitem{Ashtekar1987new}
A.~Ashtekar.
\newblock New hamiltonian formulation of general relativity.
\newblock {\em Physical Review D}, 36(6):1587, 1987.

\bibitem{Wilson1974}
K.~G. Wilson.
\newblock Confinement of quarks.
\newblock {\em Phys. Rev. D}, 10:2445--2459, Oct 1974.

\bibitem{Rovelli1988}
C.~Rovelli and L.~Smolin.
\newblock Knot theory and quantum gravity.
\newblock {\em Physical Review Letters}, 61(10):1155, 1988.

\bibitem{Rovelli1990}
C.~Rovelli and L.~Smolin.
\newblock Loop space representation of quantum general relativity.
\newblock {\em Nuclear Physics B}, 331(1):80 -- 152, 1990.

\bibitem{Rovelli1994}
C.~Rovelli and L.~Smolin.
\newblock {Discreteness of area and volume in quantum gravity}.
\newblock {\em Nucl. Phys. B}, 442:593--622, 1995.
\newblock [Erratum: Nucl.Phys.B 456, 753--754 (1995)].

\bibitem{AmelinoCamelia:2000mn}
G.~Amelino-Camelia.
\newblock {Relativity in space-times with short distance structure governed by
  an observer independent (Planckian) length scale}.
\newblock {\em Int. J. Mod. Phys. D}, 11:35--60, 2002.

\bibitem{AMELINOCAMELIA2001255}
G.~Amelino-Camelia.
\newblock Testable scenario for relativity with minimum length.
\newblock {\em Physics Letters B}, 510(1):255 -- 263, 2001.

\bibitem{Magueijo:2001cr}
J.~Magueijo and L.~Smolin.
\newblock {Lorentz invariance with an invariant energy scale}.
\newblock {\em Phys. Rev. Lett.}, 88:190403, 2002.

\bibitem{Magueijo:2002am}
J.~Magueijo and L.~Smolin.
\newblock {Generalized Lorentz invariance with an invariant energy scale}.
\newblock {\em Phys. Rev. D}, 67:044017, 2003.

\bibitem{Majid:1994cy}
S.~Majid and H.~Ruegg.
\newblock {Bicrossproduct structure of kappa Poincare group and noncommutative
  geometry}.
\newblock {\em Phys. Lett. B}, 334:348--354, 1994.

\bibitem{bergman1985everybody}
G.~Bergman.
\newblock Everybody knows what a hopf algebra is.
\newblock {\em Contemp. Math}, 43(198.5):25--48, 1985.

\bibitem{Garay:1994en}
Luis~J. Garay.
\newblock {Quantum gravity and minimum length}.
\newblock {\em Int. J. Mod. Phys. A}, 10:145--166, 1995.

\bibitem{Hossenfelder:2012jw}
S.~Hossenfelder.
\newblock {Minimal Length Scale Scenarios for Quantum Gravity}.
\newblock {\em Living Rev. Rel.}, 16:2, 2013.

\bibitem{narlikar1983quantum}
J.~V. Narlikar and T.~Padmanabhan.
\newblock Quantum cosmology via path integrals.
\newblock {\em Physics Reports}, 100(3):151--200, 1983.

\bibitem{srednicki2007quantum}
M.~Srednicki.
\newblock {\em Quantum field theory}.
\newblock Cambridge University Press, 2007.

\bibitem{PADMANABHAN198538}
T.~Padmanabhan.
\newblock Physical significance of planck length.
\newblock {\em Annals of Physics}, 165(1):38 -- 58, 1985.

\bibitem{Padmanabhan:1985jq}
T.~Padmanabhan.
\newblock {Planck length as the lower bound to all physical length scales}.
\newblock {\em Gen. Rel. Grav.}, 17:215--221, 1985.

\bibitem{Padmanabhan_1986}
T.~Padmanabhan.
\newblock The role of general relativity in the uncertainty principle.
\newblock {\em Classical and Quantum Gravity}, 3(5):911--920, sep 1986.

\bibitem{Padmanabhan_1987}
T.~Padmanabhan.
\newblock Limitations on the operational definition of spacetime events and
  quantum gravity.
\newblock {\em Classical and Quantum Gravity}, 4(4):L107--L113, jul 1987.

\bibitem{AMATI198941}
D.~Amati, M.~Ciafaloni, and G.~Veneziano.
\newblock Can spacetime be probed below the string size?
\newblock {\em Physics Letters B}, 216(1):41 -- 47, 1989.

\bibitem{Chang:2011jj}
L.~N. Chang, Z.~Lewis, D.~Minic, and T.~Takeuchi.
\newblock {On the Minimal Length Uncertainty Relation and the Foundations of
  String Theory}.
\newblock {\em Adv. High Energy Phys.}, 2011:493514, 2011.

\bibitem{KONISHI1990276}
K.~Konishi, G.~Paffuti, and P.~Provero.
\newblock Minimum physical length and the generalized uncertainty principle in
  string theory.
\newblock {\em Physics Letters B}, 234(3):276 -- 284, 1990.

\bibitem{ROVELLI1995593}
C.~Rovelli and L.~Smolin.
\newblock Discreteness of area and volume in quantum gravity.
\newblock {\em Nuclear Physics B}, 442(3):593 -- 619, 1995.

\bibitem{Ashtekar_1997}
A.~Ashtekar and J.~Lewandowski.
\newblock Quantum theory of geometry: I. area operators.
\newblock {\em Classical and Quantum Gravity}, 14(1A):A55--A81, jan 1997.

\bibitem{thiemann1998length}
T.~Thiemann.
\newblock Closed formula for the matrix elements of the volume operator in
  canonical quantum gravity.
\newblock {\em Journal of Mathematical Physics}, 39(6):3347--3371, 1998.

\bibitem{Bianchi2009length}
E.~Bianchi.
\newblock The length operator in loop quantum gravity.
\newblock {\em Nuclear Physics B}, 807(3):591 -- 624, 2009.

\bibitem{ma2010new}
Y.~Ma, C.~Soo, and J.~Yang.
\newblock New length operator for loop quantum gravity.
\newblock {\em Phys. Rev. D}, 81:124026, Jun 2010.

\bibitem{dictionary1989oxford}
Oxford~English Dictionary.
\newblock Oxford english dictionary.
\newblock {\em Simpson, JA \& Weiner, ESC}, 1989.

\bibitem{AmelinoCamelia:2008qg}
G.~Amelino-Camelia.
\newblock {Quantum-Spacetime Phenomenology}.
\newblock {\em Living Rev. Rel.}, 16:5, 2013.

\bibitem{Bosso:2017hoq}
P.~Bosso.
\newblock {\em {Generalized Uncertainty Principle and Quantum Gravity
  Phenomenology}}.
\newblock PhD thesis, Lethbridge U., 2017.

\bibitem{PhysRevLett.33.1237}
A.~W. Overhauser and R.~Colella.
\newblock Experimental test of gravitationally induced quantum interference.
\newblock {\em Phys. Rev. Lett.}, 33:1237--1239, Nov 1974.

\bibitem{Colella:1975dq}
R.~Colella, A.W. Overhauser, and S.A. Werner.
\newblock {Observation of gravitationally induced quantum interference}.
\newblock {\em Phys. Rev. Lett.}, 34:1472--1474, 1975.

\bibitem{PhysRev.187.1767}
R.~Ruffini and S.~Bonazzola.
\newblock Systems of self-gravitating particles in general relativity and the
  concept of an equation of state.
\newblock {\em Phys. Rev.}, 187:1767--1783, Nov 1969.

\bibitem{Einstein1905}
A.~Einstein.
\newblock Ãœber die von der molekularkinetischen theorie der wÃ¤rme geforderte
  bewegung von in ruhenden flÃ¼ssigkeiten suspendierten teilchen.
\newblock {\em Annalen der Physik}, 322(8):549--560, 1905.

\bibitem{Ford1995}
L.~H. Ford.
\newblock Gravitons and light cone fluctuations.
\newblock {\em Phys. Rev. D}, 51:1692--1700, Feb 1995.

\bibitem{AmelinoCamelia:1999gg}
G.~Amelino-Camelia.
\newblock {Gravity wave interferometers as probes of a low-energy effective
  quantum gravity}.
\newblock {\em Phys. Rev. D}, 62:024015, 2000.

\bibitem{Hogan:2008zw}
Craig~J. Hogan.
\newblock {Indeterminacy of Holographic Quantum Geometry}.
\newblock {\em Phys. Rev. D}, 78:087501, 2008.

\bibitem{BOSSO2018498}
P.~Bosso, S.~Das, and R.~B. Mann.
\newblock Potential tests of the generalized uncertainty principle in the
  advanced ligo experiment.
\newblock {\em Physics Letters B}, 785:498 -- 505, 2018.

\bibitem{tHooft:1996ziz}
G.~'t~Hooft.
\newblock {Quantization of point particles in (2+1)-dimensional gravity and
  space-time discreteness}.
\newblock {\em Class. Quant. Grav.}, 13:1023--1040, 1996.

\bibitem{AmelinoCamelia:2000ge}
G.~Amelino-Camelia.
\newblock {Testable scenario for relativity with minimum length}.
\newblock {\em Phys. Lett. B}, 510:255--263, 2001.

\bibitem{Lukierski:1993wx}
J.~Lukierski, H.~Ruegg, and W.~J. Zakrzewski.
\newblock {Classical quantum mechanics of free kappa relativistic systems}.
\newblock {\em Annals Phys.}, 243:90--116, 1995.

\bibitem{mattingly2005modern}
D.~Mattingly.
\newblock Modern tests of lorentz invariance.
\newblock {\em Living Reviews in relativity}, 8(1):5, 2005.

\bibitem{Liberati:2013xla}
S.~Liberati.
\newblock {Tests of Lorentz invariance: a 2013 update}.
\newblock {\em Class. Quant. Grav.}, 30:133001, 2013.

\bibitem{Ackermann:2009aa}
M.~Ackermann et~al.
\newblock {A limit on the variation of the speed of light arising from quantum
  gravity effects}.
\newblock {\em Nature}, 462:331--334, 2009.

\bibitem{AmelinoCamelia:2009pg}
G.~Amelino-Camelia and L.~Smolin.
\newblock {Prospects for constraining quantum gravity dispersion with near term
  observations}.
\newblock {\em Phys. Rev. D}, 80:084017, 2009.

\bibitem{Maccione:2007yc}
L.~Maccione, S.~Liberati, A.~Celotti, and J.~G. Kirk.
\newblock {New constraints on Planck-scale Lorentz Violation in QED from the
  Crab Nebula}.
\newblock {\em JCAP}, 10:013, 2007.

\bibitem{PhysRevLett.16.748}
K.~Greisen.
\newblock End to the cosmic-ray spectrum?
\newblock {\em Phys. Rev. Lett.}, 16:748--750, Apr 1966.

\bibitem{Abraham:2008ru}
J.~Abraham et~al.
\newblock {Observation of the suppression of the flux of cosmic rays above
  $4\times 10^{19}$eV}.
\newblock {\em Phys. Rev. Lett.}, 101:061101, 2008.

\bibitem{Stecker:2011ps}
Floyd~W. Stecker.
\newblock {A New Limit on Planck Scale Lorentz Violation from Gamma-ray Burst
  Polarization}.
\newblock {\em Astropart. Phys.}, 35:95--97, 2011.

\bibitem{Lue:1998mq}
A.~Lue, L.~Wang, and M.~Kamionkowski.
\newblock {Cosmological signature of new parity violating interactions}.
\newblock {\em Phys. Rev. Lett.}, 83:1506--1509, 1999.

\bibitem{Landsberg:2008ax}
G.~Landsberg.
\newblock {\em Collider searches for extra spatial dimensions and black holes},
  pages 99--106.
\newblock World Scientific, 2009.

\bibitem{MAGGIORE199365}
M.~Maggiore.
\newblock A generalized uncertainty principle in quantum gravity.
\newblock {\em Physics Letters B}, 304(1):65 -- 69, 1993.

\bibitem{PhysRevD.49.5182}
M.~Maggiore.
\newblock Quantum groups, gravity, and the generalized uncertainty principle.
\newblock {\em Phys. Rev. D}, 49:5182--5187, May 1994.

\bibitem{Kempf1995}
A.~Kempf, G.~Mangano, and R.~B. Mann.
\newblock {Hilbert space representation of the minimal length uncertainty
  relation}.
\newblock {\em Phys. Rev. D}, 52:1108--1118, 1995.

\bibitem{Bosso:2020aqm}
B.~Pasquale.
\newblock On the quasi-position representation in theories with a minimal
  length.
\newblock {\em ArXiv}, 5 2020.

\bibitem{bosso2017generalized}
P.~Bosso and S.~Das.
\newblock Generalized uncertainty principle and angular momentum.
\newblock {\em Annals of Physics}, 383:416--438, 2017.

\bibitem{Das2008}
S.~Das and E.~C. Vagenas.
\newblock {Universality of Quantum Gravity Corrections}.
\newblock {\em Phys. Rev. Lett.}, 101:221301, 2008.

\bibitem{Ali:2010yn}
A.~F. Ali, S.~Das, and E.~C. Vagenas.
\newblock {The Generalized Uncertainty Principle and Quantum Gravity
  Phenomenology}.
\newblock In {\em {12th Marcel Grossmann Meeting on General Relativity}}, pages
  2407--2409, 1 2010.

\bibitem{Brau:1999uv}
F.~Brau.
\newblock {Minimal length uncertainty relation and hydrogen atom}.
\newblock {\em J. Phys. A}, 32:7691--7696, 1999.

\bibitem{Ali2011}
A.~F. Ali, S.~Das, and E.~C. Vagenas.
\newblock {A proposal for testing Quantum Gravity in the lab}.
\newblock {\em Phys. Rev. D}, 84:044013, 2011.

\bibitem{Bosso2018}
P.~Bosso, S.~Das, and R.~B. Mann.
\newblock {Potential tests of the Generalized Uncertainty Principle in the
  advanced LIGO experiment}.
\newblock {\em Phys. Lett. B}, 785:498--505, 2018.

\bibitem{Ali2010}
A.~F. Ali, S.~Das, and E.~C. Vagenas.
\newblock {The Generalized Uncertainty Principle and Quantum Gravity
  Phenomenology}.
\newblock In {\em {12th Marcel Grossmann Meeting on General Relativity}}, pages
  2407--2409, 1 2010.

\bibitem{Das2011}
S.~Das and R.B. Mann.
\newblock {Planck scale effects on some low energy quantum phenomena}.
\newblock {\em Phys. Lett. B}, 704:596--599, 2011.

\bibitem{Das2014}
S.~Das, M.~P.~G. Robbins, and M.~A. Walton.
\newblock {Generalized Uncertainty Principle Corrections to the Simple Harmonic
  Oscillator in Phase Space}.
\newblock {\em Can. J. Phys.}, 94(1):139--146, 2016.

\bibitem{bushev2019testing}
P.A. Bushev, J.~Bourhill, M.~Goryachev, N.~Kukharchyk, E.~Ivanov, S.~Galliou,
  M.E. Tobar, and S.~Danilishin.
\newblock Testing the generalized uncertainty principle with macroscopic
  mechanical oscillators and pendulums.
\newblock {\em Physical Review D}, 100(6):066020, 2019.

\bibitem{Szabo:2006wx}
R.~J. Szabo.
\newblock {Symmetry, gravity and noncommutativity}.
\newblock {\em Class. Quant. Grav.}, 23:R199--R242, 2006.

\bibitem{Chaichian:2004za}
M.~Chaichian, P.P. Kulish, K.~Nishijima, and A.~Tureanu.
\newblock {On a Lorentz-invariant interpretation of noncommutative space-time
  and its implications on noncommutative QFT}.
\newblock {\em Phys. Lett. B}, 604:98--102, 2004.

\bibitem{snyder1947quantized}
H.~S. Snyder.
\newblock Quantized space-time.
\newblock {\em Physical Review}, 71(1):38, 1947.

\bibitem{Quesne:2006is}
C.~Quesne and V.M. Tkachuk.
\newblock {Lorentz-covariant deformed algebra with minimal length}.
\newblock {\em Czech. J. Phys.}, 56:1269--1274, 2006.

\bibitem{Faizal:2017map}
M.~Faizal and T.~S. Tsun.
\newblock {Topological defects in a deformed gauge theory}.
\newblock {\em Nucl. Phys. B}, 924:588--602, 2017.

\bibitem{Faizal:2014dua}
M.~Faizal, A.~F. Ali, and A.~Nassar.
\newblock {AdS/CFT Correspondence Beyond its Supergravity Approximation}.
\newblock {\em Int. J. Mod. Phys. A}, 30(30):1550183, 2015.

\bibitem{Pramanik:2014zfa}
S.~Pramanik, S.~Ghosh, and P.~Pal.
\newblock {Conformal Invariance in noncommutative geometry and mutually
  interacting Snyder Particles}.
\newblock {\em Phys. Rev. D}, 90(10):105027, 2014.

\bibitem{Deriglazov:2014yta}
A.~A. Deriglazov and A.~M. Pupasov-Maksimov.
\newblock {Frenkel electron on an arbitrary electromagnetic background and
  magnetic Zitterbewegung}.
\newblock {\em Nucl. Phys. B}, 885:1--24, 2014.

\bibitem{Pramanik:2013zy}
S.~Pramanik and S.~Ghosh.
\newblock {GUP-based and Snyder Non-Commutative Algebras, Relativistic Particle
  models and Deformed Symmetries and Interaction: A Unified Approach}.
\newblock {\em Int. J. Mod. Phys. A}, 28(27):1350131, 2013.

\bibitem{Husain:2013zda}
V.~Husain, S.~S. Seahra, and E.~J. Webster.
\newblock {High energy modifications of blackbody radiation and dimensional
  reduction}.
\newblock {\em Phys. Rev. D}, 88(2):024014, 2013.

\bibitem{Kober:2010sj}
M.~Kober.
\newblock {Gauge Theories under Incorporation of a Generalized Uncertainty
  Principle}.
\newblock {\em Phys. Rev. D}, 82:085017, 2010.

\bibitem{Hossenfelder:2006cw}
S.~Hossenfelder.
\newblock {Interpretation of quantum field theories with a minimal length
  scale}.
\newblock {\em Phys. Rev. D}, 73:105013, 2006.

\bibitem{Capozziello:1999wx}
S.~Capozziello, G.~Lambiase, and G.~Scarpetta.
\newblock {Generalized uncertainty principle from quantum geometry}.
\newblock {\em Int. J. Theor. Phys.}, 39:15--22, 2000.

\bibitem{zakrzewski1994quantum}
S.~Zakrzewski.
\newblock Quantum poincare group related to the kappa-poincare algebra.
\newblock {\em Journal of Physics A: Mathematical and General}, 27(6):2075,
  1994.

\bibitem{Kober:2011dn}
M.~Kober.
\newblock {Electroweak Theory with a Minimal Length}.
\newblock {\em Int. J. Mod. Phys. A}, 26:4251--4285, 2011.

\bibitem{Hossenfelder:2014ifa}
S.~Hossenfelder.
\newblock {The Soccer-Ball Problem}.
\newblock {\em SIGMA}, 10:074, 2014.

\bibitem{amelino2011relative}
G.~Amelino-Camelia, L.~Freidel, J.~Kowalski-Glikman, and L.~Smolin.
\newblock Relative locality and the soccer ball problem.
\newblock {\em Physical Review D}, 84(8):087702, 2011.

\bibitem{hossenfelder2013comment}
S.~Hossenfelder.
\newblock Comment on relative locality and the soccer ball problem.
\newblock {\em Physical Review D}, 88(2):028701, 2013.

\bibitem{amelino2013reply}
G.~Amelino-Camelia, L.~Freidel, J.~Kowalski-Glikman, and L.~Smolin.
\newblock Reply to comment on ``relative locality and the soccer ball problem".
\newblock {\em Physical Review D}, 88(2):028702, 2013.

\bibitem{Todorinov:2018arx}
V.~Todorinov, P.~Bosso, and S.~Das.
\newblock Relativistic generalized uncertainty principle.
\newblock {\em Annals of Physics}, 405:92 -- 100, 2019.

\bibitem{Quesne2006}
C.~Quesne and V.M. Tkachuk.
\newblock {Lorentz-covariant deformed algebra with minimal length}.
\newblock {\em Czech. J. Phys.}, 56:1269--1274, 2006.

\bibitem{ohlsson2011relativistic}
T.~Ohlsson.
\newblock {\em Relativistic quantum physics: from advanced quantum mechanics to
  introductory quantum field theory}.
\newblock Cambridge University Press, 2011.

\bibitem{Snyder:1946qz}
H.~S. Snyder.
\newblock Quantized space-time.
\newblock {\em Phys. Rev.}, 71:38--41, Jan 1947.

\bibitem{hebey2008introduction}
E.~Hebey and B.~Pausader.
\newblock An introduction to fourth order nonlinear wave equations.
\newblock {\em Rn}, 2:H2, 2008.

\bibitem{gordon1926comptoneffekt}
W.~Gordon.
\newblock Der comptoneffekt nach der schr{\"o}dingerschen theorie.
\newblock {\em Zeitschrift f{\"u}r Physik}, 40(1-2):117--133, 1926.

\bibitem{klein1986quantentheorie}
V.~O. Klein.
\newblock Quantentheorie und f{\"u}nfdimensionale relativit{\"a}tstheorie.
\newblock {\em Surveys in High Energy Physics}, 5(3):241--244, 1986.

\bibitem{Bosso:2018uus}
P.~Bosso.
\newblock {Rigorous Hamiltonian and Lagrangian analysis of classical and
  quantum theories with minimal length}.
\newblock {\em Phys. Rev. D}, 97(12):126010, 2018.

\bibitem{Das:2009hs}
S.~Das and E.~C. Vagenas.
\newblock {Phenomenological Implications of the Generalized Uncertainty
  Principle}.
\newblock {\em Can. J. Phys.}, 87:233--240, 2009.

\bibitem{Griffiths2018}
D.~J. Griffiths and D.~F. Schroeter.
\newblock {\em Introduction to Quantum Mechanics}.
\newblock Cambridge University Press, 3 edition, 2018.

\bibitem{Hill2002}
R.J.A. Hill, A.~PatanÃ¨, P.C. Main, L.~Eaves, B.~Gustafson, M.~Henini,
  S.~Tarucha, and D.G. Austing.
\newblock Measuring the energy levels and wave functions in a single quantum
  dot.
\newblock {\em Physica E: Low-dimensional Systems and Nanostructures},
  13(2):634 -- 637, 2002.

\bibitem{messiah1999quantum}
A.~Messiah.
\newblock {\em Quantum Mechanics}.
\newblock Dover books on physics. Dover Publications, 1999.

\bibitem{yin2016experimental}
L.~Yin, Y.~Zhang, J.~Qiao, S.~Li, and L.~He.
\newblock Experimental observation of surface states and landau levels bending
  in bilayer graphene.
\newblock {\em Physical Review B}, 93(12):125422, 2016.

\bibitem{wildoer1997observation}
J.W.G. Wild{\"o}er, C.J.P.M. Harmans, and H.~Van~Kempen.
\newblock Observation of landau levels at the inas (110) surface by scanning
  tunneling spectroscopy.
\newblock {\em Physical Review B}, 55(24):R16013, 1997.

\bibitem{Das:2010zf}
S.~Das, E.~C. Vagenas, and A.~F. Ali.
\newblock {Discreteness of Space from GUP II: Relativistic Wave Equations}.
\newblock {\em Phys. Lett. B}, 690:407--412, 2010.
\newblock [Erratum: Phys.Lett.B 692, 342--342 (2010)].

\bibitem{Citation2017}
R.~P. Mart\'inez-y Romero.
\newblock Relativistic hydrogen atom revisited.
\newblock {\em American Journal of Physics}, 68(11):1050--1055, 2000.

\bibitem{BOSSO2020168319}
P.~Bosso, S.~Das, and V.~Todorinov.
\newblock Quantum field theory with the generalized uncertainty principle i:
  Scalar electrodynamics.
\newblock {\em Annals of Physics}, 422:168319, 2020.

\bibitem{Bosso:2020jay}
Pasquale Bosso, Saurya Das, and Vasil Todorinov.
\newblock {Quantum field theory with the generalized uncertainty principle II:
  Quantum Electrodynamics}.
\newblock {\em Annals Phys.}, 424:168350, 2021.

\bibitem{Pons:1988tj}
J.M. Pons.
\newblock {Ostrogradski Theorem for Higher Order Singular Lagrangians}.
\newblock {\em Lett. Math. Phys.}, 17:181, 1989.

\bibitem{Woodard:2015zca}
R.~P. Woodard.
\newblock {Ostrogradsky's theorem on Hamiltonian instability}.
\newblock {\em Scholarpedia}, 10(8):32243, 2015.

\bibitem{deUrries:1998obu}
F.J. de~Urries and J.~Julve.
\newblock {Ostrogradski formalism for higher derivative scalar field theories}.
\newblock {\em J. Phys. A}, 31:6949--6964, 1998.

\bibitem{Srednicki2005QFT}
M.~A. Srednicki.
\newblock {\em Quantum field theory}.
\newblock Cambridge University Press, Cambridge, 2007.

\bibitem{Weinberg2005QFT}
S.~Weinberg.
\newblock {\em The quantum theory of fields}.
\newblock Cambridge University Press, Cambridge, 2005.

\bibitem{Das2008QFT}
A.~Das.
\newblock {\em Lectures on quantum field theory}.
\newblock World Scientific, Hackensack, NJ Singapore, 2008.

\bibitem{Halzen:1984mc}
F.~Halzen and A.~D. Martin.
\newblock {\em {Quarks and leptons: An introductory course in modern particle
  physics}}.
\newblock John Wiley \& Sons, 1 1984.

\bibitem{gerlach1922experimentelle}
W.~Gerlach and O.~Stern.
\newblock Der experimentelle nachweis der richtungsquantelung im magnetfeld.
\newblock {\em Zeitschrift f{\"u}r Physik}, 9(1):349--352, 1922.

\bibitem{Balek:2019nqk}
P.~Balek.
\newblock {Charged-hadron suppression in Pb+Pb and Xe+Xe collisions measured
  with the ATLAS detector}.
\newblock {\em Nucl. Phys. A}, 982:571--574, 2019.

\bibitem{Harikumar:2006xf}
E.~Harikumar and Victor~O. Rivelles.
\newblock {Noncommutative Gravity}.
\newblock {\em Class. Quant. Grav.}, 23:7551--7560, 2006.

\bibitem{Mannheim:2007ki}
P.~D. Mannheim.
\newblock {Conformal Gravity Challenges String Theory}.
\newblock In {\em {13th International Symposium on Particles, Strings and
  Cosmology}}, 7 2007.

\bibitem{mamat2016landau}
Jumakari Mamat, Sayipjamal Dulat, and Hekim Mamatabdulla.
\newblock Landau-like atomic problem on a non-commutative phase space.
\newblock {\em International Journal of Theoretical Physics}, 55(6):2913--2918,
  2016.

\bibitem{Dapor:2020jvc}
A.~Dapor and K.~Liegener.
\newblock {Modifications to Gravitational Wave Equation from Canonical Quantum
  Gravity}.
\newblock {\em Eur. Phys. J. C}, 80(8):741, 2020.

\bibitem{Mirshekari:2011yq}
S.~Mirshekari, N.~Yunes, and C.~M. Will.
\newblock {Constraining Generic Lorentz Violation and the Speed of the Graviton
  with Gravitational Waves}.
\newblock {\em Phys. Rev. D}, 85:024041, 2012.

\bibitem{tedesco2011fine}
L.~Tedesco.
\newblock Fine structure constant, domain walls, and generalized uncertainty
  principle in the universe.
\newblock {\em International Journal of Mathematics and Mathematical Sciences},
  2011, 2011.

\bibitem{wang2016solutions}
B.~Wang, C.~Long, Z.~Long, and T.~Xu.
\newblock Solutions of the schr{\"o}dinger equation under topological defects
  space-times and generalized uncertainty principle.
\newblock {\em The European Physical Journal Plus}, 131(10):378, 2016.

\bibitem{Arcos:2011nz}
H.I. Arcos, C.S.O. Mayor, G.~Otalora, and J.G. Pereira.
\newblock {Spin-2 fields and helicity}.
\newblock {\em Found. Phys.}, 42:1339--1349, 2012.

\bibitem{10.2307/1968645}
H.~Weyl.
\newblock Geodesic fields in the calculus of variation for multiple integrals.
\newblock {\em Annals of Mathematics}, 36(3):607--629, 1935.

\bibitem{hatfield2018quantum}
B.~Hatfield.
\newblock {\em Quantum field theory of point particles and strings}.
\newblock CRC Press, 2018.

\end{thebibliography}

\appendix
\chapter{Composition law problem}\label{app:CLP}
The arguments presented here are taken from \cite{Hossenfelder:2014ifa,amelino2011relative,hossenfelder2013comment,amelino2013reply}.
\section{A naive argument}

     Let one consider a  modified dispersion relation of the form 
\begin{equation}\label{app:modifiedDispersionRelation}
    D^{2}(p) = E^2-  \vec{p}^2 -  \eta \frac{E}{M_{\text{Pl}}} \vec{p}^2+ \cdots = m^2\,.
\end{equation}
The non-linear corrections to the Einstein's dispersion relation between energy and momentum arises from a
 choice of a non-trivial connection on momentum space, $\cal P$.  
In the case of plastic collision $A + B \rightarrow C$,  this takes the form
\begin{equation}
\label{2}
    p^{(C)}_\mu = \left(p^{(A)} \oplus p^{(B)}\right)_\mu\, ,
\end{equation}
the above relies on the assumption that there exists a nice set of coordinates on the momentum space $\cal P$, which allows for the expansion
\begin{equation}\label{appMomentumComposition}
\left(p^{(A)} \oplus p^{(B)}\right)_\mu =p^{(A)}_\mu + p^{(B)}_\mu -
\frac{1}{M_{\text{Pl}}}  \tilde{\Gamma}_\mu{}^{\alpha\beta}\, p^{(A)}_\alpha\,
p^{(B)}_\beta+ \cdots\,.
\end{equation}
where ${\tilde \Gamma}$ denotes the
 connection coefficients on momentum space evaluated
at the origin $p_{\mu}=0$. The
 connection coefficients
 on momentum space $\Gamma_\mu{}^{\alpha\beta}(0)$
has dimensions of inverse mass, so the ${\tilde \Gamma}$ are dimensionless.
One can write the dispersion relation as the standard one plus leading order corrections. This is justified in the
case of elementary particles, because even for the most energetic cosmic
rays the ratio of their energies to $M_{\text{pl}}$ is of order of $10^{-8}-10^{-9}$,
and the higher order terms can be safely neglected.

     When one considers a composite system, made
of a huge number $N$ of elementary particles. Let each elementary
particle obey the same dispersion relation shown in Eq.\eqref{app:modifiedDispersionRelation} and assume,
for simplicity, that all the particles have identical masses $m$ and
momenta $p_\mu$. The total mass of the system is therefore
$M_{\text{system}}= N\, m$ and its total momentum is $\mathbf{P}_{\text{system}\,\,\mu} =
N\, p_\mu$. Substituting this to Eq.\eqref{app:modifiedDispersionRelation} it can be easily shown that
\begin{equation}\label{app:TotalDispersionRelation}
    E_{\text{system}}^2=  \vec{\mathbf{P}}_{\text{system}}^2 + M_{\text{system}}^2 + N\eta\, \frac{E_{\text{system}}}{ M_{\text{Pl}}} \vec{\mathbf{P}}_{\text{system}}^2+ \ldots
\end{equation}
Comparing Eq.\eqref{app:modifiedDispersionRelation} with Eq.\eqref{app:TotalDispersionRelation} one notices that considering the composite system makes the QG corrections grow linearly with the number of parts $N$. 

The same argument can be applied to non-linear conservation law of the form  
\begin{equation}
    \left(p^{(A)} \oplus p^{(B)}\right)_\mu =p^{(A)}_\mu + p^{(B)}_\mu -
\frac{1}{M_{\text{Pl}}}  \tilde{\Gamma}_\mu{}^{\alpha\beta}\, p^{(A)}_\alpha\,
p^{(B)}_\beta+ \cdots\quad.
\end{equation}
One can then consider two systems $A$ and $B$ containing $N$ and $M$ particles
with identical momenta $p^A _\mu$ and $p^B _\mu$.
The total momenta of the systems can then be defined as 
\begin{subequations}
\begin{align}
\mathbf{P}^{(A)}_\mu  &= N p^A _\mu\,\\
\mathbf{P}^{(B)}_\mu  &= M p^B _\mu\,.
\label{naivedef}
\end{align}
\end{subequations}
A collision between those two systems can be described by a number of collisions between two particles of the systems $A$ and $B$. Every collision occurs between two parts of the systems, and  is governed by the dispersion relation as shown in Eq.\eqref{app:modifiedDispersionRelation}. It can be shown that
\begin{equation}\label{3naive}
\left(\mathbf{P}^{(A)}  \oplus \mathbf{P}^{(B)} \right)_\mu =\mathbf{P}^{(A)}_\mu +\mathbf{P}^{(B)}_\mu -
\frac{N}{ M_{\text{Pl}}}  \tilde{\Gamma}_\mu{}^{\alpha\beta}\, \mathbf{P}^{(A)}_\alpha\,
\mathbf{P}^{(B)}_\beta+ \cdots\,.
\end{equation}
Therefore the QG corrections again grow with the number of parts of the system.

\section{A more rigorous argument}

\subsection{The choice of coordinates on momentum space}

     The coordinates on momentum space  must be such that the origin
corresponds to the state with zero momentum and that both the modified
dispersion relation and momentum composition rules become the
standard special-relativistic ones in the limit of vanishing
momentum space curvature or $M_{\text{Pl}} \rightarrow \infty$. One such set of coordinates is  Riemann normal coordinates.  The metric geodesics from the origin are straight lines
and one finds
\begin{equation}\label{DR}
    m^2 = D^2(p) \equiv \eta^{\mu\nu}p_{\mu}p_{\nu}\, ,
\end{equation}
and therefore the dispersion relation in normal coordinates is not
modified. In this case therefore all the information about the
momentum space curvature is contained in the deformed momentum
composition rule Eq.\eqref{appMomentumComposition}. In any  other coordinate
system,  the dispersion relation, still defined
as $m^2=D^2(p)$, would take the general form Eq.\eqref{app:modifiedDispersionRelation}.

     Another set of important coordinates is the connection
normal coordinates, for which the geodesics associated with  the
connection are straight lines, even if the connection is not
metrical. In these coordinates $\hat{p}$ the addition of parallel
momenta is linear {\it i.e.} 
\begin{equation}(a\hat{p}) \hat{\oplus} (b \hat{p}) =
(a+b) \hat{p}\,, \end{equation}
where $a,b$ are any scalars. The connection for the coordinates up to first order in
$M_{\text{Pl}}$ is given by
$\hat{p}_{\mu}=F_{\mu}(p)$ where
\begin{equation} F_{\mu}({p}) = p_{\mu} +
\frac1{2M_{\text{Pl}}} \tilde{\Gamma}_{\mu}^{\alpha \beta} {p}_{\alpha}
{p}_{\beta}+\cdots\,. 
\end{equation} 
The addition in the new coordinates is given
by $\hat{p}\hat{\oplus} \hat{q} \equiv F( F^{-1}(\hat{p}) \oplus
F^{-1}(q))$ while its expansion is
\begin{equation} \hat{p}\hat{\oplus} \hat{q} =
\hat{p}_\mu + \hat{q}_\mu - \frac{1}{M_{\text{Pl}}}\,
\tilde{\Gamma}_\mu{}^{\{\alpha\beta\}}\, \hat{p}_\alpha\,
\hat{q}_\beta+ \cdots \,.
\end{equation}
 where the bracket denotes
antisymmetrization. Since only the torsion component at $p_{\mu}=0$
enters at first order one obtains the desired result. If the
connection is metrical, that is if $\nabla^{\mu}g^{\alpha\beta}=0$
then the Riemann and connection normal coordinates agree. If the
connection is non metrical the metric geodesics and connection
geodesics no longer agree. This result has phenomenological consequences worth investigating.

\subsection{A model of macroscopic bodies in collision}

     Considering an idealized set up, similar to the previous section,  ``$A$" and ``$B$" each composed of $N$ atoms. One assumes that in the course of their interaction the bodies exchange
photons. Denoting the photon's momentum by $k_\mu$ and the initial
and final momentum of the atom by $p_\mu$ and $\tilde p_\mu$,
respectively one can easily finds that for the photon emission process 
\begin{equation}\label{5}
    p_\mu = ({\tilde p} \oplus k)_\mu\,,
\end{equation}
while for photon absorption
\begin{equation}\label{6}
    (k \oplus p)_\mu = {\tilde p}_\mu\,.
\end{equation}
When one takes a closer look at the process of a single photon exchange between
the body $A$ and $B$. Assuming that the body $A$ emits the photon,
and body $B$ absorbs it one finds the relations
\begin{equation}
p_{\mu}^{A} = ({\tilde p}^A \oplus k)_\mu\, , \quad (k \oplus
p^{B})_\mu = {\tilde p}_{\mu}^{B}\, .
\end{equation}
Solving these two equations for $k$, the following
relation is recovered
\begin{equation}\label{7}
    [(\ominus {\tilde p}^A \oplus p^A) \oplus p^{B}]_\mu = {\tilde p}_{\mu}^{B}\,,
\end{equation}
where  $\ominus p$  is defined by
$(\ominus p) \oplus p =0$.  To the leading order, $\ominus p$ is given by
\begin{equation}
(\ominus p)_\mu = - p_\mu -\frac{1}{M_{\text{Pl}}}\,
\tilde{\Gamma}_\mu{}^{\alpha\beta}\, p_\alpha\, p_\beta+ \ldots\,.
\end{equation}
Eq.\eqref{7} describes the momentum dispersion relation of a single
interaction (emission and absorption) process. The composition law problem 
arises due to the fact that the same form of the dispersion relation would
hold  for macroscopic, massive bodies with initial and final total
momenta $P^{A,B}_\mu$ and $\tilde P^{A,B}_\mu$,
i.e. 
\begin{equation}\label{8}
    [(\ominus {\tilde P}^A \oplus P^A) \oplus P^{B}]_\mu = {\tilde
    P}_{\mu}^{B}\, .
\end{equation}

\subsection{The definition of the total momentum of a body}

   A  similar problem arises if one tries to find the total momentum of a composite system
\begin{equation}\label{11}
    \mathbf{P}^{A} \equiv p^{A_{1}} \oplus (p^{A_{2}}\oplus (\cdots \oplus p^{A_{N}})\cdots)\,.
\end{equation}
 In addition,
let one assume for simplicity that all the momenta of microscopic
constituents are identical. This  is a  simplistic model of a macroscopic
body, which lets one capture the main features of the composition law problem.
 Then one can choose the coordinates such that, $ \mathbf{P}^{A}_\mu$ equals just
$Np_{\mu}^{A}$. One can then easily sum up expressions Eq.(\ref{11}) over
$a$ to obtain
\begin{equation}\label{12}
    [ \mathbf{P}^{A} + \mathbf{P}^{B}]_{\mu} - [ {\tilde{\mathbf{P}}}^{A} +{\tilde{\mathbf{P}}}^{B}]_\mu =
\frac{N\,\tilde{\Gamma}_\mu^{[\alpha \beta]}}{ M_{\text{Pl}}}
  \left(\mathbf{P}_{\alpha}^{A} \mathbf{P}_{\beta}^{B}- \tilde{\mathbf{P}}_{\alpha}^{A} \tilde{\mathbf{P}}_{\beta}^{B}\right)
\end{equation}
It is easy to see that the non-linearities of the model grow with the number of parts considered.
\section{Summary}
     This appendix shows that deviations from the quadratic form of the dispersion relation leads to the non-linear addition of both momentum and energy. This is called the composition law problem, also known as the soccer ball problem. As shown in the thesis, the composition law problem can be avoided by the use of RGUP.

\chapter{Irreducible representations of the RGUP modified Poincar\'e group}\label{app:Poincare}
\section{Poincare group representations} 
     As one knows the elementary particles are classified by the irreducible representations of the Poincar\'e group \cite{Das2008QFT}. Additionally it is well known that there is a spin 2 tensor field representation which is the best candidate for a graviton particle \cite{Arcos:2011nz}. Prompting one to explore the irreducible representations of the RGUP modified Poincar\'e group.
     
One begins by defining the group for the auxiliary or low energy position $x_{0\,\mu}$ and momentum $p_{0\,\mu}$
\begin{align}
p_0^{\mu} =  -i\hbar\frac{\partial}{\partial x_{0\,\mu}}, \quad
[x_0^{\mu},p_0^{\nu}] =  i\hbar\eta^{\mu\nu}.
\end{align}
The auxiliary Lorentz generators are defined as 
\begin{equation}
\tilde{M}^{\mu\nu}=p_0^{\mu}x_0^{\nu}-p_0^{\nu}x_0^{\mu}\,.
\end{equation}
The Poincar\'e group is then defined by the following algebra 
\begin{subequations}
\begin{align}
[x_0^\mu,\tilde{M}^{\nu\rho}]&=i\hbar\left(x_0^{\nu}\delta^{\mu\rho}-x_0^{\rho}\delta^{\mu\nu}\right)\\
[p_0^\mu,\tilde{M}^{\nu\rho}]&=i\hbar\left(p_0^{\nu}\delta^{\mu\rho}-p_0^{\rho}\delta^{\mu\nu}\right)\\
 [\tilde{M}^{\mu\nu},\tilde{M}^{\rho\sigma}]&=i\hbar\left(\eta^{\mu\rho}\tilde{M}^{\nu\sigma}-\eta^{\mu\sigma}\tilde{M}^{\nu\rho}-\eta^{\nu\rho}\tilde{M}^{\mu\sigma}+\eta^{\nu\sigma}\tilde{M}^{\mu\rho}\right)\,.
\end{align}
\end{subequations}
$p_{0\mu}p_0^{\mu}$ is a Casimir operator and the dispersion relation takes the usual quadratic form.
One can define the auxiliary rotations $\tilde{J}_i$ and boosts $\tilde{K}_i$ as 
\begin{equation}
\tilde{J}_i=\frac{1}{2}\varepsilon_{imn}\tilde{M}^{mn}~\text{and}~\tilde{K}_i=\tilde{M}_{0i}\,,
\end{equation}
where $\varepsilon_{imn}$ is the Levi-Civita completely anti-symmetric tensor. The algebra for them is given by 
 \begin{subequations}
\begin{align}
[\tilde{J}_i,\tilde{J}_j]&=-i\varepsilon_{ijk}\tilde{J}^k\,, \\
[\tilde{K}_i,\tilde{K}_j]&=i\varepsilon_{ijk}\tilde{J}^k \,,\\
[\tilde{J}_i,\tilde{K}_j]&=i\varepsilon_{ijk}\tilde{K}^k\,.
\end{align} 
\end{subequations}
An observation can be made that the algebra for the auxiliary boosts and rotations is entangled. Therefore, one can define a new set of operators
\begin{equation}
\tilde{A}_i=\frac{1}{2}\left(\tilde{J}_i+i\tilde{K}_i\right)~\text{and}~\tilde{B}_i=\frac{1}{2}\left(\tilde{J}_i-i\tilde{K}_i\right)\,,
\end{equation}
The algebra expressed in them is given by the following commutation relations
 \begin{subequations}
\begin{align}
[\tilde{A}_i,\tilde{A}_j]&=-i\varepsilon_{ijk}\tilde{A}^k\,, \\
[\tilde{B}_i,\tilde{B}_j]&=-i\varepsilon_{ijk}\tilde{B}^k \,,\\
[\tilde{A}_i,\tilde{B}_j]&=0\,.
\end{align} 
\end{subequations}
As one knows, this algebra is equivalent to $SU(2)\otimes SU(2)$ algebra. The irreducible representation of which are 
\begin{itemize}
    \item $(0,0)$ scalar field\\
    \item $(1/2,0)$ and $(0,1/2)$ fermion/spinor field\\
    \item $(1/2,1/2)$ vector boson field (photon) \\
    \item $(1,1)$ spin 2 tensor field.
\end{itemize}
The last one is a candidate for the graviton particle.

A more interesting case is the the RGUP modified Poincar\'e algebra.
The physical position $x_\mu$ and momentum $p_\mu$ are functions of the auxiliary ones represented as 
 \begin{subequations}
\begin{align}
x^{\mu}&=x_0^{\mu} \,,\\
p^{\mu}&=p_0^\mu\,(1+\gamma p_0^{\rho}p_{0\rho})\,,
\end{align} 
\end{subequations}
while the Lorentz generators are defined as 
\begin{equation}
    M^{\mu\nu} = p^{\mu}x^{\nu}-p^{\nu}x^{\mu}
    = \left[1+\gamma p_0^{\rho}p_{0\,\rho}\right]\tilde{M}^{\mu\nu}\,,
\end{equation}
where it is worth mentioning that all the results are truncated to first order in the RGUP parameter $\gamma$.
The modified Poincar\'e algebra is then calculated to be %
\begin{subequations}
\begin{align}
  [x^\mu,M^{\nu\rho}] &=  i\hbar[1 +  \gamma p^{\rho} p_{\,\rho}]\left(x^{\nu}\delta^{\mu\rho}-x^{\rho}\delta^{\mu\nu}\right) + i\hbar 2\gamma p^{\mu} M^{\nu\rho}\,,\\
  [p^\mu, M^{\nu\rho}]& =  i\hbar[1 +  \gamma p^{\rho} p_{\,\rho}]\left(p^{\nu}\delta^{\mu\rho}-p^{\rho}\delta^{\mu\nu}\right),\\
 [M^{\mu\nu},M^{\rho\sigma}]& = i\hbar\left(1 +\gamma p^{\rho} p_{\,\rho}\right)\left(\eta^{\mu\rho}M^{\nu\sigma}-\eta^{\mu\sigma} M^{\nu\rho} 
  -\eta^{\nu\rho}M^{\mu\sigma}+\eta^{\nu\sigma}M^{\mu\rho}\right)\,.
\end{align}
\end{subequations}
One can show that the physical rotations $J_i$ and boosts $K_i$ are as follows 
\begin{subequations}
\begin{align}
J_i&=\frac{1}{2}\varepsilon_{imn}M^{mn}=\frac{1}{2}\left(1 +\gamma p^{\rho} p_{\,\rho}\right)\varepsilon_{imn}\tilde{M}^{mn}\,,\\
K_i&=M_{0i}=\left(1 +\gamma p^{\rho} p_{\,\rho}\right)\tilde{M}_{0i}\,.
\end{align}
\end{subequations}
The algebra for the physical rotations $J_i$ and boosts $K_i$ is given by the commutators
 \begin{subequations}
\begin{align}
[J_i,J_j]&=-i\varepsilon_{ijk}\left(1 +\gamma p^{\rho} p_{\,\rho}\right)J^k\,, \\
[K_i,K_j]&=i\varepsilon_{ijk}\left(1 +\gamma p^{\rho} p_{\,\rho}\right)J^k \,,\\
[J_i,K_j]&=i\varepsilon_{ijk}\left(1 +\gamma p^{\rho} p_{\,\rho}\right)K^k\,.
\end{align} 
\end{subequations}
Once again a new set of this time physical operators $A_i$ and $B_i$ is defined
 \begin{subequations}
\begin{align}
A_i=\frac{1}{2}\left(J_i+i K_i\right)=\frac{1}{2}\left(1 +\gamma p^{\rho} p_{\,\rho}\right)\left(\tilde{J}_i+i\tilde{K}_i\right)\,,\\
B_i=\frac{1}{2}\left(J_i-i K_i\right)=\frac{1}{2}\left(1 +\gamma p^{\rho} p_{\,\rho}\right)\left(\tilde{J}_i-i\tilde{K}_i\right)\,.
\end{align} 
\end{subequations}
The algebra formed by those operators has the following form
 \begin{subequations}
\begin{align}
[A_i,A_j]&=-i\varepsilon_{ijk}\left(1 +\gamma p^{\rho} p_{\,\rho}\right)A^k\,, \\
[B_i,B_j]&=-i\varepsilon_{ijk}\left(1 +\gamma p^{\rho} p_{\,\rho}\right)B^k \,,\\
[A_i,B_j]&=0\,.
\end{align} 
\end{subequations}
An important result is the fact that once again this algebra is equivalent to $SU(2)\otimes SU(2)$ algebra. The irreducible representation of which are 
\begin{itemize}
    \item $(0,0)$ scalar field\\
    \item $(1/2,0)$ and $(0,1/2)$ fermion/spinor field\\
    \item $(1/2,1/2)$ vector boson field (photon) \\
    \item $(1,1)$ spin 2 tensor field.
\end{itemize}
One  has to take two things in consideration. First the modified Poincar\'e algebra has both the auxiliary momentum squared $p_{0\,\mu}p_0^\mu$ and the physical momentum squared  $p_{\mu}p^\mu$ as Casimir invariants. And second the modified Poincar\'e algebra has the same irreducible representations as the unmodified case. Therefore it describes the same particles.  

\section{Summary}
     Here, one can see that the scalar, spinor and gauge fields are classified by the irreducible representation of the modified Poincar\'e group. This gives strong evidence that the RGUP modified KG and Dirac equations are the equations of motion of the scalar and spinor fields. Furthermore, it provides justification of the use of KG-like modification for the equation of motion for the gauge field. 
\chapter{Minimum length modified Lagrangians}
\section{Obtaining the Scalar Field Lagrangian} \label{app:Lagrangian}

\noindent Following the Ostrogradsky method for higher derivative Lagrangians presented in \cite{Pons:1988tj,Woodard:2015zca,deUrries:1998obu}, Eq.\eqref{Eq:DIfferentialKGEoM} is obtained by applying the Euler-Lagrange equations to the most general form of the Lagrangian
\begin{align}\label{L}
   \nonumber \mathcal{L}=\frac{1}{2}\partial_{\mu}\phi\partial^{\mu}\phi&+\gamma \left( C_1\,\partial_\mu\partial^\mu\phi\,\partial_\nu\partial^\nu\phi+C_2\,\partial_\mu\phi\,\partial^\mu\partial_\nu\partial^\nu\phi+C_3\,\partial_\nu\partial^\nu\partial^\mu\phi\,\partial_\mu\phi\right)\\\nonumber&+\gamma^2\left(C_4\, \partial_\mu\partial^\mu\partial_\nu\phi\,\partial^\nu\partial_\rho\partial^\rho\phi+C_5\, \partial_\mu\partial^\mu\partial_\nu\partial^\nu\phi\,\partial_\rho\partial^\rho\phi+C_6\, \partial_\mu\partial^\mu\partial_\nu\partial^\nu\partial_\rho\phi\,\partial^\rho\phi\right.\\
   &\left.+C_7\, \partial_\mu\partial^\mu\phi\,\partial_\nu\partial^\nu\partial_\rho\partial^\rho\phi+C_8\, \partial_\mu\phi\,\partial^\mu\partial_\nu\partial^\nu\partial_\rho\partial^\rho\phi\right)+C_9m^2\phi^2\,,
\end{align}
which has up to fourth order derivatives in order to match the number of derivatives in the Equations of motion.
The application of the Ostrogradsky method to  Eq.\eqref{L}, should  give rise to the equations of motion in Eq.\eqref{Eq:DIfferentialKGEoM}.
According to the Ostrogradsky method, the Euler-Lagrange equations for theories with higher derivatives has the form:
\begin{equation}
    \frac{dL}{dq} -\frac{d}{dt}\frac{dL}{d\dot{q}}+\frac{d^2}{dt^2}\frac{dL}{d\ddot{q}}+\ldots+(-1)^n\frac{d^n}{dt^n}\frac{dL}{d (d^nq/dt^n)}=0\,,
\end{equation}
which in the case of fields is
\begin{equation}
    \frac{\partial\mathcal{L}}{\partial\phi}-  \partial_\mu  \frac{\partial\mathcal{L}}{\partial(\partial_\mu\phi)}+   \partial_{\mu_1} \partial_{\mu_2}\frac{\partial\mathcal{L}}{\partial(\partial_{\mu_1} \partial_{\mu_2}\phi)}+\ldots
    +(-1)^m\partial_{\mu_1}\ldots\partial_{\mu_m}\frac{\partial\mathcal{L}}{\partial(\partial_{\mu_1} \ldots\partial_{\mu_m}\phi)}=0\,.
\end{equation}
One can now calculate the Euler-Lagrange equations for the Lagrangian Eq.\eqref{L}
\begin{align}
\nonumber& 2C_9 m^2c^2\phi-\partial_\mu\partial^\mu \phi +2\gamma C_2 \partial_\mu\partial^\mu\partial_\nu\partial^\nu\phi+2\gamma C_3 \partial_\mu\partial^\mu\partial_\nu\partial^\nu\phi-\gamma^2C_6\partial_\mu\partial^\mu\partial_\nu\partial^\nu\partial_\rho\partial^\rho\phi\\
\nonumber&-\gamma^2C_8\partial_\mu\partial^\mu\partial_\nu\partial^\nu\partial_\rho\partial^\rho\phi+4\gamma C_1\partial_\mu\partial^\mu\partial_\nu\partial^\nu\phi +\gamma^2C_5\partial_\mu\partial^\mu\partial_\nu\partial^\nu\partial_\rho\partial^\rho\phi+\gamma^2C_7\partial_\mu\partial^\mu\partial_\nu\partial^\nu\partial_\rho\partial^\rho\phi\\\nonumber&
+2\gamma C_2 \partial_\mu\partial^\mu\partial_\nu\partial^\nu\phi+2\gamma C_3\partial_\mu\partial^\mu\partial_\nu\partial^\nu-\gamma^2C_4\partial_\mu\partial^\mu\partial_\nu\partial^\nu\partial_\rho\partial^\rho\phi+\gamma^2C_5\partial_\mu\partial^\mu\partial_\nu\partial^\nu\partial_\rho\partial^\rho\phi\\ &+\gamma^2C_7\partial_\mu\partial^\mu\partial_\nu\partial^\nu\partial_\rho\partial^\rho\phi-\gamma^2C_7\partial_\mu\partial^\mu\partial_\nu\partial^\nu\partial_\rho\partial^\rho\phi-\gamma^2C_8\partial_\mu\partial^\mu\partial_\nu\partial^\nu\partial_\rho\partial^\rho\phi=0\,.
\end{align}
The expression above simplifies to
\begin{align}\nonumber
    2C_9m^2\phi&+\partial_\mu\partial^\mu\phi-4\gamma (C_2+C_3+C_1)\partial_\mu\partial^\mu\partial_\nu\partial^\nu\phi\\&+\gamma^2(C_6+2C_8-2C_5-C_7+C_4)\partial_\mu\partial^\mu\partial_\nu\partial^\nu\partial_\rho\partial^\rho=0\,.\label{SimplEOM}
\end{align}
One then compares Eq.\eqref{SimplEOM} to the Eq.\eqref{Eq:DIfferentialKGEoM}. The following relationship for the numerical constants
\begin{align}
    &C_9=\frac{1}{2}\\
    &C_1+C_2+C_3=\frac{1}{2}\\
    &C_6+2C_4-C_7+2C_8-2C_5=1
    \end{align}
The resulting  Lagrangian is then simplified by removing surface terms, as shown below
\begin{align}
    C_1\,\partial_\mu\partial^\mu\phi\,\partial_\nu\partial^\nu\phi&=\underbrace{C_1 \partial_\mu\left(\partial^\mu\phi\,\partial_\nu\partial^\nu\phi\right)}_{surface}-C_1\partial^\mu\phi\,\partial_\mu\partial_\nu\partial^\nu\phi\\
  \nonumber  C_4\, \partial_\mu\partial^\mu\partial_\nu\phi\,\partial^\nu\partial_\rho\partial^\rho\phi&=C_4\partial_\nu\left(\partial_\mu\partial^\mu\phi\partial^\nu\partial_\rho\partial^\rho\phi\right)-C_4\partial_\mu\partial^\mu\phi\partial_\nu\partial^\nu\partial_\rho\partial^\rho\phi\\&=\underbrace{C_4\partial_\nu\left(\partial_\mu\partial^\mu\phi\partial^\nu\partial_\rho\partial^\rho\phi\right)}_{surface}\nonumber\\
   &-\underbrace{C_4\partial_\mu\left(\partial^\mu\phi\partial_\nu\partial^\nu\partial_\rho\partial^\rho\phi\right)}_{surface}+C_4\partial^\mu\phi\,\partial_\mu\partial_\nu\partial^\nu\partial_\rho\partial^\rho\phi\,.
\end{align}
The process  was repeated for all the terms. Additionally, one uses the fact that for scalar fields $[\phi,\phi]=0$, in other words the field is commutative. Therefore ,
\begin{equation}
\partial^\mu\phi\,\partial_\mu\partial_\nu\partial^\nu\phi=\partial_\mu\partial_\nu\partial^\nu\phi\,\partial^\mu\phi\,.
\end{equation}
Applying the above equation to the Lagrangian allows for the combination of the constants $C_1$, $C_2$, and $C_3$ terms into one $C$.  In addition the $C_4\,,\ldots\,C_8$ can also be combined into one $D$.  The resulting Lagrangian is 
\begin{equation}
 \mathcal{L}=\frac{1}{2}\partial_{\mu}\phi\partial^{\mu}\phi+C_9m^2\phi^2+\gamma C\,\partial_\nu\partial^\nu\partial^\mu\phi\,\partial_\mu\phi+\gamma^2D\partial_\mu\phi\,\partial^\mu\partial_\nu\partial^\nu\partial_\rho\partial^\rho\phi\,.
\end{equation}
One can observe that the result is a higher derivative Lagrangian, therefore  the Ostrogradsky method to calculate the Equation of Motion should be used
\begin{equation}
     \frac{\partial\mathcal{L}}{\partial\phi}-  \partial_\mu  \frac{\partial\mathcal{L}}{\partial(\partial_\mu\phi)}-\partial_\mu \partial_\nu\partial^\nu \frac{\partial\mathcal{L}}{\partial(\partial_\mu\partial_\nu\partial^\nu\phi)}-\partial_\mu \partial_\nu\partial^\nu\partial_\rho\partial^\rho \frac{\partial\mathcal{L}}{\partial(\partial_\mu \partial_\nu\partial^\nu\partial_\rho\partial^\rho)}=0\,.
\end{equation}
Repeating the process once again in order to fix the new coefficients $\{C,D,C_9\}$. One derives the Equations of motion once again
\begin{align}
  \nonumber  C_9m^2\phi-\partial_\mu\partial^\mu\phi-C\gamma \partial_\mu\partial^\mu\partial_\nu\partial^\nu\phi-C\gamma \partial_\mu\partial^\mu\partial_\nu\partial^\nu\phi\qquad\qquad&\\-D\gamma^2\gamma \partial_\mu\partial^\mu\partial_\nu\partial^\nu\partial_\rho\partial^\rho\phi-D\gamma^2\gamma \partial_\mu\partial^\mu\partial_\nu\partial^\nu\partial_\rho\partial^\rho\phi&=0\\
   C_9m^2\phi-\partial_\mu\partial^\mu\phi-2C\gamma \partial_\mu\partial^\mu\partial_\nu\partial^\nu\phi-2D\gamma^2\gamma \partial_\mu\partial^\mu\partial_\nu\partial^\nu\partial_\rho\partial^\rho\phi&=0\,.
\end{align}
Comparing the above to Eq.\eqref{Eq:DIfferentialKGEoM} the parameters are fixed to be
\begin{align}
    C_9&=-\frac{1}{2}\,,\\
    C&=-1\,,\\
    D&=\frac{1}{2}\,.
\end{align}
One can easily check that the coefficients are uniquely defined.

\section{\label{app:lagrangian}Obtaining the Dirac Field Lagrangian}

 To obtain the Lagrangian shown in Eq.\eqref{Eq:DirackLagrangian}, the differential form of the Dirac equation is needed.
One substitutes the physical momentum in terms of the auxiliary one
\begin{equation}
p_\mu=p_{0\,\mu}(1+\gamma p_{0\rho}p_0^\rho)\,,
\end{equation}
in Dirac equation  Eq.\eqref{Eq:DifferentialDiracEoM}.
One gets the following expression
\begin{equation}
    \left[\tau^{\mu}p_{0\,\mu}(1+\gamma p_{0\rho}p_0^\rho)-m\right]\psi=0\,.
\end{equation}
Because $p_0$ and $x_0$ are canonically conjugated to each other, one can write the differential form of the Dirac equation as
\begin{equation}\label{app:DiffDirac}
    \left[ i\tau^\mu\partial_\mu(1+\gamma\partial_\rho\partial^\rho) -m\right]\psi=0\,.
\end{equation}
Above, one can recognise  Euler-Lagrange equation for this theory.
Therefore, the Lagrangian can be recovered by  applying  the Ostrogradsky method \cite{deUrries:1998obu,Woodard:2015zca,Pons:1988tj} to obtain the Euler-Lagrange equations for theories with higher derivatives
\begin{equation}
    \frac{dL}{dq} -\frac{d}{dt}\frac{dL}{d\dot{q}}+\frac{d^2}{dt^2}\frac{dL}{d\ddot{q}}+\ldots+(-1)^n\frac{d^n}{dt^n}\frac{dL}{d (d^nq/dt^n)}=0\,,
\end{equation}
which in the case of fields is
\begin{equation}
    \frac{\partial\mathcal{L}}{\partial\phi}-  \partial_\mu  \frac{\partial\mathcal{L}}{\partial(\partial_\mu\phi)}+   \partial_{\mu_1} \partial_{\mu_2}\frac{\partial\mathcal{L}}{\partial(\partial_{\mu_1} \partial_{\mu_2}\phi)}+\ldots
    +(-1)^m\partial_{\mu_1}\ldots\partial_{\mu_m}\frac{\partial\mathcal{L}}{\partial(\partial_{\mu_1} \ldots\partial_{\mu_m}\phi)}=0\,.
\end{equation}
The first step is to assume a general form of the Lagrangian, where the order of derivatives is determined by the order of the equation of motion Eq.\eqref{app:DiffDirac}.
Moreover, an  assumption is made that different terms will have an arbitrary numerical coefficients multiplying every term
\begin{equation}\label{generalL}
     \mathcal{L}_{\psi}=\bar\psi\left[ iC_1\tau^\mu\partial_\mu(1+C_2\gamma\partial_\rho\partial^\rho) -C_3 m\right]\psi\,.
 \end{equation}
Next step is to prove that it has Eq.\eqref{app:DiffDirac} as an equation of motion.
Applying the Ostrogratsky method, one gets the following equations of motion for the field and its complex conjugated
\begin{align}
     &C_1i\tau^\mu\partial_\mu\psi +C_1C_2\gamma\partial_\rho\partial^\rho\psi -C_3 m\psi=0,\\
     &C_1i\tau^\mu\partial_\mu\bar\psi+C_1C_2\gamma\partial_\rho\partial^\rho\bar\psi -C_3 m\bar\psi=0\,.
\end{align}
The equations of motion obtained through the Ostrogradsky method from Eq.\eqref{Eq:DirackLagrangian} need to be identical to Eq.\eqref{app:DiffDirac}, which was obtained through different means.  Therefore, the Lagrangian corresponding to the QFT spinor with minimum length will be of the form presented in Eq.\eqref{Eq:DirackLagrangian}.
This can be used to fix the value of the arbitrary coefficients. The result is an unique set of coefficients 
\begin{equation}
    C_1=C_2=C_3=1\,.
\end{equation}
Therefore, the Lagrangian corresponding to the QFT spinor with minimum length will be of the form presented in Eq.\eqref{Eq:DirackLagrangian}.
\section{Summary}
\noindent The Lagrangians for the charged scalar and Dirac fields are obtained from the RGUP modified Equations of Motion Eq.\eqref{Eq:DIfferentialKGEoM} and \eqref{Eq:DifferentialDiracEoM}. This was achieved through the use of the Ostrogradsky method for finding the Equations of Motion of higher than second order Lagrangians. Additionally, it is shown that that the numerical coefficients which fix the effective Lagrangian are uniquely defined by the Equation of Motion considered.
\chapter{Dirac Equations}
       In the following appendix, the solutions of the RGUP modified Dirac Equations are considered in order to find the form of the spinor field obeying the modified dispersion relation. A solution of the equations of motion is needed for the calculations of the scattering amplitude. 
\section{\label{app:DiracSolutions}Dirac equation solutions}

       In terms of the physical and the auxiliary momentum from the dispersion relation the Klein-Gordon (KG) equation reads as follows
\begin{equation}
     p^{\rho}p_{\rho}=  p_0^{\rho}p_{0\rho}(1+2\gamma p_0^{\sigma}p_{0\sigma}
    )=-(mc)^2 \,.
 \end{equation}
Since the KG equation written in terms of the physical momentum is unchanged, one can reasonably assume that the Dirac equation will have the same form 
  \begin{equation}\label{app:dirac}
   E\psi =\left(\vec\alpha \cdot \vec{p}+ \beta m\right)\psi\,,
 \end{equation}
where $E$ is the zeroth component and $\vec{p}$ is the spatial part of the physical momentum $p^\mu$. When one assumes that the Kline-Gordon equation is obtained upon squaring Eq.\eqref{app:dirac}. It can be shown that one has the following properties for $\vec\alpha$ and $\beta$. 

In fact,  one obtains
 \begin{equation}
   E^2\psi=\left(\vec\alpha \cdot \vec{p}+ \beta m\right)\left(\vec\alpha \cdot \vec{p}+ \beta m\right)\psi\,.
 \end{equation}
 Expanding the above one gets 
\begin{equation}
  E^2\psi=\left(\underbrace{\vec\alpha_i^2}_{1} \vec{p}_i^2+ \underbrace{\left(\vec\alpha_i\vec\alpha_j+\vec\alpha_j\vec\alpha_i\right)}_{0}\vec{p}_i\vec{p}_j+\underbrace{\left(\vec\alpha_i\beta+\beta\vec\alpha_i\right)}_{0}\vec{p}_i m+\underbrace{\beta^2}_{1} m^2\right)\psi\,.
\end{equation}
In order this equation to be equivalent to the KG equation one needs to put the following conditions for $\vec\alpha$ and $\beta$
 \begin{align}
    &\left\{\alpha_i,\,\alpha_j\right\}=\left\{\alpha_i,\,\beta\right\}=\left\{\beta,\,\beta\right\}=0\\
     &\alpha_i^2=\beta^2=1\,.
 \end{align}
 One can recall that these are the properties of Dirac matrices. In fact one can recover a representation of the Dirac matrices form $\vec\alpha$ and $\beta$ as follows 
 \begin{equation}
   \tau^\mu\equiv \left(\beta,\beta\vec\alpha \right)\,,
 \end{equation}
which in terms of the Pauli matrices has the form presented in Eq.\eqref{gamma}.

Now, by multiplying both sides of  Eq.\eqref{app:dirac} by $\beta$ and rearranging one can write 
\begin{equation}\label{app:PhysDirac}
    \left(\tau^\mu p_\mu-m\right)\psi =0\,.
 \end{equation}
One can show that 
\begin{equation}
H u=\left(\vec\alpha \cdot  \vec{p}-\beta m\right)u=E u
\end{equation}
In terms of linear operators, the Dirac equation can be written in the following form 
\begin{equation}
Hu=\left(\begin{array}{cc}
 m    & \vec\sigma\cdot\vec{p} \\
\vec\sigma\cdot\vec{p}     & -m
\end{array}\right)\left(\begin{array}{c}
     u_A  \\
     u_B
\end{array}\right)=E\left(\begin{array}{c}
     u_A  \\
     u_B
\end{array}\right)\,.\end{equation}
The above equation reduces to 
\begin{align}
    \vec\sigma\cdot\vec{p}\,u_B&=\left(E-m\right)u_A\,,\\
    \vec\sigma\cdot\vec{p}\,u_A&=\left(E+m\right)u_B\,.
\end{align}
One can easily prove that Eqs.\eqref{freefieldsolutions}
\begin{subequations}
\begin{align}
    u^{(s)}\left(p\right)=N\begin{pmatrix}
\chi^{(s)}\\
\frac{\sigma \cdot p}{E+m}\chi^{(s)}
\end{pmatrix}, E>0\\
    u^{(s+2)}\left(p\right)=N\begin{pmatrix}
\frac{-\sigma \cdot p}{E+m}\chi^{(s+2)}\\
\chi^{(s+2)}
\end{pmatrix}, E<0\,.
\end{align}
\end{subequations}
are the respective solutions for a free Dirac field. Therefore,  they can be used when calculating the transition amplitude. Note that $\vec{p}$ and $E$ are the physical momentum and energy, each of which can be expressed as a function of the auxiliary variables $\vec{p}_0$ and $E_0$.
\section{Summary}
       In the appendix presented above it was shown that two of the solutions of the RGUP modified Dirac equation coincide with the usual solutions with a modified dispersion relation. It is worthy of note that the solutions presented above are not the only solutions of the third order differential equation Eq.\eqref{Eq:DiracEoM}. However, for the purposes of the perturbative calculations of the scattering amplitudes, a single solution for the Dirac field is needed and the one found here is sufficient. 
\chapter{Ostrogradsky method and Vacuum Instability}\label{app:Ostr}
The Ostrogradsky method is a mathematical constructions allowing one to work with higher derivative Lagrangians \cite{Pons:1988tj,Woodard:2015zca,deUrries:1998obu}. One of the major issues with it, is the instability of the Hamiltonian, arising when one considers higher than second order derivative. The instability for the particular case presented in this thesis is discussed in the following appendix. 
\section{Review of Ostrogradsky method and its instability}

      One begins assuming the most general expression for the Lagrangian and the Lagrangian density
\begin{subequations}
\begin{align}
  \label{lagrangian}  L=&L(\phi,\dot\phi,\ddot\phi,\ldots,\overset{(n)}{\phi})\,,\\
  \label{density}  \mathcal{L}=&\mathcal{L}(\phi,\partial_{\mu_1}\phi,\partial_{\mu_1}\partial_{\mu_2}\phi,\ldots,\partial_{\mu_1}\ldots\partial_{\mu_n}\phi)
\end{align}
\end{subequations}
Equations of motion for high derivatives theories are obtained using the generalization of the Euler-Lagrange equations
\begin{subequations}
\begin{equation}
    \frac{dL}{dq} -\frac{d}{dt}\frac{dL}{d\dot{q}}+\frac{d^2}{dt^2}\frac{dL}{d\ddot{q}}+\ldots+(-1)^n\frac{d^n}{dt^n}\frac{dL}{d (d^nq/dt^n)}=0\,,
\end{equation}
\begin{equation}
    \frac{\partial\mathcal{L}}{\partial\phi}-  \partial_\mu  \frac{\partial\mathcal{L}}{\partial(\partial_\mu\phi)}+   \partial_{\mu_1} \partial_{\mu_2}\frac{\partial\mathcal{L}}{\partial(\partial_{\mu_1} \partial_{\mu_2}\phi)}+\ldots
    +(-1)^m\partial_{\mu_1}\ldots\partial_{\mu_m}\frac{\partial\mathcal{L}}{\partial(\partial_{\mu_1} \ldots\partial_{\mu_m}\phi)}=0\,.
\end{equation}
\end{subequations}
The generalized coordinates for the Lagrangian in Eq.\eqref{lagrangian} are defined as follows 
\begin{align}
  q_1 &\equiv  \phi\,, \cr
            q_i &\equiv 
     \overset{(i-1)}\phi
      \quad\quad (i=2,...,n)\,.
\end{align}
The generalized momenta corresponding to the coordinates are defined as derivatives to the Lagrangian with respect of the generalized coordinates
\begin{align}
  p_n &\equiv  
      \frac{\partial L}{\partial{\overset{(n)}{\phi}}} \,, \cr
            p_i &\equiv 
      \frac{\partial L}{\partial{\overset{(i)}\phi}}  -\frac{d}{dt}p_{i+1}
      \quad\quad (i=1,...,n-1)\,.
\end{align}
One notices that there is difference in the way generalized momentum is defined for the higher order derivative terms. Mainly the  therms subtraction the total time derivative of the generalized momentum of the term  one order higher. 

One now has all the parts needed to perform the Legendre transformation, which is defined as follows
\begin{equation}
    H[q_i;p_i]=\sum_{i=1}^{m}\; p_i{\bf{\dot \phi}_i} - L[q_i;p_i]\,.
\end{equation}
One can apply similar arguments and calculations to field theory. Beginning from the Lagrangian density Eq.\eqref{density}, defining the generalized coordinates as follows 
\begin{equation}
   \phi\quad\text{and}\quad \phi_{\mu_1,\ldots,\mu_i}\equiv\partial_{\mu_1}\ldots\partial_{\mu_i}\phi \quad\quad (i=1,...,n-1)\,,
\end{equation}
the corresponding generalized momenta are therefore defined as 
\begin{align}
    \pi^{{\mu}_1\cdots{\mu}_n}&\equiv 
   \frac{\partial{\mathcal{L}}}{\partial{\phi}_{{\mu}_1\cdots{\mu}_n}}\,,\cr
    \pi^{{\mu}_1\cdots{\mu}_i}&\equiv 
   \frac{\partial{\cal L}}{\partial{\phi}_{{\mu}_1\cdots{\mu}_i}}
   -\partial_{\mu_{i+1}} \pi^{{\mu}_1\cdots{\mu}_i{\mu_{i+1}}}
    \quad\quad (i=1,...,n-1)\,.
\end{align}
The Hamiltonian corresponding to a theory defined by Eq.\eqref{density} is obtained through the generalized Legendre transformation. 
\begin{align}\label{LegandreHdensity}
   \mathcal{H} [\phi,\phi_\mu,...,{\phi}_{{\mu}_1\cdots{\mu}_{m-1}};
                                   \pi^\mu,...,&\pi^{{\mu}_1\cdots{\mu}_m}]
  =\pi^\mu\phi_\mu +\cdots
   + \pi^{{\mu}_1\cdots{\mu}_{m-1}}{\phi}_{{\mu}_1\cdots{\mu}_{m-1}}\cr
   &+\pi^{{\mu}_1\cdots{\mu}_m}{\bar\phi}_{{\mu}_1\cdots{\mu}_m}
   -{\cal L}[\phi,\phi_{\mu},...,{\bar\phi}_{{\mu}_1\cdots{\mu}_m}]\,.
\end{align}
Due to the nature of Ostrogradsky method, the Hamiltonian and its density have terms linear in momentum for all but the highest order derivative. Which means that the Hamiltonian and its density can not be positively defined. This is called the Ostrogradsky instability and the theory contains a decaying vacuum and an infinite number of ghost fields named Ostrogradsky ghosts.
\section{Ostrogradsky instability in our theory}
\subsection{Real Scalar field}
       One begins from the equation of motion for the real scalar field, obtained from the modified dispersion relation which corresponds to the Casimir operator of the modified Poincar\'e group defined in \cite{Todorinov:2018arx} 
 \begin{equation}\label{ModKG}
     p_0^{\rho}p_{0\rho}(1+2\alpha\gamma^2p_0^{\sigma}p_{0\sigma})=-(mc)^2 \,.
 \end{equation}
  The Lagrangian for a real scalar field is derived by assuming the most general form of the Lagrangian, and applying the Ostrogradsky method to obtain its equations of motion and comparing to Eq.\eqref{ModKG}.
  The form of the Lagrangian is
\begin{equation}\label{RealL}
    \mathcal{L}_{\phi,\mathbb{R}}=\frac{1}{2}\partial_{\mu}\phi\partial^{\mu}\phi-\frac{1}{2}m^2\phi^2+\gamma \,\partial_\nu\partial^\nu\partial^\mu\phi\,\partial_\mu\phi
    \,.
\end{equation}
One can define the generalized coordinates as 
\begin{equation}\label{generalizedfieldcoordinates1}
    q_1\equiv\phi\quad     q_2\equiv\partial_\mu\partial^\mu\phi\,.
\end{equation}
The generalized momenta are then derived as
\begin{subequations}
\begin{align}
     \pi^{{\mu}\rho}_\rho&=
   \frac{\partial\mathcal{L}}{\partial(\partial_\mu\partial_\rho \partial^\rho \phi)}=2\gamma\partial^\mu \phi\,,\\
   \pi^{\mu}&=
   \frac{\partial\mathcal{L}}{\partial(\partial_\mu \phi)}-\partial_\sigma\partial^\sigma\pi^{\mu\rho}_\rho=\partial^\mu \phi\,.
\end{align}
\end{subequations}
The Hamiltonian density is then obtained using Eq.\eqref{LegandreHdensity}
\begin{subequations}
\begin{align}
    \mathcal{H}&=\pi^\mu\partial_\mu\phi+\pi^{\mu\rho}_\rho\partial_\sigma\partial^\sigma\partial_\mu\phi-\frac{1}{2}\pi^\mu\pi_\mu -\pi^{\mu\rho}_\rho\partial^\sigma\partial_\sigma\pi_\mu+\frac{1}{2}m\phi^2\\
    &= \frac{1}{2}\pi^\mu\pi_\mu+\frac{1}{2}m^2\phi^2\,.
\end{align}
\end{subequations}
One can easily see that the above is positive definite. However one can see, also that the correction terms cancel, the Quantum Gravity corrections are hidden in the dispersion relation of $\phi$. The field $\phi$ is solution of Eq.\eqref{ModKG} which has four different solutions all containing  Quantum Gravity corrections. The calculation above is done in the framework of De Donder Weyl Covariant Hamiltonian
Formulation of Field Theory \cite{10.2307/1968645}.

Considering the same definition of generalized coordinates presented in Eq.\eqref{generalizedfieldcoordinates1}, one can derive the field momenta outside of the De Donder Weyl formulation as 
\begin{subequations}
\begin{align}
     \pi^{\rho}_\rho&=
   \frac{\partial\mathcal{L}}{\partial(\partial_\rho \partial^\rho \dot\phi)}=2\gamma\dot\phi\,,\\
   \pi&=
   \frac{\partial\mathcal{L}}{\partial( \dot\phi)}-\partial_\sigma\partial^\sigma\pi^{\rho}_\rho=\dot\phi\,.
\end{align}
\end{subequations}
 Performing the Legendre transformation the Hamiltonian density is obtained as
 \begin{align}\label{realscalarfield}
    \mathcal{H}&=\pi\dot\phi+\pi^{\rho}_\rho\partial_\sigma\partial^\sigma\dot\phi-\frac{1}{2}\pi\pi -\pi^{\rho}_\rho\partial^\sigma\partial_\sigma\pi+\frac{1}{2}m\phi^2+\frac{1}{2}\left(\nabla \phi\right)\cdot\left(\nabla \phi\right)+\gamma\left(\nabla \phi\right)\cdot\left(\partial_\sigma\partial^\sigma\nabla \phi\right)\\
    &= \frac{1}{2}\pi\pi+\frac{1}{2}m\phi^2+\frac{1}{2}\left(\nabla \phi\right)\cdot\left(\nabla \phi\right)+\gamma\left(\nabla \phi\right)\cdot\left(\partial_\sigma\partial^\sigma\nabla \phi\right)\,.
\end{align}
One can not  be sure if he Hamiltonian density is positive definite. This is due to the fact that one is not sure of the sign of the term proportional to $\gamma=\frac{\gamma_0}{(M_{\mathrm{Pl}}\,c)^2}$. If one assumes that $\gamma_0$ is of the order one, the coefficient is $\gamma\sim 10^{-38}$. Therefore, for energies smaller than the Planck energy $E<E_{Pl}$, the leading order terms in  Eq.\eqref{realscalarfield} are several orders of magnitude bigger than the quantum gravity corrections, which means that the Hamiltonian density on the whole is positive definite and does \textit{not} have Ostrogradsky instability. Conclusion about the positive definition of the Hamiltonian for energies bigger than the Planck energy $E>E_{Pl}$ cannot be drawn. Making the QFT presented here an effective theory.

\subsection{Complex scalar field}
      The procedure of obtaining the Lagrangian density for the complex scalar field is similar to the one presented in the previous section. The resulting Lagrangian is
\begin{equation}\label{complexL}
 \mathcal{L}_{\phi,\mathbb{C}} =  \left(\partial_{\mu}\phi\right)^\dagger \partial^{\mu}\phi -  m^2 \phi^\dagger \phi + 2\gamma \left[\left(\partial_\nu \partial^\nu \partial^\mu \phi \right)^\dagger \partial_\mu \phi + \partial_\nu \partial^\nu \partial^\mu \phi \left(\partial_\mu \phi \right)^\dagger \right] \,.
\end{equation}
The generalized coordinates are also similarly defined 
\begin{subequations}\label{generalizedcomplexcoordinates}
\begin{align}
     &q_1\equiv\phi\quad     q_2\equiv\partial_\mu\partial^\mu\phi\,,\\
     &q^\dagger_1\equiv\phi^\dagger\quad     q^\dagger_2\equiv\partial_\mu\partial^\mu\phi^\dagger\,.
\end{align}
\end{subequations}
According to the Ostrogradsky method their corresponding generalized momenta are 
\begin{subequations}
\begin{align}
     \pi^{{\mu}\rho}_\rho&=
   \frac{\partial\mathcal{L}}{\partial(\partial_\mu\partial_\rho \partial^\rho \phi)}=2\gamma\partial^\mu \phi^\dagger \,,\\
   \pi^{\mu}&=
   \frac{\partial\mathcal{L}}{\partial(\partial_\mu \phi)}-\partial_\sigma\partial^\sigma\pi^{\mu\rho}_\rho=\partial^\mu \phi^\dagger\,,\\
     \pi^{{\mu}\rho\dagger}_\rho &=
  \frac{\partial\mathcal{L}}{\partial(\partial_\mu\partial_\rho \partial^\rho \phi)}=2\gamma\partial^\mu \phi\,,\\
   \pi^{\mu\dagger}&=
   \frac{\partial\mathcal{L}}{\partial(\partial_\mu \phi)}-\partial_\sigma\partial^\sigma\pi^{\mu\rho\dagger}_\rho=\partial^\mu \phi\,.
\end{align}
\end{subequations}
Performing the Legendre transformations prescribed by Ostrogradsky, one obtains the Hamiltonian density
\begin{align}
   \nonumber \mathcal{H}&=\pi^\mu\pi_\mu^\dagger+\pi^{\mu\dagger}\pi_\mu+\pi^{\mu\rho}_\rho\partial_\sigma\partial^\sigma\pi_\mu^\dagger+\pi^{\mu\rho\dagger}_\rho\partial_\sigma\partial^\sigma\pi_\mu\\\nonumber&-\pi^\mu\pi_\mu^\dagger-\pi^{\mu\rho}_\rho\partial_\sigma\partial^\sigma\pi_\mu^\dagger-\pi^{\mu\rho\dagger}_\rho\partial_\sigma\partial^\sigma\pi_\mu+m^2\phi\phi^\dagger\\
    &=\pi^\mu\pi_\mu^\dagger+m^2\phi\phi^\dagger
\end{align}
Again one can see that in the framework of De-Donder Weyl Covariant Hamiltonian Formulation of Field Theory the Hamiltonian density is always positively defined. Similarly to the case of the real scalar field, the Quantum Gravity corrections are hidden in the dispersion relation of the field $\phi$ and its complex conjugate $\phi^\dagger$.
Analogously to the previous section, one has the same generalized coordinates Eq.\eqref{generalizedcomplexcoordinates}. Without considering the De-Donder Weyl framework, the generalized momenta are as follows
\begin{subequations}
\begin{align}
     \pi^{\rho}_\rho&=
   \frac{\partial\mathcal{L}}{\partial(\partial_\rho \partial^\rho \dot\phi)}=2\gamma\dot\phi^\dagger \,,\\
   \pi&=
   \frac{\partial\mathcal{L}}{\partial( \phi)}-\partial_\sigma\partial^\sigma\pi^{\rho}_\rho=\dot\phi^\dagger\,,\\
     \pi^{\rho\dagger}_\rho &=
   \frac{\partial\mathcal{L}}{\partial(\partial_\rho \partial^\rho \dot\phi)}=2\gamma\dot\phi\,,\\
   \pi^{\dagger}&=
  \frac{\partial\mathcal{L}}{\partial(\dot\phi)}-\partial_\sigma\partial^\sigma\pi^{\rho\dagger}_\rho=\dot\phi\,.
\end{align}
\end{subequations}
Performing the Legendre transformation in this case  yield the following
\begin{align}\label{complexscalarfield}
   \nonumber \mathcal{H}&=2\pi^{\dagger}\pi+\pi^{\rho}_\rho\partial_\sigma\partial^\sigma\pi^\dagger+\pi^{\rho\dagger}_\rho\partial_\sigma\partial^\sigma\pi-\pi \pi^\dagger-\pi^{\rho}_\rho\partial_\sigma\partial^\sigma\pi^\dagger-\pi^{\rho\dagger}_\rho\partial_\sigma\partial^\sigma\pi+m^2\phi\phi^\dagger\\
   \nonumber &+m^2\phi \phi^\dagger+\left(\nabla \phi\right)\cdot\left(\nabla \phi^\dagger\right)+2\gamma\left(\nabla \phi\right)\cdot\left(\partial_\sigma\partial^\sigma\nabla \phi^\dagger\right)+2\gamma\left(\nabla \phi^\dagger\right)\cdot\left(\partial_\sigma\partial^\sigma\nabla \phi\right)\\
    &=\pi \pi^\dagger+m^2\phi \phi^\dagger+\left(\nabla \phi\right)\cdot\left(\nabla \phi^\dagger\right)+2\gamma\left(\nabla \phi\right)\cdot\left(\partial_\sigma\partial^\sigma\nabla \phi^\dagger\right)+2\gamma\left(\nabla \phi^\dagger\right)\cdot\left(\partial_\sigma\partial^\sigma\nabla \phi\right)\,.
\end{align}
Once again the issue of the sign of the Hamiltonian density depends on the terms proportional to $\gamma=\frac{\gamma_0}{(M_{\mathrm{Pl}}\,c)^2}$. Therefore, for energies smaller than the Planck energy $E<E_{Pl}$, the leading order terms in  Eq.\eqref{complexscalarfield} are several orders of magnitude bigger than the quantum gravity corrections, which means that the Hamiltonian density on the whole is positive definite and does \textit{not} have Ostrogradsky instability. 

\subsection{Dirac field}
      To obtain the Lagrangian shown below in Eq.\eqref{Dlag} one needs its equations of motion, \textit{i.e.} the Dirac equation. The Dirac equation can be obtained from Eq.\eqref{ModKG} by using the standard method. The modified Dirac equation has the following form
\begin{equation}\label{Dirac}
    \left[ i\tau^\mu\partial_\mu(1+\gamma\partial_\sigma\partial^\sigma) -m\right]\psi=0\,.
\end{equation}
Therefore, the Lagrangian for a spinor field can be derived by assuming a most general form of the Lagrangian, and applying the Ostrogradsky method to obtain its Equations of Motion and comparing it to the above Eq.\eqref{Dirac}.
  The form of the Lagrangian is
 \begin{equation}\label{Dlag}
      \mathcal{L}_{\psi}=\bar\psi\left[ i\tau^\mu\partial_\mu(1-\gamma\,\partial_\sigma\partial^\sigma) -m\right]\psi\,.
 \end{equation}
  One then defines the generalized coordinates as 
  \begin{equation}\label{generalizedfieldcoordinates}
    q_1\equiv\psi\,,\quad     q_2\equiv\partial_\mu\partial^\mu\psi\,.
\end{equation}
Their corresponding generalized momenta are 
\begin{subequations}
\begin{align}
     \pi^{\rho}_\rho&=
   \frac{\partial\mathcal{L}}{\partial(\partial_\rho \partial^\rho \dot\psi)}=-i\gamma\bar\psi\tau^0 \,, \\
   \pi&=
   \frac{\partial\mathcal{L}}{\partial( \dot\psi)}-\partial_\sigma\partial^\sigma\pi^{\rho}_\rho=i\bar\psi\tau^0\,.
\end{align}
\end{subequations}
One can then use the generalized Legendre transformations to recover the Hamiltonian density
\begin{align}\label{DiracField}
   \nonumber \mathcal{H}&=\pi\dot\psi-\gamma\pi\partial_\sigma\partial^\sigma\dot\psi-\pi\dot\psi+\gamma\pi\partial_\sigma\partial^\sigma\dot\psi+\bar\psi\left[ i\tau^a\nabla_a(1-\gamma\,\partial_\sigma\partial^\sigma) -m\right]\psi\\
    &=\bar\psi\left[ i\tau^a\nabla_a(1-\gamma\,\partial_\sigma\partial^\sigma) -m\right]\psi\,.
\end{align}
Similar to the scalar field cases, the question of the sign of the Hamiltonian density depends on the sign of the terms proportional to $\gamma=\frac{\gamma_0}{(M_{\mathrm{Pl}}\,c)^2}$. One can not draw solid conclusions for the sign of this terms. However, just like previous cases, if the energy of the system is smaller than the Planck energy $E<E_{Pl}$ the correction term is smaller than the leading order making the Hamiltonian density positive definite. Therefore, there is no Ostrogradsky instability.
\subsection{Gauge field} 

       Using the Ostrogradsky method \cite{Pons:1988tj,Woodard:2015zca,deUrries:1998obu}, one can recover the Lagrangian for the modified electrodynamics
\begin{equation}
    \mathcal{L}_{A}=-\frac{1}{4}F^{\mu\nu}F_{\mu\nu}=-\frac{1}{4}F^{\mu\nu  }_0 F_{\mu\nu 0}-\frac{\gamma }{2}F_{\mu\nu 0} \partial_\rho \partial^\rho F^{\mu\nu}_0\,,
\end{equation}
writing the Lagrangian in terms of the gauge field
\begin{equation}\label{AmuLagrangian}
    \mathcal{L}_{A}=-\frac{1}{2}\left[\partial^\mu A^\nu\partial_\mu A_\nu-\partial^\nu A^\mu\partial_\mu A_\nu \right]-\gamma \left[\partial^\mu A^\nu\partial_\sigma\partial^\sigma\partial_\mu A_\nu-\partial^\nu A^\mu\partial_\sigma\partial^\sigma\partial_\mu A_\nu \right]\,.
\end{equation}
The Ostrogradsky method is employed to find the equations of motion in order to make sure that its equations of motion match with the Eq.\eqref{Eq:GaugeFieldEOM}. Observation is made that the Lagrangian contains only first and third derivatives. Therefore only these terms are used in obtaining the equation of motion. Due to freedom of gauge choice, Lorentz gauge $\partial_\mu A^\mu=0$ is chosen and used to obtain the equations of motion. 
\begin{align}
  \nonumber -  \partial_\mu  \frac{\partial\mathcal{L}}{\partial(\partial_\mu A_\nu)}
   -\partial_\sigma\partial^\sigma\partial_\mu\frac{\partial\mathcal{L}}{\partial(\partial_\sigma\partial^\sigma\partial_{\mu} A_\nu)}&=0\,,\\
i.e.,~~  \nonumber -\left[-\frac{1}{2}2 \partial_\mu\partial^\mu A^\nu -\gamma\partial_\sigma\partial^\sigma\partial_{\mu} A^\nu)\right]-\left[\gamma\partial_\sigma\partial^\sigma\left(\partial^\mu A^\nu\right)\right]&=0\,,\\\label{EoM}
i.e.,~~   \partial_\mu\partial^\mu A^\nu +2\gamma\partial_\sigma\partial^\sigma\partial_{\mu}\partial^{\mu} A^\nu&=0\,.
\end{align}
As one can read from the above equation Eq.\eqref{EoM} matches Eq.\eqref{Eq:GaugeFieldEOM}. Therefore, Eq.\eqref{AmuLagrangian} is the Lagrangian of Electromagnetic gauge field that allows for minimum length.

The Ostrogradsky ghosts arise when one has higher derivative theories and the Hamiltonian is unbounded from below. This is due to the fact that only the highest derivative term is quadratic in momentum while all the others are linear. To check if this is the case for the considered theory, one needs to do the Legendre transformation as prescribed by the Ostrogradsky formalism. The $\partial_\mu A_\nu$ and $\partial_\sigma\partial^\sigma\partial_\mu A_\nu$ are chosen to be treated as our generalized coordinates, with their corresponding generalized momentum as follows
\begin{subequations}
\begin{align}
     \pi^{{\nu}\rho}_\rho&=
   \frac{\partial\mathcal{L}}{\partial(\partial_\rho \partial^\rho\partial_0 A^\nu)}=-2\gamma[\partial^0 A^\nu-\partial^\nu A^0]\,,\\
   \pi^{\nu}&=
   \frac{\partial{\mathcal{L}}}{\partial(\partial_0 A^\nu)}-\partial_\sigma\partial^\sigma\pi^{\nu\rho}_\rho= -[\partial^0 A^\nu-\partial^\nu A^0]\,.
\end{align}
\end{subequations}
Then one finds the expression for the Lagrangian density in terms of the generalized position and momentum
\begin{align}
    \mathcal{L}_{A}=&-\frac{1}{2}\left[\partial^\mu A^\nu\partial_\mu A_\nu-\partial^\nu A^\mu\partial_\mu A_\nu \right]-\gamma \left[\partial^\mu A^\nu\partial_\sigma\partial^\sigma\partial_\mu A_\nu-\partial^\nu A^\mu\partial_\sigma\partial^\sigma\partial_\mu A_\nu \right]\\
=&-\frac{1}{2} \pi^{\nu\rho}_{\rho}\partial_\sigma\partial^\sigma\pi_{\nu}-\frac{1}{2}\pi^{\nu}\pi_{\nu}\\&-\frac{1}{2}\left[\nabla^a A^\nu \nabla_a A_\nu-\partial^\nu A^a\nabla_a A_\nu+2\gamma(\nabla^a A^\nu\partial_\sigma\partial^\sigma\nabla_a A_\nu -\partial^\nu A^a\partial_\sigma\partial^\sigma\nabla_a A_\nu)\right]
\end{align}
This case utilizes the Weyl gauge, also known as the Hamiltonian or temporal gauge discussed and used in \cite{Das2008QFT,hatfield2018quantum}
\begin{equation}
    A^0=0\,\quad\, \nabla\cdot\vec A=0 \,.
\end{equation}
 This is also known as an incomplete gauge as it is not manifestly Lorentz covariant
 and it requires longitudinal photons with a constraint on the states.
 The subsequent results are Lorentz covariant. However, this choice of gauge eliminates the negative-norm ghost. This is commonly used in conformal field theories as it sacrifices explicit Lorentz invariance in favour of scale invariance. 
 
Using the Legendre transformation as prescribed by the Ostrogradsky method, the Hamiltonian density for the modified Electrodynamics is obtained as follows
\begin{align}\label{Hdensity}
\nonumber\mathcal{H}=& \pi^{{\nu}\rho}_\rho\partial_\sigma\partial^\sigma\partial_0 A_\nu+\pi^{\nu}\partial_0 A_\nu-\mathcal{L}\\\nonumber
=&\gamma  \pi^{a}\partial_\sigma\partial^\sigma\pi_{a}+\frac{1}{2}\pi^{a}\pi_{a}+\frac{1}{2}\underline{(1+2\gamma)\pi_{\nu}\partial^\nu A^0}\\&+\frac{1}{2}\left[\nabla^a A^\nu \nabla_a A_\nu-\partial^\nu A^a\nabla_a A_\nu+2\gamma(\nabla^a A^\nu\partial_\sigma\partial^\sigma\nabla_a A_\nu -\partial^\nu A^a\partial_\sigma\partial^\sigma\nabla_a A_\nu)\right]
\end{align}
The underlined term  is the only term linear in momentum, however it is equal to zero due to our gauge fix.
One needs to see if  $\gamma  \pi^{a}\partial_\sigma\partial^\sigma\pi_{a}$ is positively defined. Using Eq.\eqref{EoM}
\begin{equation}
 \partial_0\left[    \partial^0 A^\nu +2\gamma\partial_\sigma\partial^\sigma\partial^{0} A^\nu\right]= 0\,.
\end{equation}
This means that the term inside the square brackets is either zero or a constant. Additionally, one can draw the conclusion that $ \partial_\mu\partial^\mu A^\nu$ and the $2\gamma\partial_\sigma\partial^\sigma\partial_{\mu}\partial^{\mu} A^\nu$ have opposite signs. Therefore, on its own the $\gamma  \pi^{a}\partial_\sigma\partial^\sigma\pi_{a}$  changes sign. 

As per the spatial part, in the Weyl gauge the term reads 
\begin{align}
\frac{1}{4}F^{ab}F_{ab}+\frac{\gamma }{2}F^{ab}\Box F_{ab}&\\=
\frac{1}{4}\left(-\varepsilon_{ijk}B^k\right)\left(-\varepsilon_{ijl}B^l\right)+\frac{\gamma }{2}\left(-\varepsilon_{ijk}B^k\right)\Box\left(-\varepsilon_{ijl}B^l\right)&\\=
\frac{1}{2}B_k B^k+\gamma B_k\Box B^k\,.
\end{align}
One can see that the first term in the equation above is positively defined due to it being quadratic. Applying the same line of reasoning that we did for momentum term, one can not draw solid conclusions on the sign of the correction term. 

One can notice that the sign changing terms in Eq.\eqref{Hdensity}  are both proportional to $\gamma=\frac{\gamma_0}{(M_{\mathrm{Pl}}\,c)^2}$. If one assumes that $\gamma_0$ is of the order of one, the coefficient is $\gamma\sim 10^{-38}$. Therefore for energies smaller than the Planck energy $E<E_{Pl}$, the leading order terms in  Eq.\eqref{Hdensity} are several orders of magnitude bigger than the quantum gravity corrections. One can then conclude that the Hamiltonian density is positively defined for energies smaller than Planck energy $E<E_{Pl}$. In conclusion the proposed gauge theory is an effective field theory constructed for phenomenological studies of low energy effects of minimum measurable length on Quantum field theories. Therefore, all applications must be considered in the $E<E_{Pl}$  domain. 

\section{Summary}
      In this appendix the author has shown that the QFT with minimum length formulated in this thesis has no Ostrogradsky instability, when considered in the domain $E<E_{Pl}=M_{Pl}c$. This is a sign that the formulation is an effective theory. The fact that the theory is an effective field theory is further solidified by the fact that effects like the non-commutativity of space-time at high energy are not considered. Fact worth mentioning is that, a domain in which the theory has a positive definite Hamiltonian density at all, is due to the asymmetry of the derivatives in the correction  terms. Additionally all applications considered  are well in that domain.
      
Another feature of this formulation of Quantum Field Theory with minimum length worth mentioning is that, when one takes the limit $\gamma\rightarrow 0$ all the results  predicted by the usual formulation of Quantum Field Theory are recovered. 
In addition, one can notice that there are repeating patterns in the calculations for fields of different spins. This can be taken as a sign of the robustness of the method used.

\newpage

\end{document}